\documentclass[11pt,a4paper]{book}
\usepackage{graphicx}
\usepackage{epsfig}
\usepackage{psfrag}
\usepackage{bbm}
\usepackage{amsmath}
\usepackage{exscale}
\author{Dissertation zur Erlangung des Doktorgrades\\
vorgelegt von\\Joachim Tabaczek}
\title{Simulations of Discrete Random Geometries:\\
Simplicial Quantum Gravity\\and\\Quantum String Theory}
\begin{document}

\frontmatter

\maketitle

\noindent
I investigate two discrete models of random geometries, namely simplicial
quantum gravity and quantum string theory.

In four-dimensional simplicial quantum gravity, I show that the addition of
matter gauge fields to the model is capable of changing its phase structure
by replacing the branched polymers of the pure gravity model with a new phase
that has a negative string susceptibility exponent and a fractal dimension
of four. Some of the results are derived from a strong coupling expansion of
the model, a technique which is used here for the first time in this context.

In quantum string theory, I study a discrete version of the {\em IIB}
superstring. I show that the divergences encountered in the discretization
of the bosonic string are eliminated in the supersymmetric case. I give
theoretical arguments for the appearance of one-dimensional structures in
the region of large system extents that manifest as a power-law tail in
the link length distribution; this is confirmed by numerical simulations
of the model. I also examine a lower-dimensional version of the {\em IKKT}
matrix model, in which a similar effect can be observed.

\tableofcontents

\mainmatter

\chapter{Introduction}

If we wanted to give a very general description of what we know about
fundamental physics today, it would look something like this: There are
four basic forces that shape the universe, namely electromagnetism, strong
and weak interaction, and gravity. Each of these can individually be cast
into a theoretical formalism that has been thoroughly tested and confirmed
experimentally. Furthermore, electromagnetism, the weak interaction, and
the strong interaction can all be joined in the more general formalism of
the standard model of particle physics. The same, however, cannot be said
about gravity.

Quantum theory and the theory of general relativity, even though they
are both well tested in their respective areas of applicability,
nevertheless seem to be incompatible with each other in a fundamental
way that has so far defied all attempts to overcome the differences.
This is not in itself a contradiction because the areas of applicability
of both theories do not overlap, at least not in those parts of the
universe that lend themselves to observation; generally speaking,
quantum theory is important only on very small scales, whereas gravity
effects are so weak that they can normally be observed only on very large
scales. (In fact, gravity can be noticed at all only because, contrary to
the other forces, it is both universally attractive and, by virtue of the
graviton being massless, a long-range force.) This could be taken as a
hint that both theories may be only effective models of a more general
underlying theory that would consistently describe all four interactions.
This theory, however, has yet to be found.

The problem is one that has been known for a long time, and the attempts
to solve it are manifold. Invariably, however, given the inherent randomness
of quantum theory and the equally inherent geometrical structure of general
relativity, all these different approaches have one thing in common: they
are all, in one way or another, theories of random geometries. In the models
that will be of interest in this thesis, this fact manifests itself in a path
integral over fluctuating spaces, described either by external coordinates
(assuming a flat embedding space) or an internal metric. In particular, our
focus will be on Euclidean quantum gravity, where the action of general
relativity is put into a path integral, with the integration to be taken over
all possible metrics of a four-dimensional space-time; and on quantum string
theory, where a path integral is taken over all possible world-sheets of the
string.

However, the solution, and in some cases the very definition, of a path
integral over surfaces is a difficult and mostly still unfinished problem.
One reason is that any mathematical description of a surface, whether in
terms of external or internal coordinates, is invariant under a
reparametrization of these coordinates. This invariance has to be fixed in
some way to ensure that only physically distinct surfaces are counted in
the integral; unfortunately, there exists as yet no recipe for how to do
this in three or more dimensions. In superstring theory, where the surfaces in
question are two-dimensional, the path integral can be defined as an
integral over the conformal factor, but it still cannot be solved except
in an embedding space of dimension $d \le 1$. Alternatively, superstrings
can be studied in the critical dimension $d = 10$, where the surface integral
decouples from the rest of the theory; however, one then has to find a way
of arguing away the excess dimensions, a task that so far has not been
conclusively solved, various compactification prescriptions notwithstanding.

A different possibility for treating an integral over metrics lies in
trying to find an appropriate regularization. In particular, what we will
be interested in is a discretization of the surfaces in question, $i.\,e.$
we want to replace a continuous space-time or world-sheet with one built from
finite elementary pieces. Such a discretization has already been attempted
for both of the models in question; however, in both cases problems have
emerged that prevent us from taking the model to a well-defined continuum
limit. The aim of this thesis will be to examine possible modifications of
these models in an attempt to improve their behaviour.

The remainder of this text is divided into four parts. The first of these is
a general introduction to the practical issues of how to discretize an
integral over metrics. Each of the following two parts deals with one of the
mentioned models, respectively simplicial quantum gravity and quantum string
theory. The results of both studies will be summarized in the final part.

\chapter{Dynamical triangulations}

This chapter describes the basic techniques that will allow us to perform
simulations of random geometries on a computer, independently of the actual
model that we want to study. In principle, this is about taking an integration
measure $\int {\cal D} g_{\mu\nu}$ over fluctuating metrics, discretizing
it in terms of a fluctuating lattice, and putting it into a Monte Carlo
algorithm. The integrand -- $i.\,e.$ the action, which holds the physical
content of the model -- will be inserted later.

\section{Simplicial manifolds}

First of all, we need to build the lattice itself, and define a metric
on it. {\em A priori}, there is nothing to prevent us from choosing an
ordinary hypercubic lattice \cite{hyper}; in fact, simulations of random
geometries can and have already been successfully performed in this way
\cite{morehyper, evenmorehyper}. However, for purposes of simulating models
on fluctuating lattices, hypercubic structures are actually not the simplest
possible choice. Instead, we are led to consider triangulations, lattices
built from triangles or their higher-dimensional generalizations.

\subsection{Definition of a simplicial manifold}

Define a {\em $d$-dimensional simplex} or {\em $d$-simplex}
$\langle i_1 ... i_{d+1} \rangle$ as a set of $d + 1$ points
that are pairwise connected by links.
Specifically, a 0-simplex $\langle i \rangle$ is defined as a single
point or vertex; a 1-simplex $\langle ij \rangle$ is a line; a 2-simplex
$\langle ijk \rangle$, a triangle; and a 3-simplex $\langle ijkl \rangle$,
a tetrahedron. A 4-simplex $\langle ijklm \rangle$, the highest-dimensional
construct we will be needing, is a four-dimensional analogue of the
tetrahedron that can no longer be easily visualized.

\begin{center}
\includegraphics[height=1.8cm]{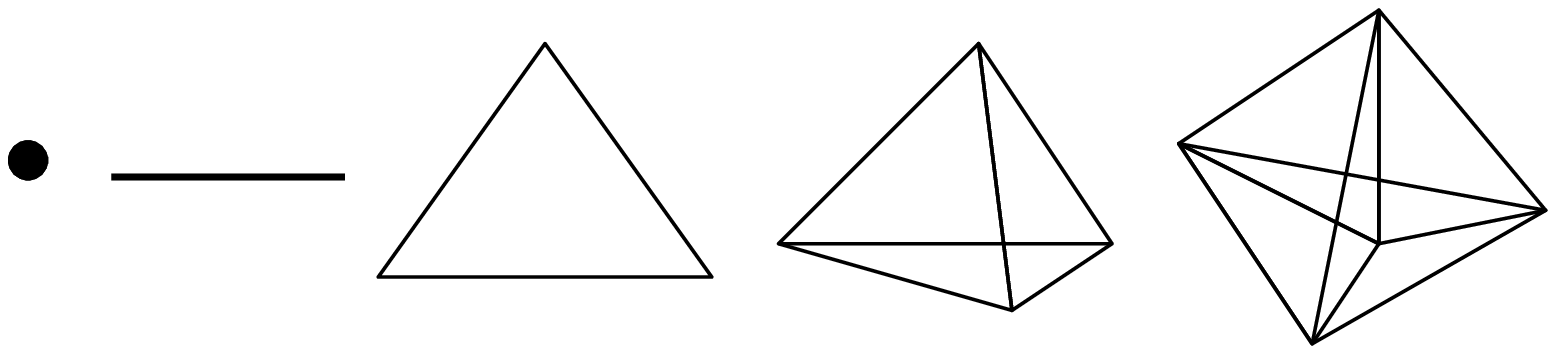}
\end{center}

\noindent
Any subset $\langle i_{\alpha_1} \ldots i_{\alpha_{k+1}} \rangle$ of
vertices of a given $d$-simplex $\langle i_1 \ldots i_{d+1} \rangle$,
along with all connecting links between these vertices, forms a
{\em $k$-dimensional subsimplex} or {\em $k$-subsimplex} of
$\langle i_1 \ldots i_{d+1} \rangle$.
In particular, the $(d - 1)$-subsimplices of a $d$-simplex are called
its {\em faces}. Obviously, any $d$-simplex contains exactly
$\binom{d + 1}{k + 1}$ $k$-subsimplices. In particular, a $d$-simplex
always has $d + 1$ faces.

Two simplices are said to be {\em connected} if they share at least one
subsimplex. We will be particularly interested in simplices that are
connected by a common face, but in general any kind of connection is
possible; we could join several tetrahedra at a common vertex, or even
connect simplices of different dimensions, such as gluing a triangle
to a tetrahedron so that they share a common link. Any set of
connected simplices is called a {\em simplicial complex}.

\begin{center}
\includegraphics[height=1.8cm]{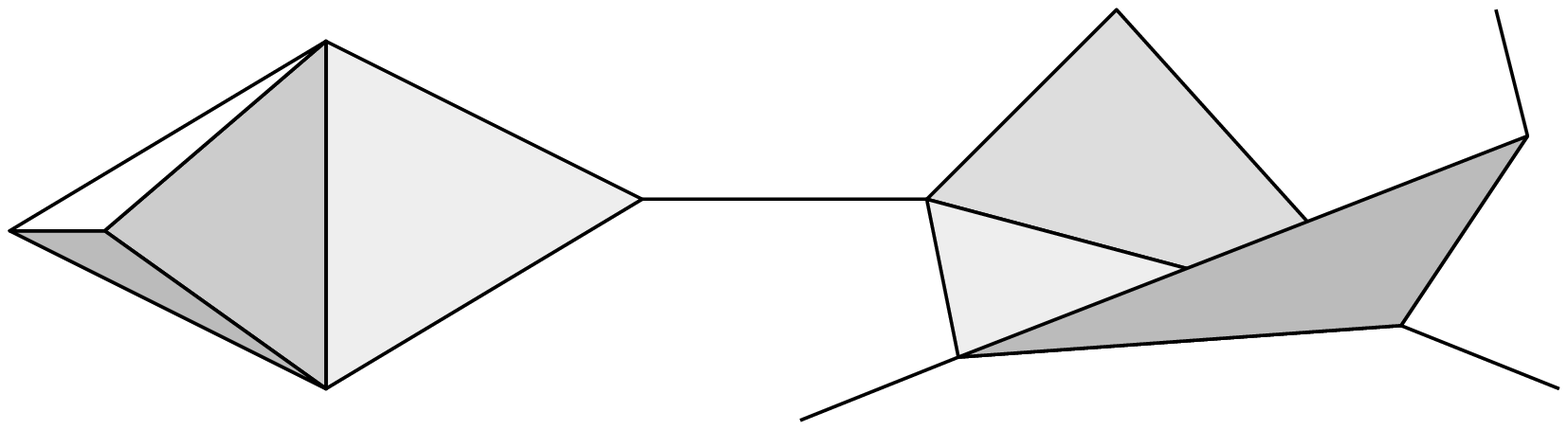}
\end{center}

\noindent
In principle, a simplicial complex is already serviceable as a lattice;
at any rate, we can define fields and derivatives on it \cite{summerdavid}.
However, it does not bear much resemblance to any kind of physical space;
in particular, we can in most cases not even assign a dimension to it.
If we want a simplicial complex to function as the discrete analogue to
some $d$-dimensional space, we need to impose some sort of manifold condition.

To this end, first define the {\em $d$-dimensional neighbourhood} of a given
subsimplex $\langle i_1 \ldots i_{k+1} \rangle$ as the set of all
$d$-simplices that contain $\langle i_1 \ldots i_{k+1} \rangle$.
For example, on a tetrahedron the two-dimensional neighbourhood of a
vertex consists of the three triangles that meet at this vertex, while
its three-dimensional neighbourhood is just the tetrahedron itself.
The size of the highest-dimensional neighbourhood that exists around
a given subsimplex is called that subsimplex's {\em order}
$o_{i_1 \ldots i_{k+1}}$.

With this, we can now define a {\em $d$-dimensional simplicial manifold}
or {\em triangulation}, in analogy to a continuum manifold, as a simplicial
complex $S$ with the additional requirement that the $d$-dimensional
neighbourhood of any vertex $\langle i \rangle \in S$ be homeomorphic to a
$d$-dimensional ball.

\begin{center}
\includegraphics[height=2.5cm]{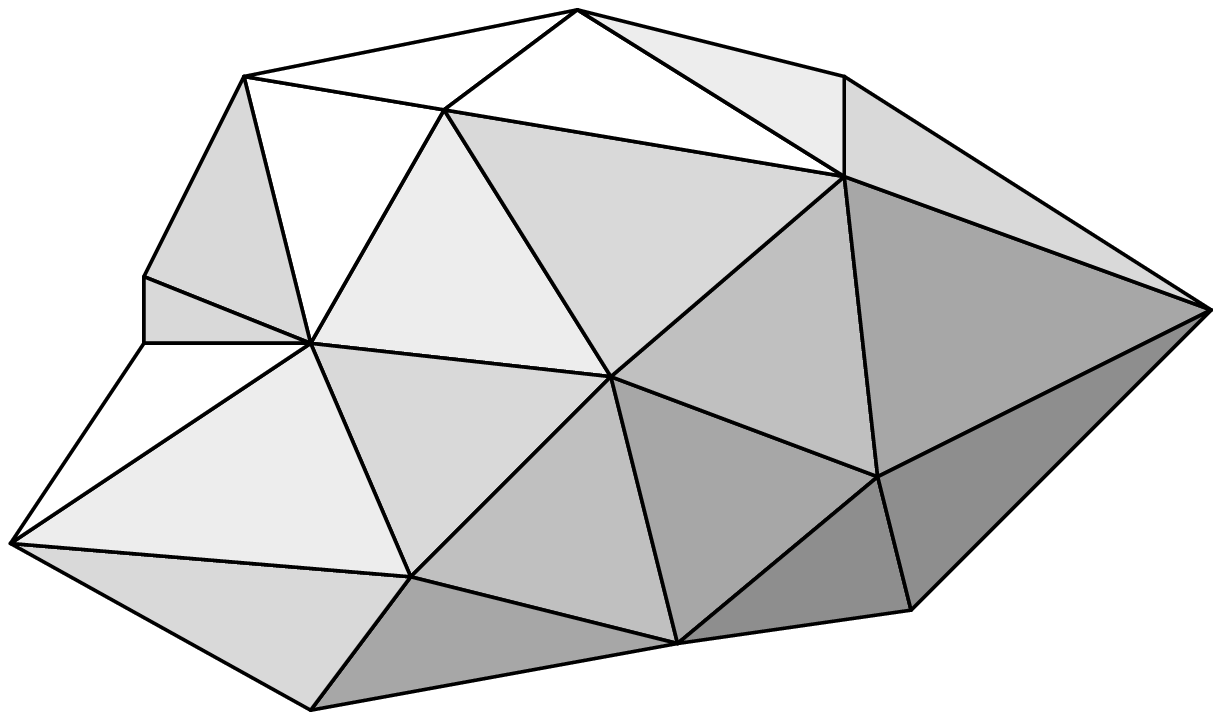}
\end{center}

\noindent
In $d \le 2$ dimensions, this condition can be fulfilled simply by
taking a number of $d$-simplices and gluing them together along their
faces in such a way that each face belongs to exactly two $d$-simplices.
In more than two dimensions, this process in general creates only a
{\em pseudo-manifold}, where not every vertex's neighbourhood is
homeomorphic to a ball. Fortunately, however, the method by which we
will create configurations in the numerical algorithm is guaranteed to
produce actual simplicial manifolds as long as we can provide a starting
configuration of this kind.

\subsection{The discrete metric}

Once we know how to build the lattice, the next step is to define a metric
on it in such a way that we can choose a well-defined integration measure
${\cal D} g_{\mu\nu}$. Usually, this is done by demanding the following:

\begin{itemize}

\item The metric is flat inside the $d$-simplices.

\item The metric remains continuous when moving from one $d$-simplex
to another by crossing a face.

\item The faces are flat linear subspaces of the $d$-simplices that
contain them.

\end{itemize}

\noindent
In this way, the curvature on the lattice can be non-zero only in the
$(d - 2)$-subsimplices -- the vertices on a two-dimensional triangulation,
or the triangles in $d = 4$. Such a metric is called, for obvious reasons,
{\em piecewise flat}.

\subsubsection{The integration measure}

What, now, remains of an integral over metrics in such a piecewise
flat space? In principle, there are still two things whose variation
can change the metric: the connectivity of the simplices, and the
lengths of the links. In general, we would therefore replace the
integral over metrics by a sum over triangulations and an integral
over link lengths,
\begin{equation}
\int {\cal D} g_{\mu\nu}
\to \sum_T W (T) \prod_{\langle ij \rangle} {\cal D} l_{ij}
\label{completedisc}
\end{equation}
where the weights $W (T)$ and the measure ${\cal D} l_{ij}$ have still
to be defined.

In practice, one does not actually use the general form (\ref{completedisc}),
but restricts oneself to varying either the connectivity or the link
lengths. The latter method, integration over link lengths on a fixed
triangulated lattice, is the one closest to the original idea by Regge
\cite{regge}, and is therefore known as {\em Regge calculus}, or alternatively
{\em fixed triangulations}.\footnote{It should be mentioned that Regge
suggested this method as a way of treating classical gravity only; the idea
of using it to regularize a path integral over fluctuating surfaces was not
formed until much later \cite{quantumregge}.} Unfortunately, it has a number
of problems; among other things, reparametrization invariance is still a
serious difficulty in the choice of the measure ${\cal D} l_{ij}$, and so
far, all attempts to apply Regge calculus to two-dimensional simplicial
quantum gravity have produced results that are in variance to those of the
analytic solution of the continuum theory that exists in this case
\cite{badregge, morebadregge}.

The other method, and the one we will use here, is called {\em dynamical
triangulations} \cite{dynatri, moredynatri, evenmoredynatri} and consists
of summing over all possible connectivities
while keeping all link lengths fixed and, for simplicity, equal to each
other. This leaves us with the responsibility of deciding on the weights
$W (T)$, which have to be chosen in such a way as to avoid any overcounting
of equivalent configurations due to reparametrizations. Fortunately, any
two different connectivities automatically represent physically distinct
metrics \cite{connmetric}, so the only thing we have to worry about is the
possible existence of different parametrizations of the same connectivity.

This possibility depends on exactly how we describe a triangulation. If
we use {\em labeled triangulations}, where each vertex, link, triangle,
and so on gets a specific `name' attached to it, then the number of
different possible descriptions of a given connectivity is
$n_0! n_1! \ldots n_d!$, where $n_i$ is the number of $i$-simplices on
the triangulation. Each of these labelings can be thought of as a different
parametrization of the same geometrical structure, and we would therefore
divide out the number of all possible labelings in the sum over
triangulations; in other words, for labeled triangulations we would choose
the weights as $W (T) = \frac{1}{n_0! n_1! \ldots n_d!}$.

Many possible labelings of a given connectivity can be mapped onto
each other by a permutation of their labels, which means that they all
correspond to the same {\em unlabeled} configuration. In fact, one might
naively assume that there should be exactly one unlabeled configuration
for each distinct connectivity, giving us a constant factor of
$n_0! n_1! \ldots n_d!$ between labeled and unlabeled triangulations.
However, this is true only if the configuration is not symmetric. If there
exist two vertices that are both connected to exactly the same neighbouring
vertices, then these two can be exchanged without altering the connectivity;
but the exchange still gives us a new unlabeled configuration, because now
we are actually exchanging vertices and not just labels. In general, if we
denote the number of possible vertex exchanges by the {\em symmetry factor}
$C(T)$, with $C(T) = 1$ for a completely non-symmetric configuration, we
can write the number of labeled configurations that correspond to the same
unlabeled one as $\frac{n_0! n_1! \ldots n_d!}{C(T)}$. Thus, we would for
unlabeled triangulations change the weights to
$W(T) = \frac{1}{n_0! n_1! \ldots n_d!}
\frac{n_0! n_1! \ldots n_d!}{C(T)} = \frac{1}{C(T)}$.

Since all we really care about is the geometrical structure of a
configuration and not some arbitrary choice of labels, we will use
unlabeled triangulations.\footnote{This might appear inconsistent with
the fact that I use sets of labels $\langle i_1 \ldots i_{d+1} \rangle$
to represent simplices. However, this is only a convenience of notation.
It is always implied that any different labeling of the same simplex is
to be regarded as equivalent.} Our discretization prescription for an
integral over metrics should therefore be
\begin{equation}
\int {\cal D} g_{\mu\nu} \to \sum_T \frac{1}{C(T)}
\label{substi}
\end{equation}
So far, we have not yet specified which set of triangulations ${\cal T}$
we intend to use. Trying to actually sum over {\em all} triangulations,
including all possible topologies, does not seem feasible because the
number of distinct triangulations is estimated to grow factorially with
the number of simplices, whereas we cannot expect the action, in any
interesting model, to provide more than an exponential suppression of
configurations. To obtain a well-defined sum over triangulations we
should therefore restrict ourselves to a set that is at least exponentially
bounded. This seems to be the case if we restrict ourselves to just one
topology \cite{topodisc}. We will again go for the simplest case first
and study only configurations with spherical topology (no `holes'). For
simulations of models with different topologies, see for example
\cite{toposimp, moretoposimp}.

Of course, it should be clear from the preceding discussion that the
substitution (\ref{substi}) is not the only possible discretization of
an integral over metrics, nor have we shown yet that it is actually a good
one. For example, we do not know whether our chosen set of discrete metrics
really provides us with a good representation of the complete space of
continuous metrics. For now, we simply deal with a particular discrete model,
and still have to show that it reproduces the original theory in a suitable
continuum limit. Justification for the hope that this might indeed be the
case comes mainly from simulations of two-dimensional simplicial quantum
gravity, where both the continuum model and the triangulated version can be
solved analytically, and the results shown to agree with each other. More
on this in chapter 3.

\section{Numerical simulations}

Even though the replacement of the integral over metrics by a sum over
triangulations greatly simplifies matters, it is in most cases still not
enough for an analytic treatment of the model in question. Instead,
numerical methods are used to extract information about it. In this
section, I will shortly describe the basics of the Monte Carlo algorithm
in general, and its application to triangulations in particular.

\subsection{Monte Carlo algorithms}

Generally speaking, a Monte Carlo simulation is a numerical method for
estimating the value of an integral, typically with many degrees of freedom,
that cannot be solved exactly. In its simplest version, it consists of
randomly sampling the integration space, evaluating the function in question
in the chosen points, and using the average of these values as an
approximation for the real integral. In other words, we estimate an integral
of some function $f$ over an integration space $I$ by
\begin{equation}
\frac{1}{A} \int_{I} d^d x \, f(x)
\approx \frac{1}{N (I_0)} \sum_{x \in I_0} f (x)
\end{equation}
where $I_0$ is the random sample of the integration space, $N (I_0)$
is the number of points in this sample, and $A \equiv \int_I d^d x$
is the volume of the integration space (assuming, of course, that it
is finite).

Using this formula on the special case of a discrete sum over
triangulations weighted with an action $S(T)$ gives us
\begin{equation}
\frac{1}{N ({\cal T})} \sum_{T \in {\cal T}} e^{-S(T)}
\approx \frac{1}{N ({\cal T}_0)} \sum_{T \in {\cal T}_0} e^{-S(T)}
\end{equation}
Likewise, we can estimate the expectation value of some observable
${\cal O}$ as
\begin{equation}
\langle {\cal O} \rangle
\equiv \frac{\sum_{T \in {\cal T}} O(T) \ e^{-S(T)}}
{\sum_{T \in {\cal T}} e^{-S(T)}}
\approx \frac{\sum_{T \in {\cal T}_0} O(T) \ e^{-S(T)}}
{\sum_{T \in {\cal T}_0} e^{-S(T)}}
\label{obsinmonte}
\end{equation}
where $N ({\cal T})$ and $N ({\cal T}_0)$ drop out because of the
normalization factors.

\subsubsection{Importance sampling}

So far, we have assumed that the random sample is chosen from a flat
distribution, $i.\,e.$ each point in the configuration space is picked
up with equal possibility. While this is a valid approach that eventually
leads to correct results, it also tends to produce large statistical
errors, especially if we have an exponential weight as in (\ref{obsinmonte}).
In this case, only a small part of the configuration space is actually
significant for the result, which means that the algorithm will pick up
unimportant configurations almost all of the time, leading to extremely
long simulation times.

The way around this is to pick each configuration not from
a flat distribution, but with a probability that is given by its weight,
$p(T) \sim e^{-S(T)}$. This ensures that those configurations that
contribute the most to the
overall sum are also the most likely to be picked, whereas configurations
with a weight near zero will be ignored most of the time. The estimate of
an observable's expectation value $\langle {\cal O} \rangle$ simplifies
in this case to
\begin{equation}
\langle {\cal O} \rangle \approx \frac{1}{N ({\cal T}_0)}
\sum_{T \in {\cal T}_0} {\cal O} (T)
\end{equation}
However, even for an only moderately complicated action it is in general
not possible to simply generate this distribution. What can be done instead
is to choose configurations from a {\em Markov chain} that has $p(T)$ as
its static distribution.

A Markov chain describes the evolution of a statistical model whose state
at any given point $t$ in the chain depends only on its state at the
previous point, $t - 1$. Thus, a Markov chain is fully characterized by
a set of transition probabilities $P (A \to B)$ for all pairs of
configurations $(A, B)$, plus a set of starting probabilities $p (A)$.
We also demand that the following three conditions be fulfilled:

\begin{itemize}

\item The chain is {\em irreducible}. This means that if we denote by
$P^i (A \to B)$ the probability of the chain reaching a configuration
$B$ from $A$ in exactly $i$ steps, then there exists for any pair of
configurations $(A, B)$ a finite $i$ such that $P^i (A \to B) > 0$.

\item The chain is {\em aperiodic}. This means that for any configuration
$A$ and any finite number $i$, there exists a non-vanishing probability
of the chain returning to $A$ after exactly $i$ steps, $P^i (A \to A) > 0$.

\item The transition probabilities are {\em stationary}. This means
that there exists a probability distribution $\pi$ such that
\begin{equation}
\sum_A \pi (A) P (A \to B) = \pi (B)
\label{stationary}
\end{equation}
is true for all configurations $B$.

\end{itemize}

\noindent
If this is assured, it can be shown that the $i$-step transition
probabilities $P^i (A \to B)$ converge to the stationary distribution
$\pi (B)$ independently from where we start,
\begin{equation}
\lim_{i \to \infty} P^i (A \to B) = \pi (B) \quad \forall A
\end{equation}
Thus, coming back to our specific task, we can use a Markov chain to
create a set of randomly chosen configurations $A$ weighted by $e^{-S(A)}$
if we can just choose the transformation probabilities $P (A \to B)$
in such a way that we have a stationary distribution $\pi (A) \sim e^{-S(A)}$.
A sufficient condition for this can be found in the {\em detailed balance
equation}
\begin{equation}
\pi (A) P (A \to B) = \pi (B) P (B \to A)
\label{detbal}
\end{equation}
which implies (\ref{stationary}), as can be seen by taking the sum over
all $A$ on both sides.

One possible choice of $P (A \to B)$ that fulfils (\ref{detbal}) is the
following:
\begin{equation}
P (A \to B) = \min \left( 1, \frac{\pi (B)}{\pi (A)} \right)
= \min \left( 1, e^{S(A) - S(B)} \right)
\label{metropolis}
\end{equation}

\subsubsection{The Metropolis algorithm}

To summarize, we now have the following recipe for `measuring' the
expectation value of an observable ${\cal O}$ in a given triangulated model:

\begin{enumerate}

\item Choose an arbitrary triangulation as a starting point.

\item Propose a random transformation to a different triangulation;
accept or reject it with a probability given by (\ref{metropolis}).
Repeat this step sufficiently often so that the chain comes close to
its stationary distribution.\footnote{How do we know what `sufficiently
often' means? There is no mathematically exact criterion for this, but
we can give a good estimate of the required number of steps by measuring
the integrated autocorrelation time $\tau$ while running the simulation. If
we perform a number of transformations equal to a few times $\tau$, we
can be reasonably confident that any dependence on the starting
configuration is gone, which in turn can be taken to mean that we have
come close enough to the stationary distribution.}

\item Calculate ${\cal O}$.

\item Repeat steps 2 and 3 until we have a reasonably large sample
to estimate $\langle {\cal O} \rangle$.

\end{enumerate}

\noindent
This procedure is known as the {\em Metropolis algorithm}.

\subsection{Transformations}

The one thing we still need to implement a simulation of dynamical
triangulations is a set of transformations that change the connectivity
of the simplices. It is clear from the discussion above that to be of any
use in a Metropolis algorithm, this set has to be {\em ergodic}, $i.\,e.$
it must be possible to reach any configuration from any other by applying
a finite number of these transformations. From a purely practical
viewpoint, we would also like this set to be as small and simple as possible.

The transformations that are best suited to these purposes are the
so-called {\em $(p, q)$ moves}. Their general definition is the following:
On a $d$-dimensional simplicial manifold, choose $p$ $d$-simplices
that are connected to each other in such a way that they form part of a
$d$-dimensional minimal sphere. (A minimal sphere is the smallest possible
simplicial manifold of a given dimension; it consists of $d + 1$ simplices
that are pairwise connected by a face, so that they form the surface of
a $(d + 1)$-simplex. For example, a two-dimensional minimal sphere is the
surface of a tetrahedron.) Then replace this complex by the $q = d - p + 2$
simplices that would complete the minimal sphere. Obviously, in this way
one arrives at $d + 1$ different possible moves, which together can be
shown to be ergodic, at least in dimensions $d \le 4$
\cite{ergomoves}.\footnote{The actual proof is rather long and depends
on showing the equivalence of $(p, q)$ moves with another kind of
transformations called Alexander moves that are known to be ergodic
\cite{alexmoves}.} Note that move $(p, q)$ is the inverse of $(q, p)$,
$i.\,e.$ using move $(q, p)$ on exactly those simplices created by move
$(p, q)$ leads back to the original configuration.

To better illustrate this rather abstract definition, here is what the
$(p, q)$ moves look like in detail in the cases that interest us, namely
two and four dimensions.

\subsubsection{The $(p, q)$ moves in two dimensions}

In $2d$, we have three moves: $(1, 3)$, $(2, 2)$, and $(3, 1)$.

\begin{center}
\includegraphics[height=1.5cm]{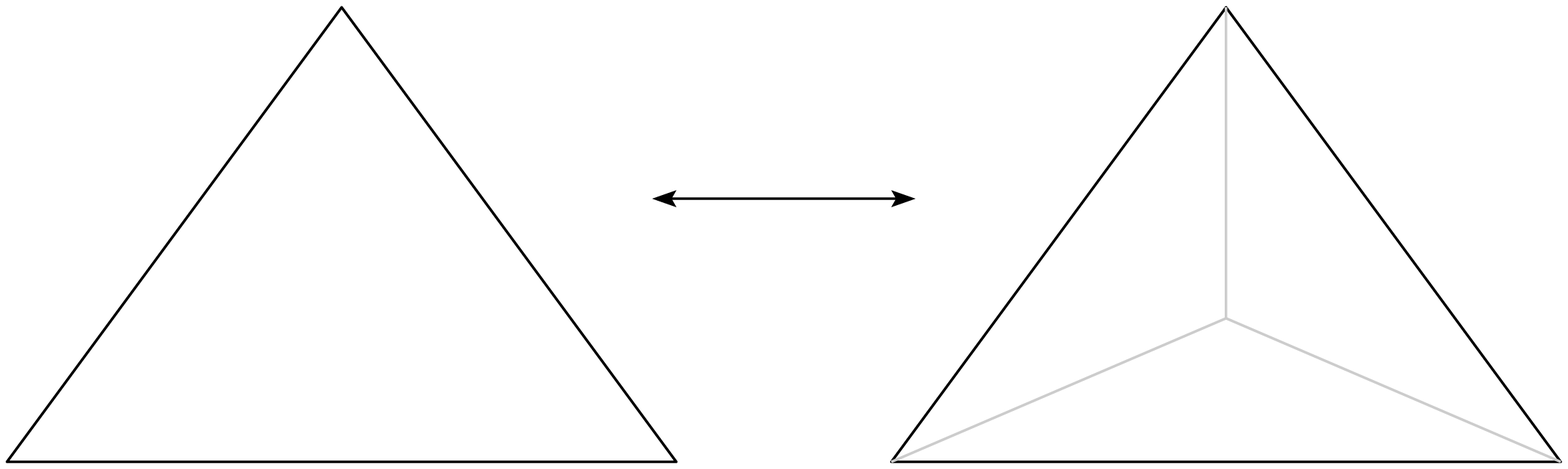} \\
{\small $\langle ijk \rangle \ \longleftrightarrow \
\langle ijl \rangle \langle ikl \rangle \langle jkl \rangle$}
\end{center}

\noindent
$(1, 3)$: Take any triangle, insert a new vertex in its centre, and connect
this point to all the vertices on the triangle. This creates three new
triangles, while the original one is removed from the configuration.
$(3, 1)$: The reverse of $(1, 3)$; take three triangles that share a common
vertex of order 3, then remove this vertex and all links connected with it,
thus replacing the three triangles with one new 2-simplex.

\begin{center}
\includegraphics[height=2cm]{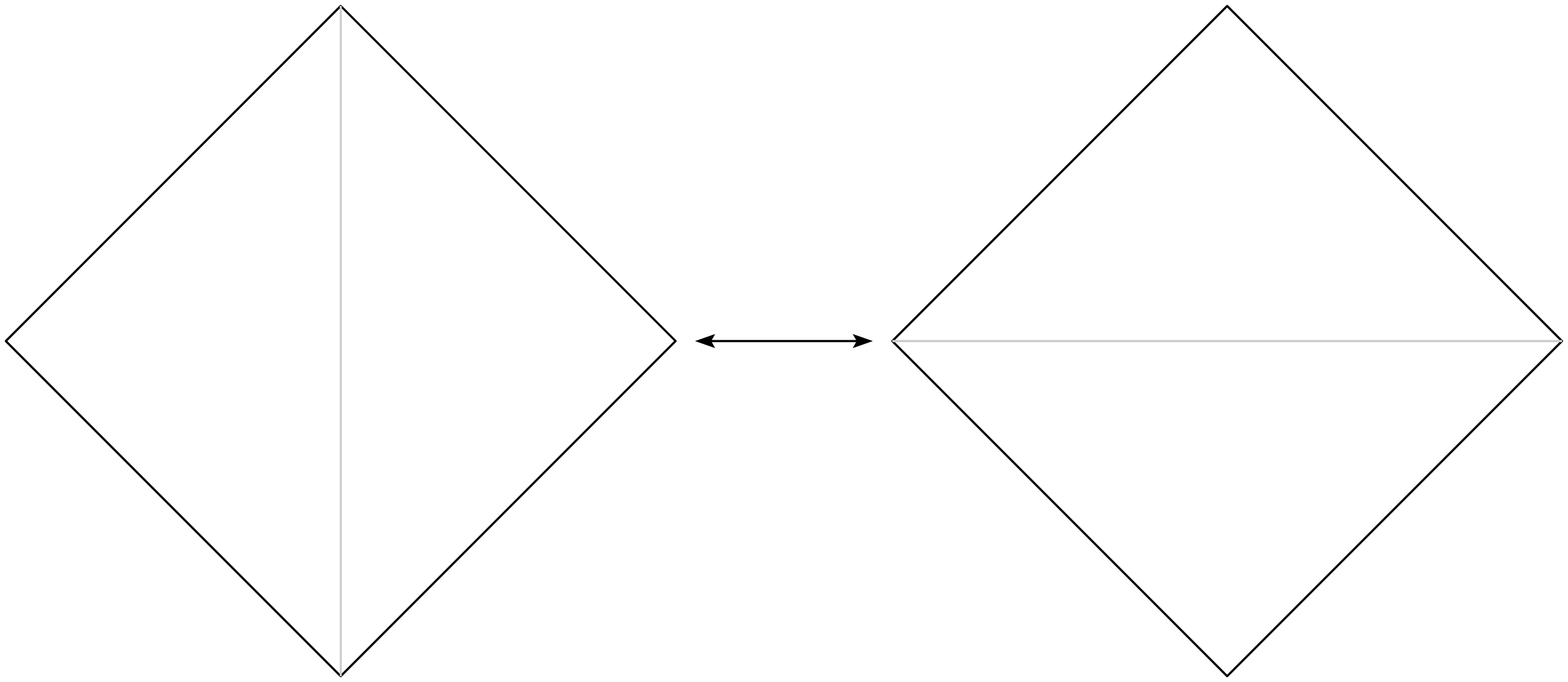} \\
{\small $\langle ijk \rangle \langle ijl \rangle \ \longleftrightarrow \
\langle ikl \rangle \langle jkl \rangle$}
\end{center}

\noindent
$(2, 2)$: Take two triangles that share a common link, remove this link,
and replace it with a new one connecting the other two vertices. This move
is its own inverse. Note that in a canonical ensemble of two-dimensional
triangulations, this move is ergodic by itself, $i.\,e.$ we can reach any
configuration with $n_2$ triangles from any other that has the same number
of triangles by repeated application of move $(2, 2)$ only
\cite{evenmoredynatri, ergotwo}.

Move $(2, 2)$ is also the first case where we encounter the phenomenon of
so-called {\em degenerate manifolds}, meaning configurations where two
different simplices or sub-simplices are built from the same set of vertices,
$i.\,e.$ lying on top of each other. In this particular case, it is possible
that the two vertices being connected already do have a link in common, so
that after the move they would be connected by two different links. While
this is not necessarily a problem -- simulations including degenerate
manifolds can and have been performed, and apparently lead to similar results
\cite{degenerate, moredegenerate} -- we would nevertheless like to avoid
these configurations. This means that before performing this move, we have
to check whether there already exists a link connecting the two vertices in
question; if so, the move has to be rejected by the algorithm.

\subsubsection{The $(p, q)$ moves in four dimensions}

In $4d$, we have five moves. These are somewhat harder to illustrate than
their two-dimensional counterparts, but the analogies should be obvious.

\begin{center}
\includegraphics[height=2cm]{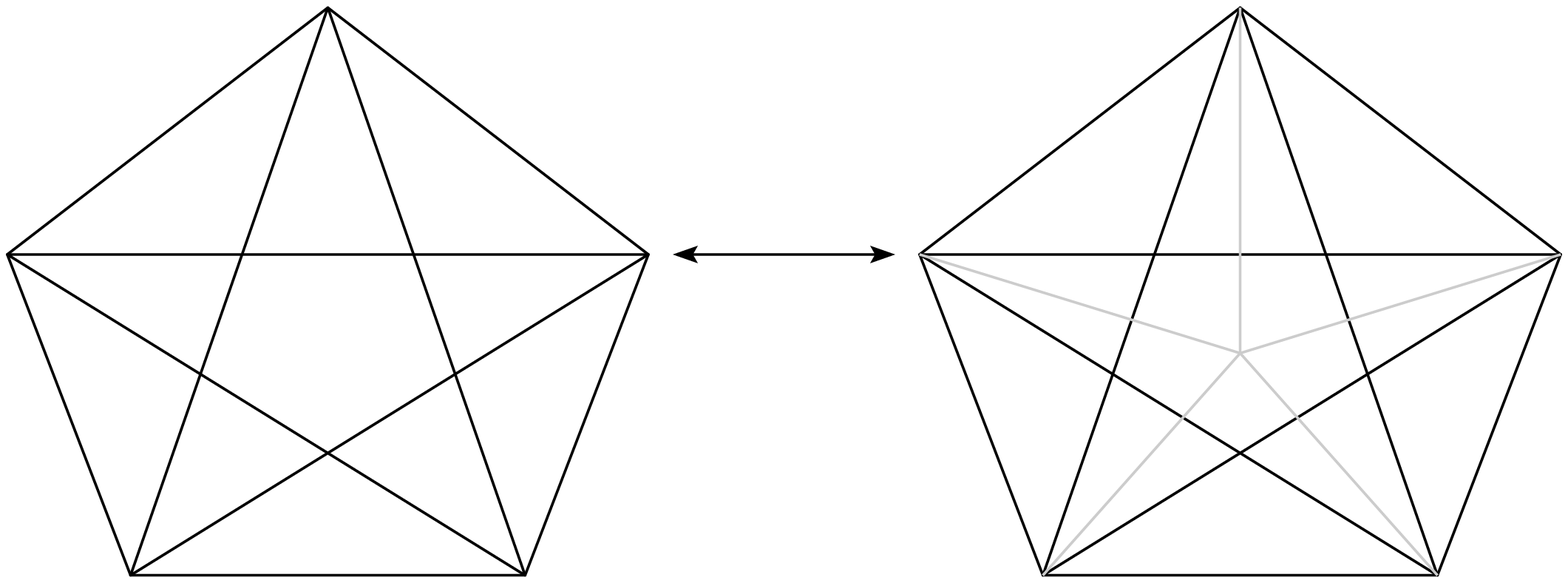} \\
{\small $\langle ijklm \rangle \ \longleftrightarrow \
\langle ijkln \rangle \langle ijkmn \rangle
\langle ijlmn \rangle \langle iklmn \rangle \langle jklmn \rangle$}
\end{center}

\noindent
$(1, 5)$: Choose a 4-simplex, insert a new vertex in its centre, and
connect it with the other vertices, thus destroying the original 4-simplex
and creating five new ones. $(5, 1)$: The inverse of this, where we
take five 4-simplices joined at a vertex of order 5 and remove this vertex,
thus destroying five 4-simplices and creating one.

\begin{center}
\includegraphics[height=2cm]{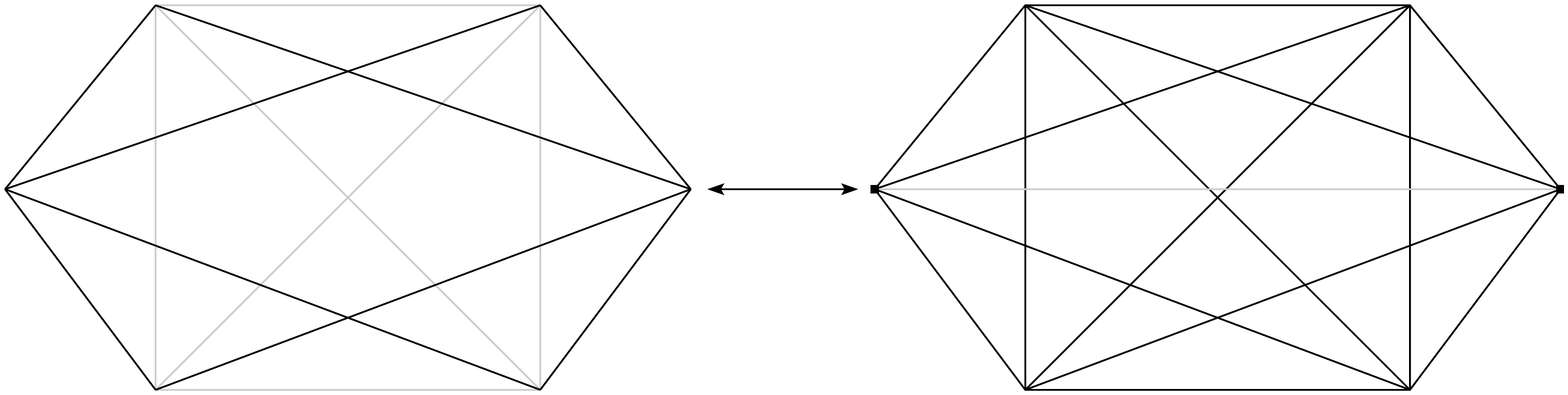} \\
{\small $\langle ijklm \rangle \langle ijkln \rangle \ \longleftrightarrow \
\langle ijkmn \rangle \langle ijlmn \rangle \langle iklmn \rangle
\langle jklmn \rangle$}
\end{center}

\noindent
$(2, 4)$: Choose two 4-simplices that have a common face, take the two
vertices that are not part of this face, and connect them by a link. This
destroys the original 4-simplices (and thus also the tetrahedron that formed
their common face) and creates four new ones. With this move, we again have
to test for a double link to avoid creating a degenerate configuration.
$(4, 2)$: The inverse, where we choose four 4-simplices joined at a link of
order 4, remove the link, and replace it by a tetrahedron joining two new
4-simplices. Again, we have to test whether there is already a tetrahedron
present that contains the same four vertices.

\begin{center}
\includegraphics[height=2cm]{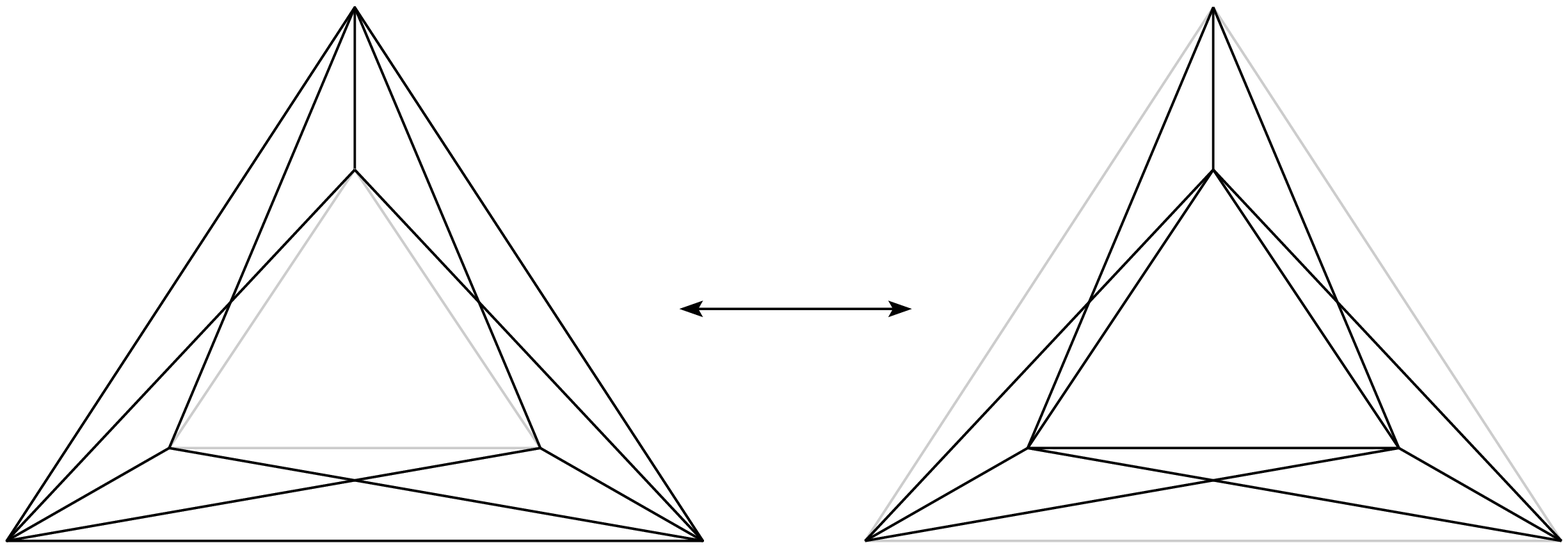} \\
{\small $\langle ijklm \rangle \langle ijkln \rangle \langle ijkmn \rangle
\ \longleftrightarrow \
\langle ijlmn \rangle \langle iklmn \rangle \langle jklmn \rangle$}
\end{center}

\noindent
$(3, 3)$: Choose three 4-simplices that share a common triangle of order
3, remove this triangle, and create a new one from the three vertices that
were not part of the original triangle. This also destroys the original
three 4-simplices and creates three new ones. Again, we have to test for
the presence of a triangle in the place where we are trying to create the
new one. Like $(2, 2)$, this move is its own inverse.

\chapter{The strong coupling expansion in simplicial quantum gravity}

Simplicial quantum gravity is an attempt to quantize Einstein's theory
by taking a path integral over all physically distinct $d$-dimensional
Riemannian surfaces, weighted with the Einstein-Hilbert action of general
relativity. Given the inherent difficulties of defining such an integral over
fluctuating geometries, the model is then discretized by means of dynamical
triangulations in an attempt to make sense of it. This approach can be
directly shown to be valid in $d = 2$ dimensions, which gives us hope that it
might be made to work in the physically interesting case of $d = 4$ dimensions
as well. The situation is more complicated there, however, and as I will
show there are reasons to believe that the addition of matter fields could
be essential for the four-dimensional model to work.

The aim of this chapter is twofold. For one thing, we want to introduce an
alternative method for the evaluation of the partition function, namely the
strong coupling expansion. This method has already been used in the context
of two-dimensional simplicial quantum gravity \cite{ratiomet, moreratiomet},
but has to be adapted to work in the more complicated case of four dimensions.
Secondly, we want to apply this method to study the changes in the model
if it is coupled to matter fields. Also, I will compare the results of the
strong coupling expansion to those found in Monte Carlo studies of the same
model.

\section{Simplicial quantum gravity}

The equations of motion of general relativity are
\begin{equation}
R_{\mu\nu} - g_{\mu\nu} \left( \frac{1}{2} R - \lambda \right)
= 8 \pi G \, T_{\mu\nu}
\end{equation}
where $g_{\mu\nu}$ is the metric tensor, $R_{\mu\nu}$ is the Ricci tensor,
$R \equiv R^\mu_\mu$ is the scalar curvature, $T_{\mu\nu}$ is the
energy-momentum tensor, $\lambda$ is the cosmological constant, and $G$ is
Newton's constant.

These equations can be derived from the action principle using the
Einstein-Hilbert action
\begin{equation}
S_{EH} = \frac{1}{8 \pi G} \int d^d x \sqrt{-|g|}
\left( \frac{1}{2} R - \lambda \right) + S_{matter}
\label{einsthil}
\end{equation}
where $S_{matter}$ is the action for the matter fields, $|g|$ is the
determinant of the metric tensor, and $d$ is the dimensionality of space-time.
(Obviously, to describe our physical universe we should set $d = 4$, but in
principle the model can be defined in any dimension; as we will see, we can
learn much about the dynamics of the four-dimensional model by studying the
two-dimensional one first.)

We can try to quantize this theory by taking the path integral
\begin{equation}
Z = \int {\cal D} g_{\mu\nu} \ e^{-i S_{EH}}
\label{pathint}
\end{equation}
where the integration is to be performed over all physically distinct
metrics with given boundary conditions. As written, this is for now just
a formal expression; even disregarding the question of how to define the
integration measure, we find that we have some difficulties to deal with.

For one thing, we see that for a Lorentzian space-time --
where the metric has a signature $(-+++)$ -- the Einstein-Hilbert action
(\ref{einsthil}) is real, making the integrand of (\ref{pathint})
oscillatory and thus in general rendering the integral ill-defined
\cite{wickrot}. This problem is not unique to the model discussed here;
it is well-known in quantum field theory, where it can be dealt with by
performing a Wick rotation on the metric, replacing the time coordinate by
an imaginary quantity $t \to -i \tau$ and thus giving the metric a Riemannian
signature $(++++)$. In our case, this would result in the following modified
versions of the Einstein-Hilbert action
\begin{equation}
S_{EH} = - \frac{1}{8 \pi G} \int d^d x \sqrt{|g|}
\left( \frac{1}{2} R - \lambda \right) + S_{matter}
\label{einsthil'}
\end{equation}
and the path integral
\begin{equation}
Z = \int {\cal D} g_{\mu\nu} \ e^{-S_{EH}}
\label{pathint'}
\end{equation}
This ansatz is commonly called `Euclidean' quantum gravity, even though
`Riemannian' would be more accurate. One can hope that once a solution
to the Euclidean model has been found, it might be possible to
continue it analytically into the region of Lorentzian metrics. An
example of a case where this method apparently works comes from
semi-classical calculations of the spectrum of Hawking radiation
\cite{hawkrad}, where a treatment of the model in both the Euclidean and
the Lorentzian sector leads to exactly the same thermal radiation
distribution \cite{morehawkrad}.

Even if we accept the Wick rotation as valid, however, we are immediately
confronted with a new problem, namely that the action (\ref{einsthil'})
is not positive definite; in fact, we can
make it arbitrarily large and negative just by choosing a suitably extreme
curvature, and thus cause the path integral to blow up. There is no
obvious cure for this problem, and it therefore seems that, even if we
knew how to choose the measure ${\cal D} g_{\mu\nu}$, the integral would
still have to be regularized in some way.

In this situation, a discretization in terms of dynamical triangulations
looks very promising, since it not only provides us with a definition of
the measure but also, by way of the fixed link lengths, introduces an
upper limit on the curvature, which in turn makes the action bounded
from below.\footnote{One might worry about the effects of an artificial
bound like this on the results of numerical simulations of the model,
since it might seem that the system should simply stick to the upper limit
and thus prevent us from observing the model's dynamics. However, it was
shown that this is actually not the case \cite{nostick, morenostick}.
Specifically, it was found that the distribution of the integrated
curvature exhibits a peak that is near to but still clearly below the
upper bound.} The crucial question, of course, is whether it is possible
to take the resulting model to a continuum limit in which the original
theory is recovered.

\subsection{Discretization of pure gravity}

We will start with a discussion of the model without matter fields,
$T_{\mu\nu} = 0$ (`pure gravity'). Since we already know how we want to
discretize the integral over metrics, all that remains to be done is to
find the discrete version of the action. This was first done in
\cite{discrefirst, morediscrefirst}; here I will only briefly describe
the necessary steps.

\subsubsection{The curvature on a simplicial manifold}

First, we need a definition of curvature on a $d$-dimensional
triangulation. From our choice of the discrete metric, we know that the
curvature can exist only on the $(d - 2)$-subsimplices. For each of these
subsimplices $\langle i_1 \ldots i_{d-1} \rangle$ we can define
$R_{i_1 \ldots i_{d-1}}$, in analogy to the continuum definition, as the
change of direction of a vector after parallel transport around an area that
includes the subsimplex itself but no others. The generic choice for such an
area is the {\em dual polygon} of $\langle i_1 \ldots i_{d-1} \rangle$,
which can be found by going around the subsimplex and connecting the
centres of all $d$-simplices in its neighbourhood.

\begin{center}
\includegraphics[height=3.5cm]{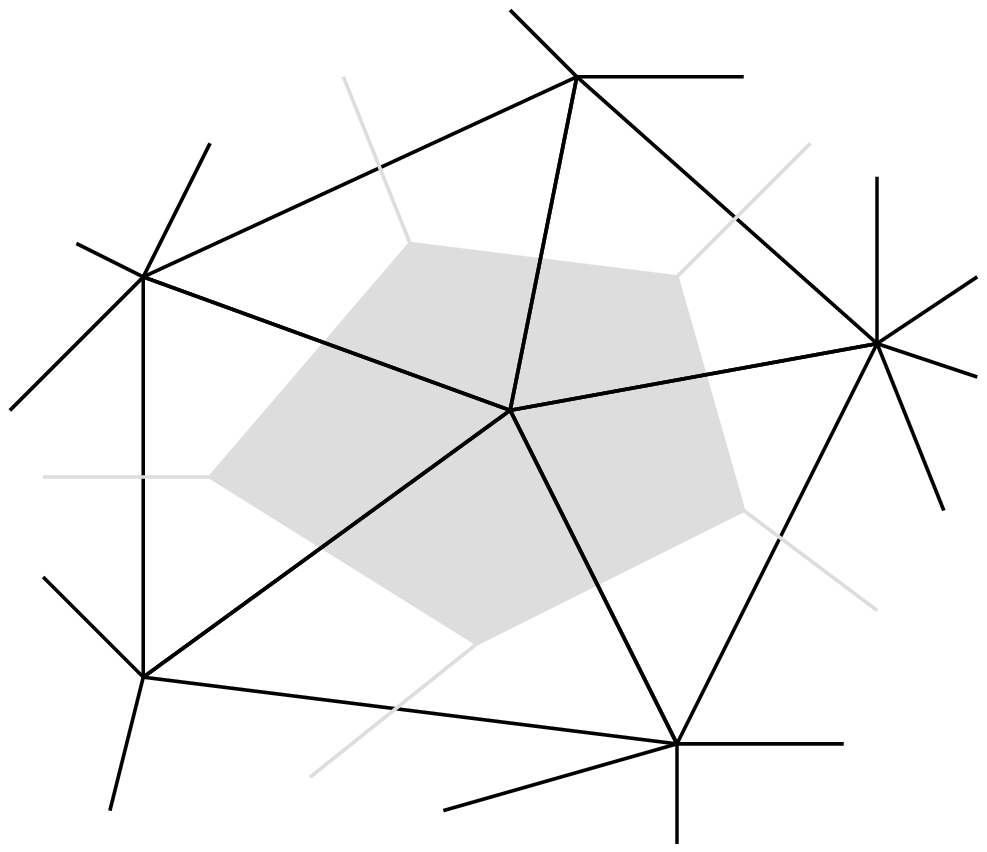} \\
{\small The dual polygon of a vertex on a two-dimensional simplicial manifold.}
\end{center}

\noindent
The change of direction of a vector is then given by the deficit angle
\begin{equation}
\delta_{i_1 \ldots i_{d-1}} = 2 \pi - \hspace{-0.3cm}
\sum_{jk: \langle i_1 \ldots i_{d-1} jk \rangle} \hspace{-0.3cm}
\theta^{i_1 \ldots i_{d-1} j}_{i_1 \ldots i_{d-1} k}
\end{equation}
where $\theta^{i_1 \ldots i_{d-1} j}_{i_1 \ldots i_{d-1} k}$ is the dihedral
angle formed by the two faces $\langle i_1 \ldots i_{d-1} j \rangle$ and
$\langle i_1 \ldots i_{d-1} k \rangle$.

For the special case of dynamical triangulations, this formula becomes
particularly simple: because all links are of equal length, the angle between
two faces of a simplex is just a constant, $\theta_d \equiv \arccos 1/d$,
and we have
\begin{equation}
\delta_{i_1 \ldots i_{d-1}} = 2 \pi - \hspace{-0.3cm}
\sum_{jk: \langle i_1 \ldots i_{d-1} jk \rangle} \hspace{-0.3cm} \theta_d \
= 2 \pi - \theta_d o_{i_1 \ldots i_{d-1}}
\end{equation}
where $o_{i_1 \ldots i_{d-1}}$ is the order of
$\langle i_1 \ldots i_{d-1} \rangle$ as defined in section 2.1.1.
Accordingly, the curvature of a $(d - 2)$-subsimplex can be defined as
\begin{equation}
R_{i_1 \ldots i_{d-1}} \equiv
\frac{2 \pi - \theta_d o_{i_1 \ldots i_{d-1}}}{V^{dual}_{i_1 \ldots i_{d-1}}}
\end{equation}
where $V^{dual}_{i_1 \ldots i_{d-1}}$ is the volume of the dual polygon.
Note that in two dimensions we have $\theta_2 = \pi/3$, which gives us
$R_{i_1 \ldots i_{d-1}} = 0$ whenever $o_{i_1 \ldots i_{d-1}} = 6$,
$i.\,e.$ a two-dimensional triangulation is locally flat around any
vertex of order 6. This is consistent with the well-known fact that we
can completely fill a flat plain with regular hexagons. In four dimensions,
we find an angle $\theta_4 = 1.318116 \ldots$ that is not an integer
divisor of $2 \pi$, which means that it is impossible to make a
four-dimensional triangulation locally flat. However, we can still have
triangulations that are almost flat, if the average triangle order is equal
to $2 \pi / \theta_4$ and the variance is small.

\subsubsection{The partition function of simplicial quantum gravity}

Now all the necessary tools for discretizing the integral (\ref{pathint'})
are assembled; we just have to put the building blocks together.
We replace
\begin{eqnarray}
\int d^d x \sqrt{g} R & \to & \hspace{-0.3cm}
\sum_{\langle i_1 \ldots i_{d-1} \rangle} \hspace{-0.2cm}
V^{dual}_{i_1 \ldots i_{d-1}}
\frac{2 \pi - \theta_d o_{i_1 \ldots i_{d-1}}}{V^{dual}_{i_1 \ldots i_{d-1}}}
= 2 \pi n_{d-2} - \theta_d \frac{d (d + 1)}{2} n_d \label{replaR} \\
\int d^d x \sqrt{g} \lambda & \to & \hspace{-0.3cm}
\sum_{\langle i_1 \ldots i_{d+1} \rangle} \hspace{-0.2cm}
V_{i_1 \ldots i_{d+1}} \lambda = V_d \lambda n_d \label{replaV} \\
\int {\cal D} g_{\mu\nu} & \to & \sum_T \frac{1}{C(T)}
\end{eqnarray}
where in (\ref{replaR}) we used the fact that every $d$-simplex has
a number of $(d - 2)$-subsimplices equal to
$\binom{d + 1}{d - 1} = \frac{d (d + 1)}{2}$, so that
\begin{equation}
\sum_{\langle i_1 \ldots i_{d-1} \rangle} o_{i_1 \ldots i_{d-1}}
= \frac{d (d + 1)}{2} n_d
\end{equation}
and in (\ref{replaV}) we noted that on dynamical triangulations every
$d$-simplex has a fixed volume that we denote by $V_d$.

All in all, we find as the partition function of pure simplicial
quantum gravity in $d$ dimensions
\begin{equation}
Z (\kappa_d, \kappa_{d-2})
= \sum_T \frac{1}{C(T)} \, e^{\kappa_{d-2} n_{d-2} - \kappa_d n_d}
\label{simplipart}
\end{equation}
where $\kappa_{d-2} \sim \frac{1}{G}$ and
$\kappa_d \sim \frac{\lambda + const.}{G}$ are the discrete versions
of the model's two coupling constants. Alternatively, we can also
consider the canonical partition function
\begin{equation}
Z (n_d, \kappa_{d-2}) = \sum_T \frac{1}{C(T)} \, e^{\kappa_{d-2} n_{d-2}}
\end{equation}
where the sum now runs over all triangulations of a given size $n_d$.

\subsection{The structure of space-time}

Before going on to discuss the properties of (\ref{simplipart}), we first
need to define some suitable observables that we can use to describe the
overall `shape' of a given space-time.\footnote{I should note that using
the word `space-time' to describe a manifold with Riemannian signature is
of course something of a misnomer, since there is no direction that could
be identified as a time coordinate. Nevertheless, I will continue to refer
to it as such, since a four-dimensional space-time is what we are
ultimately trying to represent.}

\subsubsection{The Hausdorff dimension}

One way of describing the `real' dimensionality of space-time (as opposed
to the canonical dimension $d$) is by the {\em fractal dimension} or
{\em Hausdorff dimension} $d_H$. To define it on a simplicial manifold,
we first need some concept of geodesic distance between simplices. For two
$d$-simplices $\langle i_1 \ldots i_{d+1} \rangle$ and
$\langle j_1 \ldots j_{d+1} \rangle$, define
$d^{i_1 \ldots i_{d+1}}_{j_1 \ldots j_{d+1}}$ as the length (number of
links) of the shortest path that connects the two simplices on the dual
lattice (the lattice formed by the dual polygons of all $d$-simplices
on the triangulation).

\begin{center}
\includegraphics[height=4cm]{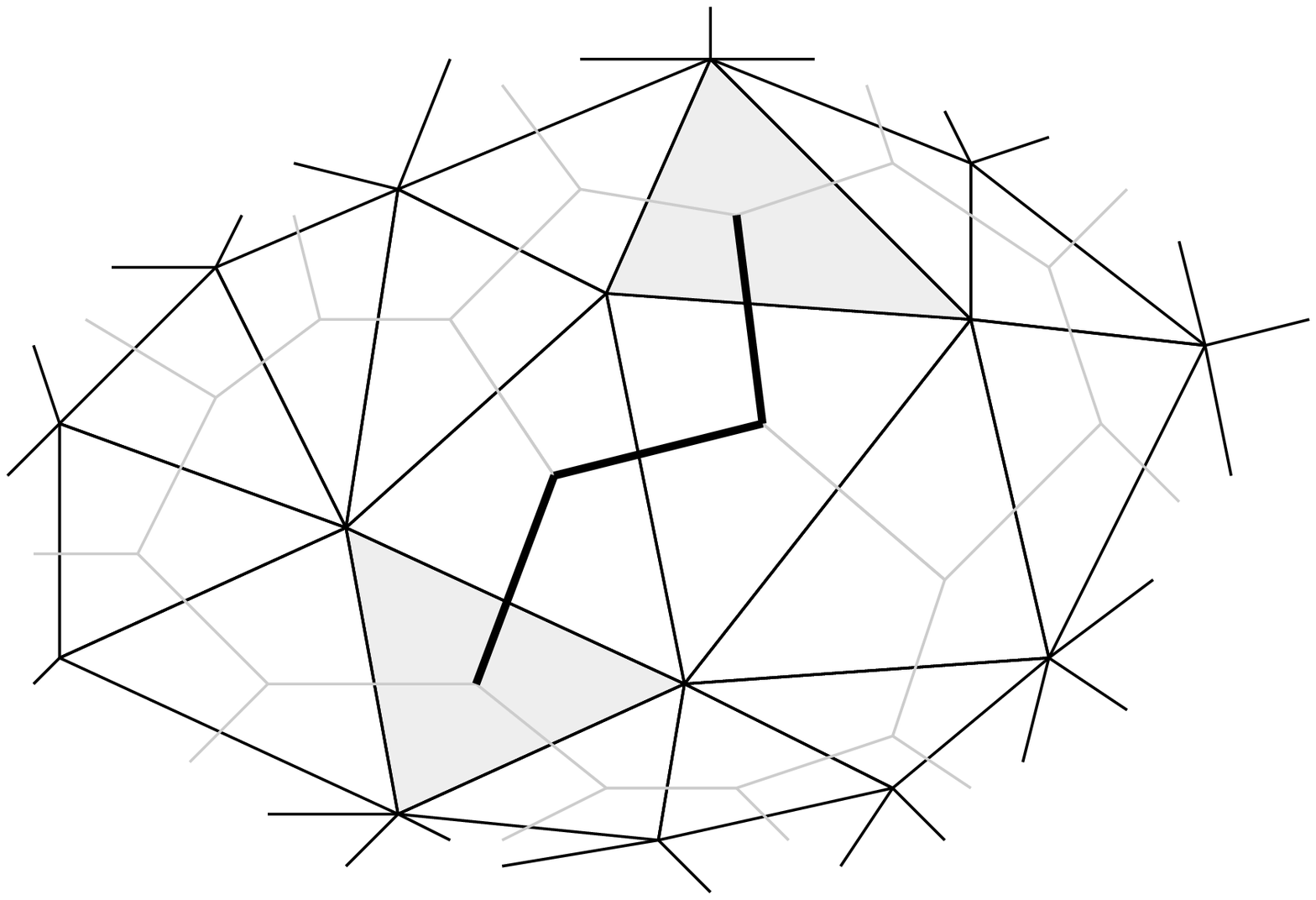} \\
{\small Two triangles at distance 3 from each other.}
\end{center}

\noindent
Next define a $d$-dimensional sphere of radius $r$ around a simplex
$\langle i_1 \ldots i_{d+1} \rangle$ as the set of all $d$-simplices
whose geodesic distance from the center is less than or equal to $r$,
\begin{equation}
S^r_{i_1 \ldots i_{d+1}} \equiv \left\{ \langle j_1 \ldots j_{d+1} \rangle:
d^{i_1 \ldots i_{d+1}}_{j_1 \ldots j_{d+1}} \le r \right\}
\end{equation}
The Hausdorff dimension is now defined by the relation between the volume
of a $d$-dimensional sphere (which is, of course, proportional to the
number of simplices contained in it) and its radius,
\begin{equation}
V \sim r^{d_H}
\end{equation}
For a smooth space-time, $d_H = d$, but in a fractal universe, $d_H$ can
be very different from the canonical dimension. It is not entirely clear
what we should expect from a sensible model of quantum gravity -- obviously,
we should have $d_H = 4$ at large scales, since that is what we can observe
in our universe; but at small scales, quantum effects could very well change
$d_H$. If so, however, it seems reasonable that they should make space-time
more fractal, not less, so a measured value of $d_H < 4$ can probably be
regarded as a bad sign.

\subsubsection{The string susceptibility exponent}

The other observable that we will be most interested in is the
{\em string susceptibility exponent} $\gamma$, which describes the
singular behaviour of the grand-canonical partition function,
\begin{equation}
Z (\kappa_d, \kappa_{d-2}) \sim Z_{analytic} (\kappa_d, \kappa_{d-2})
+ \left( \kappa_d - \kappa_d^c \right)^{2 - \gamma}
+ less \ singular \ terms
\label{suscepdef}
\end{equation}
where the critical value of $\kappa_d$ is in two dimensions a constant,
whereas in four dimensions it depends, as a monotonously increasing function,
on the choice of the other coupling constant,
$\kappa_d^c = \kappa_d^c (\kappa_{d-2})$. If the susceptibility exponent
exists ($i.\,e.$ if $| \gamma | < \infty$), then (\ref{suscepdef}) implies
an asymptotic behaviour of the canonical partition function
\begin{equation}
Z (n_d, \kappa_{d-2}) \sim e^{\kappa_d^c n_d} \, n_d^{\gamma - 3}
\end{equation}
The maximal value of the susceptibility exponent that can be expected is
$\gamma = \frac{1}{2}$, which corresponds to a sort of mean field value and
describes surfaces shaped like branched polymers (see below). Furthermore,
there exists a theorem \cite{gammapos, moregammapos} saying that any value
of $\gamma > 0$ automatically implies $\gamma = 1/2$.\footnote{This theorem
was later shown to have a few loopholes, but the only alternatives to
$\gamma = 1/2$ discovered so far are likewise uninteresting \cite{gravbarr}.}
Therefore, we would for a physical space-time expect to find $\gamma \le 0$.

\subsection{The two-dimensional case}

Now that we have the necessary tools, we can proceed to a discussion
of the properties of the model (\ref{simplipart}). I will first give a brief
overview of the situation in two dimensions, where the model simplifies
considerably, and actually becomes analytically solvable.

For one thing, the Einstein-Hilbert action in this model becomes a
topological invariant, courtesy of the Gauss-Bonnet theorem which tells us
that on any two-dimensional simplicial manifold we have
\begin{equation}
\int d^2 x \sqrt{|g|} R = 4 \pi \, (1 - h)
\end{equation}
where $h$ is the genus (number of holes) of the manifold.\footnote{Of course,
this means that classically the model becomes trivial, since Einstein's
equations will be automatically fulfilled. However, as we will see the
path integral $(\ref{pathint'})$ nevertheless has non-trivial properties
due to quantum effects.} Among other things, this means that the action is
no longer unbounded from below.

The other big simplification in two dimensions is that the metric tensor,
due to reparametrization invariance and its own symmetry, has only one
independent parameter left.\footnote{More precisely, this is true only
if the metric is not singular.} This means that we can choose to write
it in the {\em conformal gauge}
\begin{equation}
g_{\mu\nu} (x) \equiv \hat{g}_{\mu\nu} e^{\Phi (x)}
\end{equation}
where the $\hat{g}_{\mu\nu}$ are constants, and integrate only over the
field $\Phi$. The resulting model is known as {\em Liouville theory}.

The string susceptibility exponent in this model can be calculated
analytically, and comes out as $\gamma = - \frac{1}{2}$. What is more, we
can even generalize the model by including various types and numbers of
(conformally invariant) matter fields, and still get an exact value of
$\gamma$. As it turns out, this value depends only on the central charge
$c$ of these fields. The calculation can be performed for all $c \le 1$,
and yields \cite{kpz, morekpz, evenmorekpz}
\begin{equation}
\gamma = \frac{1}{12} \left( c - 1 - \sqrt{(c - 25)(c - 1)} \right)
\label{kpzscale}
\end{equation}
This behaviour is known as {\em KPZ scaling}, and for as long as it
applies, the model is said to be in the {\em Liouville phase}.

Beyond the so-called `$c = 1$ barrier', the formula (\ref{kpzscale})
breaks down, and we leave the Liouville phase to enter a regime where the
fractal structure of the universe degenerates into that of a branched
polymer: a tree structure that is built from many long but very thin tubes,
with a susceptibility exponent $\gamma = \frac{1}{2}$ and a Hausdorff
dimension $d_H = 2$. This structure can be explained as the condensation
of so-called `spikes'; the free energy of a spike can be estimated from
the effective action for the conformal factor as \cite{freespike,
morefreespike, evenmorefreespike}
\begin{equation}
F_{spike} \sim (1 - c) \ln \frac{l}{a}
\end{equation}
where $l$ is the length of the spike and $a$ is a cut-off length introduced
to prevent singularities of the metric. Obviously, for $c > 1$ $F_{spike}$
can become arbitrarily large and negative if we just take a sufficiently
long spike. It therefore seems clear that this, the {\em elongated phase}
or {\em branched polymer phase}, is physically meaningless; apparently, a
well-defined theory of two-dimensional Euclidean quantum gravity can exist
only if it does not contain `too much' matter.

If we now look at the discretized version of the model, we find that it,
too, simplifies to the point where an analytic treatment is possible. On a
two-dimensional simplicial manifold with a fixed number of triangles, the
number of vertices is a topological invariant,
$n_0 = \frac{1}{2} n_2 + 2 - 2 h$, a convenience that is no longer
available in higher dimensions. As a result, we find in analogy to the
Gauss-Bonnet theorem that the total curvature becomes once more a constant,
\begin{equation}
\sum_{\langle i \rangle} V^{dual}_i R_i = 2 \pi n_0 - \pi n_2
= 4 \pi \, (1 - h)
\end{equation}
The resulting model can be shown to be equivalent to a solvable
one-matrix model \cite{summerdavid}. More specifically, the dual graph
of each two-dimensional triangulation with $n_2$ triangles can be interpreted
as the Feynman diagram belonging to a term of order $n_2$ in the
matrix model's series expansion, and vice versa.

The solution of this matrix model, when taken to its planar limit,
shows that the two-dimensional triangulated model
does indeed have a critical point where a continuum limit can be
constructed. At this point, one also finds a string susceptibility
exponent $\gamma = - \frac{1}{2}$, which agrees with the pure gravity value
found in Liouville theory.

\begin{figure}
\begin{center}
\psfrag{Matter}{\small{$c$}}
\psfrag{beta}{\small{$\beta$}}
\psfrag{One}{\small{1}}
\psfrag{Branched}{\small{Branched polymer phase}}
\psfrag{BDetails}{\small{$\gamma = \frac{1}{2} \quad d_H = 2$}}
\psfrag{Crumpled1}{\small{Collapsed}}
\psfrag{Crumpled2}{\small{phase}}
\psfrag{CDetails1}{\small{$\gamma < 0$}}
\psfrag{CDetails2}{\small{$d_H = \infty$}}
\psfrag{Liouville}{\small{Liouville phase}}
\psfrag{LDetails1}{\small{$\gamma = \frac{1}{12}(c-1 - \sqrt{(c-25)(c-1)})$}}
\psfrag{LDetails2}{\small{$2 < d_H \le 4$}}
\includegraphics[height = 8cm]{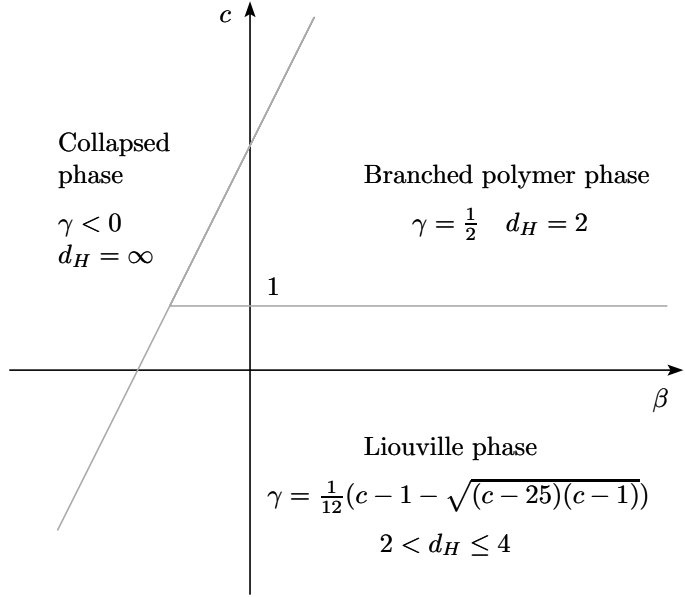}
\end{center}
\vspace{-0.5cm}
\caption{\label{twophased}Schematic phase structure of two-dimensional
simplicial quantum gravity, where the parameters are the central charge
of the matter fields $c$ and the measure term exponent $\beta$.}
\end{figure}

Finally, we can of course also perform numerical simulations of the
discretized model. This has been done both for pure gravity and for
various values of $c \ne 0$, as well as for a modified partition
function where an additional factor $\prod_{\langle i \rangle} o_i^\beta$
is included. This extra factor will be discussed in detail later;
for now, we can just regard it as the result of a slightly different
discretization procedure. The results of the simulations are indeed
found to agree with the analytic calculations (in those cases where
the calculations can be performed); from them, we can construct a phase
structure as depicted in figure \ref{twophased}.

As we can see, both the Liouville phase and the branched polymer
phase of pure gravity are reproduced for the appropriate values of $c$.
Apart from that, we see that by a combination of sufficiently extreme
values of $c$ and $\beta$ we can drive the system into a third phase.
In this, the {\em collapsed phase}, the geometry of the discretized model
degenerates to what is often called `pancake configurations':
two of the vertices become {\em singular} in the sense of developing
an order that grows linearly with the volume of the triangulations, so
that almost all other vertices are connected to these two. The result is a
universe that is basically a compressed disc, with an extremely large negative
curvature (tending towards infinity with growing $n_2$) in the two
singular vertices; it seems clear that this phase is of no more physical
relevance than branched polymers are.

To quickly summarize things, we can say that two-dimensional simplicial
quantum gravity correctly reproduces the results of the continuum theory,
which gives us some hope for the four-dimensional case. We also see that
the matter content of the theory can drastically affect its behaviour,
both in the discrete case and in the continuum; specifically, if matter
fields with a central charge $c > 1$ are included in the theory, it
degenerates into an ensemble of branched polymers. Alternatively, we can,
through a combination of matter fields and a modified measure, push the
system into a third phase where the universe collapses into a compressed,
singular disc.

\subsection{The four-dimensional case}

Finally, let us move on to the case that really interests us, namely the
four-dimensional one. Here, no analytical results exist, neither for the
continuum model nor for its discretized version. However, the
discretized partition function has been extensively studied in
numerical simulations, and its general properties are by now well-known.

The pure gravity model as defined by (\ref{simplipart}) has two different
phases, separated by a critical value $\kappa_2^c$ of the coupling constant
$\kappa_2$. When looked at more closely, both of these phases appear very
familiar indeed -- they are simply four-dimensional versions of the
collapsed and branched polymer phases that we already encountered in two
dimensions, with $d_H = 2$, $\gamma = \frac{1}{2}$ in one phase and
$d_H \to \infty$, $\gamma \to - \infty$ in the other.\footnote{The
collapsed phase is usually called {\em crumpled} in the four-dimensional
case due to its more complicated geometry. It still has two singular
vertices, but they are now connected by a link, which invalidates the
`collapsed disc' picture.} The mechanism that creates these phases is by
now well-understood and can be explained in terms of a simple mean-field
model \cite{bcboxes}, which reinforces the view that both of these phases
are merely artefacts of the discretization. The problem is that this time,
we do not find anything like the Liouville phase, where the model might
be expected to reflect continuum physics.

For some time, an alternative possibility was hoped to be found in the
phase transition. If this could be shown to be of second or higher order,
then the correlation length at the critical point would diverge and one
could hope to observe continuum behaviour. Indeed, for a long time the
transition was believed to be of second order, and measurements of the
Hausdorff dimension near the critical point seemed to indicate a value
of $d_H \approx 4$. More detailed studies revealed, however, the existence
of a small latent heat, implying that the transition is actually of first
order, if weakly so \cite{firstorder, morefirstorder}.

In other words, we have to conclude that the four-dimensional version of
pure simplicial quantum gravity as defined by (\ref{simplipart}) does not
have an interesting continuum limit.

\section{Modifications of the model}

If we nevertheless still believe that the idea of simplicial quantum
gravity is fundamentally sound -- and its success in two dimensions
certainly seems to support this view -- then we now face the challenge
of finding a suitable alteration of the pure gravity model that could
improve its critical behaviour. One possibility is that our method of
discretizing the model (\ref{pathint'}) is just too naive after all, and
that an appropriate modification of the partition function, such as the
inclusion of higher order curvature terms \cite{highcurv}, might soften
the transition to the point where it actually does become of second order.
Another, more intriguing idea that we will follow here is that a theory of
pure quantum gravity simply might not exist at all, and that the presence
of matter of some kind is necessary to formulate a well-defined theory.

\subsection{Matter fields}

As we saw in two dimensions, the matter content of the theory can radically
change its behaviour, driving it from the Liouville phase to the branched
polymer phase (or vice versa) when passing the $c = 1$ barrier. It therefore
seems possible that a similar effect might happen in four dimensions as well,
hopefully even creating a four-dimensional analogue to the Liouville phase.

As mentioned in section 3.1.3, the collapse of the two-dimensional model
into branched polymers is attributed to the creation of spiky configurations
whose free energy can be shown to grow with the logarithm of their length,
$F_{spike} \sim (1 - c) \ln \frac{l}{a}$. It has been suggested
\cite{fourspike, morefourspike, evenmorefourspike} that the same kind
of mechanism might also be responsible for the occurrence of branched
polymers in the four-dimensional case.

Strictly speaking, the conformal gauge cannot be used in four dimensions
since we now have more than one free parameter. However, we can at least
consider it as a first approximation, assuming that the `transverse'
components of the metric enter the calculation only as a correction. Then
the free energy of a spike can be calculated as
\begin{equation}
F_{spike} \sim \left( \frac{n_S + \frac{11}{2} n_F + 62 n_V - 28}{360}
+ Q_{grav}^2 - 4 \right) \ln \frac{l}{a}
\label{freespike}
\end{equation}
where $n_S$, $n_F$, and $n_V$ are the numbers of scalar, fermion, and
vector fields, respectively, and $Q_{grav}^2 \approx 3.9$ is an estimate
of the transverse gravitons' contribution based on a one-loop calculation.
Again we see that, as long as the prefactor
is negative, spikes will dominate since we can make $l$ arbitrarily
large. Contrary to the two-dimensional case, however, it now looks as
if we should {\em add} matter fields to make the theory more stable, due
to the different sign in the free energy. Also contrary to what we saw
before, the prefactor is negative in the pure gravity case, $i.\,e.$ for
$n_S = n_F = n_V = 0$. This seems to suggest that pure gravity might
indeed be ill-defined in four dimensions, and that we should add matter
fields to the partition function to obtain a sensible model.

From the prefactors of $n_S$, $n_F$, and $n_V$ in (\ref{freespike}), it
seems clear that the various types of matter fields should
have very different effects on the model. In particular, scalar fields
should produce almost no change unless added in very large numbers; this
agrees with earlier observations from numerical simulations \cite{scalarfail}.
On the other hand, we should expect a strong reaction of the model to the
addition of vector fields, which have the largest prefactor.

What we will try, then, is to change the model by adding a variable
number of $U(1)$ gauge fields (or actually their non-compact counterparts).
The corresponding discrete action was deduced in \cite{disgauge,
moredisgauge}; on dynamical triangulations it simplifies to
\begin{equation}
S_M = \sum_{\langle ijk \rangle} o_{ijk}
\left( A_{ij}^\mu + A_{jk}^\mu + A_{ki}^\mu \right)^2
\end{equation}
where $\mu = 1, \ldots, n_V$. The fields are defined on the links of the
triangulation; since these are oriented, we have $A_{ij}^\mu = - A_{ji}^\mu$.

The canonical partition function is now
\begin{equation}
Z (n_4, \kappa_2) = \sum_{T} \frac{1}{C(T)}
\int \sideset{}{'}\prod_{\langle ij \rangle} dA_{ij}
e^{-\kappa_4 n_4 + \kappa_2 n_2 - \sum_{\langle ijk \rangle} o_{ijk}
\left( A_{ij}^\mu + A_{jk}^\mu + A_{ki}^\mu \right)^2}
\label{partmatter}
\end{equation}
Here, the prime on the product over links indicates that the model now
has a number of zero modes coming from the fact that we can simultaneously
shift all fields around a given vertex $\langle i \rangle$ by a constant,
$A_{ij}^\mu \to A_{ij}^\mu + \epsilon$, without changing the action.
We can deal with these zero modes by considering only a maximal tree for
each triangulation (which corresponds to skipping the integration over
$n_0$ field variables).

\subsection{Measure terms}

In (\ref{partmatter}), the integration over matter fields is taken over
the link variables, $i.\,e.$ the individual fields themselves. But as can
be seen from the action, the physical quantities are actually not these
but the plaquette values $A_{ij}^\mu + A_{jk}^\mu + A_{ki}^\mu$. If we
assume that the individual fields are only weakly correlated to each other,
then we can as a first approximation regard the plaquettes as independent,
and simply replace the integration over fields by an integration over
plaquette values. We could then immediately perform the matter integrals
to get
\begin{equation}
Z (n_4, \kappa_2) = \sum_{T} \frac{1}{C(T)}
\prod_{\langle ijk \rangle} o_{ijk}^\beta \ e^{-\kappa_4 n_4 + \kappa_2 n_2}
\label{partmeasure}
\end{equation}
where $\beta \sim - n_V / 2$. This partition function, {\em if} it should
turn out to be essentially equivalent to (\ref{partmatter}), would have
a number of practical advantages for numerical simulations; for one thing,
they should be easier and quicker to perform, since no field updates have
to be made, for another, we can choose non-integer values of $\beta$,
which is obviously not possible in the case of $n_V$.

\section{The strong coupling expansion}

In the original work that this chapter is based on \cite{strongex,
morestrongex}, the partition functions (\ref{partmatter}) and
(\ref{partmeasure}) were studied using two different methods, namely numerical
simulations as described in chapter 2 and a non-statistical method known as
the strong coupling expansion. Since this latter method is used here for the
first time in the context of four-dimensional simplicial gravity, I will
describe it in some detail. It relies on calculating exactly the contributions
of some of the smallest configurations to the partition function and extracting
from these an estimate of the desired observables, using a suitable method
to reduce the inevitable finite size effects as much as possible. Obviously,
this scheme can work only if the first few terms of the partition function
are really the important ones -- we cannot expect it to produce meaningful
results if the coupling is too weak, $i.\,e.$ in the crumpled phase.
Fortunately, the most interesting developments will be seen to take place
in the branched polymer phase, where the strong coupling expansion provides
very good results.

\subsection{Calculation of the series terms}

The first thing we need to calculate the series is some reliable method of
constructing all distinct triangulations for a given volume. In two dimensions
this can be done simply by using the Feynman rules on the equivalent matrix
model, but the same is not possible in four dimensions.

Instead, we can exploit the properties of the $(p, q)$ moves, which we know
to (a) be ergodic and (b) change the number of 4-simplices
by always the same amount no matter what configuration they are used on.
In other words, each configuration with a given $n_4$ must be accessible
in one of these ways: using move $(1, 5)$ on a configuration
with $n_4 - 4$ simplices; using $(2, 4)$ on a configuration with $n_4 - 2$
simplices; using $(3, 3)$ on some other triangulation of the same size;
or using $(4, 2)$ or $(5, 1)$ on a correspondingly larger configuration.

With this in mind, we can formulate the following step-by-step prescription
for constructing all triangulations of a given size $n_4$, assuming only
that all smaller configurations have already been identified:

\begin{itemize}

\item Go through all configurations with $n_4 - 4$ simplices, and
successively apply move $(1, 5)$ to each of them in all possible ways
(which in this case means, to all vertices). Store all resulting
triangulations.

\item Go through all configurations with $n_4 - 2$ simplices, and successively
apply move $(2, 4)$ to each of them in all possible ways. In this case,
this means using it on all links. Again, store everything created in this
way.

\item Now we have a long list of configurations; unfortunately, not all of
them are distinct, since any given triangulation can usually be created from
smaller ones in more than one way. To eliminate the surplus copies, we need
some sort of distinguishing characteristic -- an observable that takes a
different value for each distinct geometry. There is no {\em a priori}
obvious choice for such a characteristic; in practice, we simply choose one
that seems suitably complicated, and change it whenever it turns out to be
no longer sufficient. This means, of course, that we need some reliable
method of double-checking the results, in order to judge the efficiency of
our current choice of characteristic.

\item To this end, we now calculate the symmetry factor $C(T)$ in two
different ways, one of which is always correct whereas the other depends
on whether our distinguishing characteristic has been chosen properly.

First of all, note that the relation between the symmetry factors of
two configurations $T_1$ and $T_2$ can also be expressed as the relation
between the number of different ways we can use the appropriate $(p, q)$
move to transform $T_1$ into $T_2$ and the number of ways we can use that
move's inverse to transform $T_2$ back into $T_1$:
\begin{equation}
\frac{C(T_1)}{C(T_2)} = \frac{N (T_1 \to T_2)}{N (T_2 \to T_1)}
\end{equation}
Thus, we can calculate the symmetry factor of a newly created configuration
by counting the number of ways it can be reached from a smaller system, as
well as the number of ways of going back. However, this will work only if
we have correctly distinguished all distinct triangulations; otherwise,
we will in some cases overcount $N (T_1 \to T_2)$, and the corresponding
symmetry factors will come out wrong.

The other method consists of calculating all symmetry factors directly,
by counting the number of equivalent re-labelings of vertices for each
triangulation. In practice, this means taking a configuration, re-labeling
the vertices in all possible ways, and checking for each permutation whether
it leads back to the original labeling. This may seem like a daunting task
at first, given that the number of possible permutations grows like
$n_0!$; however, we can defuse this problem by noting that only labels of
vertices with the same order have to be permuted among themselves, since
exchanging the labels of vertices with different orders leads to these
labels then occurring more or less often in the re-labeled version than in
the original one, which means the two cannot be the same. Except for
a very few highly symmetric configurations, this means that in most cases
we only face permutations of at most three or four vertex labels among
themselves, which is easily done.

Once we have calculated all the symmetry factors in both ways, we can
compare them to each other for all configurations. Disagreement for
any configuration means that our chosen characteristic failed to distinguish
between two or more configurations, which sends us back to the drawing
board to come up with a better choice.

\item Once we can be sure that we have correctly identified all
configurations that can be constructed `from below', we can make the
next step by using move $(3, 3)$ on all of
them in all possible ways ($i.\,e.$, on all triangles). As before,
we use our distinguishing characteristic to reduce the resulting list
to one of only distinct triangulations, and calculate the symmetry factors
to make sure that our choice of characteristic is still a good one. We
then compare all the `new' configurations to the original list and drop
all those that turn out to already have been constructed earlier. If any
genuinely new triangulations remain, they are added to the list, and we
again use move $(3, 3)$ in all possible ways on these new configurations.
The cycle is repeated until no more new configurations turn up.

\item Now we face the one fundamental uncertainty about the whole process,
namely the possibility of configurations that can only be reached `from
above', $i.\,e.$ those to which moves $(5, 1)$, $(4, 2)$, and $(3, 3)$
cannot be applied. In two dimensions, no such `irreducible' configurations
exist, but the same has not been proven in four dimensions. In fact,
we can never be sure of having found all of them unless we already know
all larger configurations, since in theory it is possible that such a
triangulation could be reached only by making a long excursion to
extremely large $n_4$ before going back down.

Since we cannot construct these configurations from any we already have,
we hunt for them in Monte Carlo simulations, which of course gives us
the additional uncertainty of possibly missing some configurations
simply due to the randomness of the method. Fortunately, we can estimate
the required length of the Monte Carlo runs to pick up all configurations
with a reasonable confidence, since we already know more or less how many
configurations there are in total -- we do not expect to see large numbers
of `irreducible' configurations, and the results prove us correct.

\end{itemize}

\noindent
Once all configurations and their symmetry factors have been found, the
rest becomes easy; we can calculate terms of the partition function with
any number of gauge fields and/or any power of the measure term by
running all the configurations through a simple Maple script.

Apart from the consistency check on the distinguishing characteristic
as described above, we also performed extensive tests on the program
as a whole by constructing all configurations up to $n_4 = 18$ by hand
and comparing them to the results of the program.

\subsection{The ratio method}

Once we have calculated a sufficient number of terms in the strong coupling
expansion, the next step is to find a way of extracting physical information
out of them.

Our primary aim is to show whether matter fields and/or a modified measure
can fundamentally alter the model. We are therefore not too interested in
the finer details of the geometry, but rather its overall shape. Thus, our
most important observables will again be the susceptibility exponent
$\gamma$ and the Hausdorff dimension $d_H$. The question is how we can
measure these in a way that allows us to control the finite size effects.

The prescription we will use here to calculate $\gamma$ is called the
{\em ratio method}; it has already been successfully used for the same
purpose in the simpler case of two dimensions \cite{ratiomet, moreratiomet}.
To describe it, I will assume that the finite size corrections are of
order $o(n_4^{-1})$. Then the partition function is expected to behave as
\begin{equation}
Z (n_4, \kappa_2) \sim e^{\kappa_4^c (\kappa_2) n_4} n_4^{\gamma - 3}
\left( 1 + o (n_4^{-1}) \right)
\label{suscepart}
\end{equation}
We will have to find an estimate of $\kappa_4^c (\kappa_2)$ before we can
deal with $\gamma$. To do so, define
\begin{equation}
C_{n_4}^1 (\kappa_2) \equiv \frac{Z (n_4, \kappa_2)}{Z (n_4 - 1, \kappa_2)}
\end{equation}
Inserting (\ref{suscepart}), we find to second order
\begin{eqnarray}
C_{n_4}^1 & = & e^{\kappa_4^c} \left( \frac{n_4}{n_4 - 1} \right)^{\gamma - 3}
\left( 1 + o (n_4^{-2}) \right) \nonumber \\
& = & e^{\kappa_4^c} \left( 1 + \frac{\gamma - 3}{n_4} + o (n_4^{-2}) \right)
\left( 1 + o (n_4^{-2}) \right) \nonumber \\
& = & e^{\kappa_4^c} \left( 1 + \frac{\gamma - 3}{n_4} + o (n_4^{-2}) \right)
\end{eqnarray}
$i.\,e.$ $C_{n_4}^1$ gives us an estimate of $\kappa_4^c$ that is correct
to order $o(n_4^{-1})$. Next, define
\begin{equation}
C_{n_4}^{p + 1} \equiv \frac{1}{p}
\Big( n_4 C_{n_4}^p - (n_4 - p) C_{n_4 - 1}^p \Big)
\end{equation}
Assuming that $C_{n_4}^p = e^{\kappa_4^c} (1 + o (n_4^p))
= e^{\kappa_4^c} ( 1 + c_p/n_4^p + d_p/n_4^{p+1} + \ldots )$
has already been shown, we see
\begin{eqnarray}
C_{n_4}^{p+1} & = & \frac{1}{p} \left( n_4 e^{\kappa_4^c}
\left( 1 + \frac{c_p}{n_4^p} + \frac{d_p}{n_4^{p+1}} \right)
- (n_4\!-\!p) e^{\kappa_4^c} \left( 1 + \frac{c_p}{(n_4\!-\!1)^p}
+ \frac{d_p}{(n_4\!-\!1)^{p+1}} \right) \right) \nonumber \\
& = & e^{\kappa_4^c} \left( 1 + \frac{c_p}{p n_4^{p-1}} + \frac{d_p}{p n_4^p}
- \frac{(n_4\!-\!p) c_p}{p (n_4^p\!-\!p n_4^{p - 1}\!+\!\ldots)}
- \frac{(n_4\!-\!p) d_p}{p (n_4^{p + 1}\!-\!(p\!+\!1) n_4^p\!+\!\ldots)}
\right) \nonumber \\
& = & e^{\kappa_4^c} \left( 1 + \frac{d_p}{p n_4^p} \left( 1
- \frac{n_4 - p}{n_4 - p - 1} \right) + o (n_4^{-(p+1)}) \right) \nonumber \\
& = & e^{\kappa_4^c} \left( 1 + o (n_4^{-(p+1)}) \right)
\end{eqnarray}
so the series $C_{n_4}^p$ gives us better and better estimates of $\kappa_4^c$.

Similarly, we can now find an estimate of $\gamma$ by defining
\begin{equation}
D_{n_4}^1 \equiv n_4 \left( C_{n_4}^1 - C_{n_4}^2 \right)
\end{equation}
for which, inserting the expressions for $C_{n_4}^1$ and $C_{n_4}^2$ from
above, we find
\begin{eqnarray}
D_{n_4}^1 & = & n_4 \left( (1 - n_4) C_{n_4}^1 + (n_4 - 1) C_{n_4 - 1}^1
\right) \nonumber \\
& = & n_4 \left( (1\!-\!n_4) e^{\kappa_4^c} \left( 1 + \frac{\gamma\!-\!3}{n_4}
+ \frac{d_1}{n_4^2} \right) + (n_4\!-\!1) e^{\kappa_4^c} \left( 1
+ \frac{\gamma\!-\!3}{n_4\!-\!1} + \frac{d_1}{(n_4\!-\!1)^2} \right) \right)
\nonumber \\
& = & e^{\kappa_4^c} (\gamma - 3) \left( 1 + \frac{(1 - n_4) d_1}{n_4}
+ \frac{n_4 d_1}{n_4 - 1} \right) \nonumber \\
& = & e^{\kappa_4^c} (\gamma - 3) \left( 1 + o (n_4^{-1}) \right)
\end{eqnarray}
Defining the higher order terms as before,
\begin{equation}
D_{n_4}^{p + 1} \equiv \frac{1}{p}
\Big( n_4 D_{n_4}^p - (n_4 - p) D_{n_4 - 1}^p \Big)
\end{equation}
we can again show that each term $D_{n_4}^p$ gives us an estimate of
$e^{\kappa_4^c} (\gamma - 3)$ that is correct up to order $o(n_4^{-p})$.
Finally, we can estimate $\gamma$ itself through
\begin{equation}
\frac{D_{n_4}^p}{C_{n_4}^p} + 3 = \gamma \left( 1 + o (n_4^{-p}) \right)
\end{equation}
As it turns out, calculation of the Hausdorff dimension $d_H$ from the
strong coupling expansion is far more difficult than extracting the
susceptibility exponent; here we find that the Monte Carlo simulations
actually lead to much better results.

\subsection{Results}

The calculation of the series terms is limited by several factors as $n_4$
is increased, among them the growing demands of $CPU$ time and storage space
and the ever larger difficulties in finding a distinguishing characteristic
that still works on all configurations. Despite this, we managed to push
the calculation up to $n_4 = 38$ and about $10^6$ configurations. The
numbers $N (n_4, n_2)$ and weights $Z (n_4, n_2)$ of all configurations
created in this way are collected in table \ref{expanded}.

\begin{table}
\begin{center}
\begin{tabular}{|c c|c|c||c c|c|c|}
\hline
$n_4$ & $n_2$ & $N (n_4, n_2)$ & $Z (n_4, n_2)$ &
$n_4$ & $n_2$ & $N (n_4, n_2)$ & $Z (n_4, n_2)$ \\
\hline
6     & 20    & 1      & 1                      &
30    & 74    & 139    & 73860                  \\
10    & 30    & 1      & 3                      &
      & 76    & 1276   & 672821                 \\
12    & 34    & 1      & 5                      &
      & 78    & 1208   & 564000                 \\
14    & 40    & 1      & 15                     &
      & 80    & 143    & 46376                  \\
16    & 44    & 2      & 63$\frac{3}{4}$        &
32    & 80    & 3886   & 2351430                \\
18    & 48    & 3      & 110                    &
      & 82    & 5943   & 3327045                \\
      & 50    & 3      & 95                     &
      & 84    & 1700   & 817306$\frac{7}{8}$    \\
20    & 52    & 2      & 225                    &
34    & 84    & 11442  & 7502430                \\
      & 54    & 7      & 693                    &
      & 86    & 26337  & 16396680               \\
22    & 58    & 15     & 2460                   &
      & 88    & 13231  & 7545780                \\
      & 60    & 7      & 690                    &
      & 90    & 922    & 411255                 \\
24    & 62    & 34     & 8182$\frac{1}{2}$      &
36    & 88    & 27765  & 18929925               \\
      & 64    & 34     & 7312$\frac{1}{2}$      &
      & 90    & 112097 & 74395157               \\
26    & 66    & 50     & 17865                  &
      & 92    & 85734  & 54240610               \\
      & 68    & 124    & 39645                  &
      & 94    & 15298  & 8742976$\frac{2}{3}$   \\
      & 70    & 30     & 5481                   &
38    & 92    & 71295  & 50097510               \\
28    & 70    & 89     & 41650$\frac{5}{7}$     &
      & 94    & 458083 & 315706725              \\
      & 72    & 415    & 182820                 &
      & 96    & 490598 & 328515075              \\
      & 74    & 217    & 77057$\frac{1}{7}$     &
      & 98    & 153773 & 97507410               \\
      &       &        &                        &
      & 100   & 6848   & 3781635                \\
\hline
\end{tabular}
\end{center}
\caption{\label{expanded}The numbers of existing triangulations
$N (n_4, n_2)$ and their pure gravity weights $Z (n_4, n_2)$ for
$n_4 = 6, \ldots, 38$. The weights have been normalized so that
$Z (6, 20) \equiv 1$.}
\end{table}

When trying to use the ratio method on these weights to extract $\gamma$,
we find a noticeable oscillation with growing $n_4$. In fact, we can
clearly distinguish between two separate series, one that is made up
from all configurations with $n_4 = 4k$ and one that contains all those
with $n_4 = 4k + 2$; both series seem to converge toward the same value,
except that one is coming from above and the other from below.
The reason for this can be traced back to asymmetric jumps in the maximal
integrated curvature when $n_4$ is increased \cite{weaklim}. Namely, on
dynamical triangulations we have a general constraint on the number of
triangles, $n_2 \le 2.5 n_4 + 5$ \cite{maxcurv}, and we also know that the
number of triangles must be even. This gives us an actual maximum number
of triangles $n_{2, max} = 2.5 n_4 + 5$ in the $n_4 = 4k + 2$ series, but
only $n_{2, max} = 2.5 n_4 + 4$ in the $4k$ series. From (\ref{replaR}),
we see that this means the maximal integrated curvature jumps by
$6 c_4 - 6$ when going from a $4k$ series term to the next higher
$4k + 2$ series term, but only by $4 c_4 - 6$ when going from a
$4k + 2$ term to the next higher $4k$ term. But as we will see, the
system in the large $\kappa_2$ phase is always close to the maximal
integrated curvature; hence the oscillations. As it turns out that the
finite size effects are much less pronounced for the $4k + 2$ series, we
restrict ourselves to using these terms.

\subsubsection{Effects of the matter fields}

\begin{figure}
\begin{center}
\includegraphics[height=7.5cm]{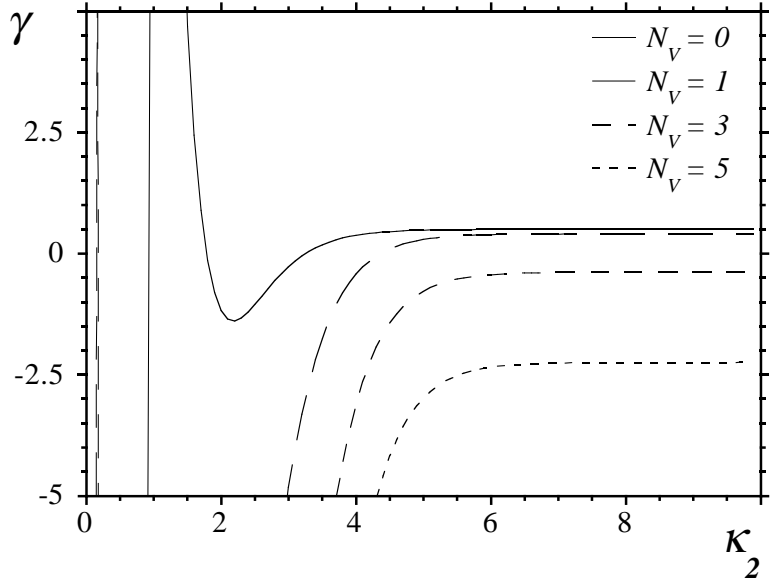}
\end{center}
\caption{\label{kappagamma}The string susceptibility exponent $\gamma$,
estimated from the strong coupling expansion, as a function of
$\kappa_2$ for various numbers of vector fields.}
\end{figure}

The results for $\gamma$ as a function of $\kappa_2$ with a varying
number of vector fields are presented in figure \ref{kappagamma}. For
pure gravity and large values of $\kappa_2$, we see the susceptibility
exponent approaching the branched polymer value $\gamma = \frac{1}{2}$,
as expected. For smaller values of the coupling constant, the estimate
becomes unstable and meaningless, which is not surprising given the
limitations of what is called, after all, the strong coupling expansion.
(Below the transition, $i.\,e.$ in the crumpled phase, $\gamma$ is not
defined in any case.)

Adding a single vector field does not seem to have much of an effect
on the model, at least in the region where we can trust the strong
coupling expansion. But starting with three vector fields, we see a
remarkable change in the string susceptibility exponent as it moves
away from the branched polymer value and becomes negative. This mirrors
the behaviour of two-dimensional simplicial gravity, where the
susceptibility exponent goes from $\frac{1}{2}$ to negative values as
the system crosses the $c = 1$ barrier and moves into the Liouville phase.
It would seem, then, that there exists a related third phase -- which we
will call {\em crinkled phase} -- in the four-dimensional model also,
but only if we add a sufficient number of vector fields. This, of course,
is just what we hoped and expected to find.

\begin{figure}
\begin{center}
\psfrag{vector}{\small{$n_V$}}
\psfrag{beta}{\small{$\beta$}}
\psfrag{kappa}{\small{$\kappa_2$}}
\psfrag{Branched}{\small{Branched polymer phase}}
\psfrag{BDetails}{\small{$\gamma = \frac{1}{2} \quad d_H = 2$}}
\psfrag{Crumpled1}{\small{Crumpled}}
\psfrag{Crumpled2}{\small{phase}}
\psfrag{CDetails1}{\small{$\gamma = -\infty$}}
\psfrag{CDetails2}{\small{$d_H = \infty$}}
\psfrag{Crinkled}{\small{Crinkled phase}}
\psfrag{CrDetails}{\small{$\gamma < 0 \quad d_H \approx 4$}}
\includegraphics[height = 8cm]{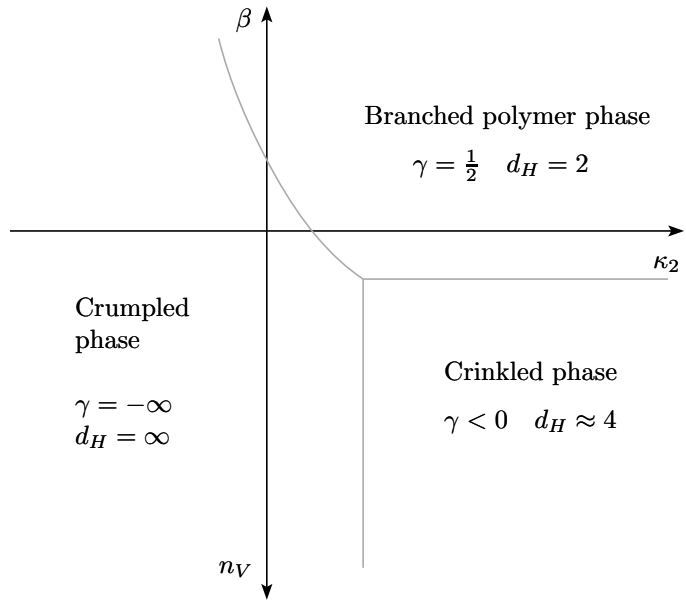}
\end{center}
\vspace{-0.5cm}
\caption{\label{fourphased}Schematic phase structure of four-dimensional
simplicial quantum gravity with either $n_V$ gauge fields or a
modified measure exponent $\beta$.}
\end{figure}

The existence of the crinkled phase is supported by results of the Monte
Carlo simulations, where the system also exhibits a negative susceptibility
exponent once the number of vector fields grows beyond 2, with the measured
value of $\gamma$ compatible with the results of the strong coupling
expansion -- for example, for $n_V = 3$ and $\kappa_2 = 4.5$ we find in the
largest systems we simulated ($n_4 = 16000$) a susceptibility exponent
$\gamma = -0.30(6)$, to be compared with a value $\gamma = -0.38$ extracted
from the strong coupling series (although at a larger value of $\kappa_2$,
since the series is not yet reliable at $\kappa_2 = 4.5$). The Monte Carlo
simulations also show a related change in the Hausdorff dimension, which
goes from the branched polymer value $d_H = 2$ for pure gravity to
$d_H = 3.97(15)$ for $n_V = 3$ and $\kappa_2 = 4.5$ -- exactly what one
would be hoping to see in a theory of four-dimensional gravity.

\subsubsection{Effects of the measure term}

\begin{figure}
\begin{center}
\includegraphics[height=7.5cm]{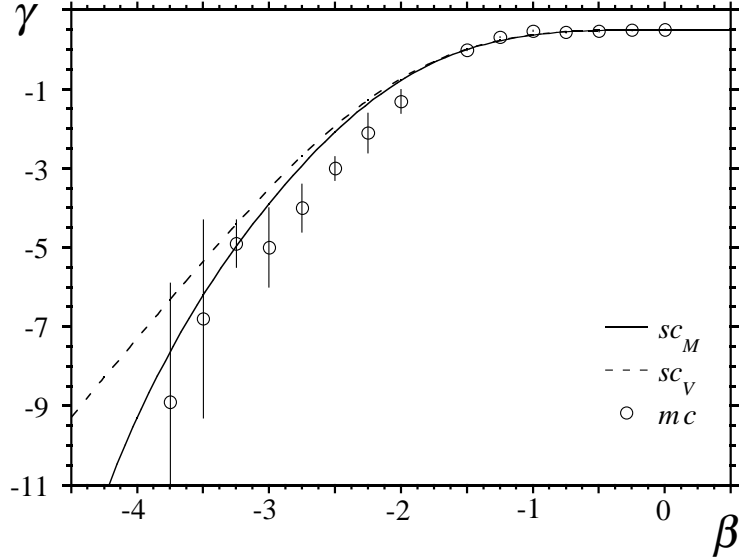}
\end{center}
\caption{\label{betagamma}The string susceptibility exponent $\gamma$ as a
function of $\beta$ for $\kappa_2 = 10$ (solid line). For comparison,
$\gamma$ as a function of $n_V$ is also shown, where the identification
$\beta \equiv - \frac{n_V}{2} - \frac{1}{4}$ is made on the $x$ axis
(dashed line). Finally, results for $\gamma$ as a function of $\beta$
from Monte Carlo simulations at $\kappa_2 = 3$ are also included (circles).}
\end{figure}

We can use the strong coupling expansion with equal ease on the model
with a modified measure, calculating the string susceptibility exponent
for a varying (negative) exponent $\beta$. Figure \ref{betagamma} shows the
results for $\kappa_2 = 10$. For small absolute $\beta$, the modification
of the measure has essentially no effect, and the susceptibility exponent
retains its branched polymer value of $\gamma = \frac{1}{2}$. Once we decrease
$\beta$ below a certain point, however, $\gamma$ starts to drop and
soon becomes negative, exactly as before.

We can even show the relation between $\beta$ and $n_V$ by superimposing
on the figure a plot of $\gamma$ from the model with matter fields,
setting $\beta = - \frac{n_V}{2} - \frac{1}{4}$ on the $x$ axis; we see
that up to about $n_V = 5$, the two curves lie practically on top of
each other. Overall, the conclusion seems to be that both matter fields
and the kind of modified measure described here lead to the same kind
of effective action, which in turn changes the behaviour of the model
by replacing the branched polymer phase with the crinkled phase.

\subsubsection{The nature of the crinkled phase}

Although the crinkled phase has many similiarities to the Liouville phase,
on closer inspection it also reveals some worrying features that do not appear
in the lower-dimensional case. For one thing, the total curvature in the
crinkled phase is very near to its upper bound, which is given by the
fact that $n_0 \le \frac{1}{4} n_4$. But this bound is just a technicality
of the discretization, created by the fact that we are using triangulations
to regularize space-time; it does not exist in the continuum theory. This
might lead us to question our results for the Hausdorff dimension and the
susceptibility exponent, since they might have been influenced by the
artificial cut-off of the upper bound. Another point is that the largest
vertex order -- which in the crumpled phase grows linearly with the volume
and thus causes that phase's problems -- grows considerably faster in the
crinkled phase than in the Liouville phase, although it does remain
sub-linear. Finally, the transition point $\kappa_2^c$ from crumpled to
crinkled phase changes with $n_4$ in a way that makes it difficult to decide
whether it will converge in the large $n_4$ limit; if it does not, the
crinkled phase will disappear in the continuum.

On the other hand, we found that just as with the branched polymer phase,
the total curvature, while close to the upper bound, does not stick to it;
instead, the distribution develops a peak at some point below this bound
and then starts to decrease again. From this point of view, the influence
of the bound on the measurements of $d_H$ and $\gamma$ should be small.
The largest vertex order, while indeed growing faster than logarithmically
as happens in the Liouville phase, does remain sub-linear and thus will
become unimportant in the large $n_4$ limit; one might argue that the details
of its growth do not matter as long as it is ensured that no singular
points will remain in the continuum. And finally, while it is possible
that $\kappa_2^c$ might move to infinity with $n_4 \to \infty$, it is
just as possible that it might converge to a finite value; for the
moment, the numerical data simply does not allow for a definite prediction
one way or the other.

There also exists a mean-field argument based on the `balls in boxes'
concept \cite{boxedballs, moreboxedballs, evenmoreboxedballs} that
seems to support the view of the crinkled phase being artificial,
since it correctly describes many general features of both the crinkled
and the branched polymer phase based solely on the existence of the
upper bound on the curvature. On the other hand, it also makes some
predictions that do not agree with what is observed in the Monte Carlo
simulations, such as a first order transition from crumpled to crinkled
phase and a latent heat that grows with $n_V$, whereas the simulations
show a transition of third or even higher order and a latent heat that
decreases with $n_V$. These might be taken as just minor failings of
an otherwise valid approximation, or as a sign that we should look for
something more than a mean-field argument to explain the crinkled phase.

The final conclusion must be that the nature of the crinkled phase is
for now uncertain, and requires further studies before one could make a
more definite statement. In particular, one should attempt to push the
simulations to larger values of $n_4$ so as to decide what happens to
the transition point.

\section{Discussion}

In the pursuit of this project, we followed two different goals: the
study of a modified model of four-dimensional simplicial quantum gravity,
in the hope that these modifications might improve the model's critical
behaviour; and a test of the applicability of the strong coupling
expansion to such a four-dimensional model.

With respect to the modified model, we find that the addition of an
appropriate number of matter gauge fields does indeed cause a major
change in the behaviour of the model. Essentially, the branched polymer
phase of the pure gravity model vanishes and is replaced by a new
`crinkled phase'. The extended tree structure of branched polymers is
no longer present in this phase, nor does it have the singular vertices
of the crumpled phase; instead, it exhibits a number of similarities to
the Liouville phase of two dimensions, such as a negative string
susceptibility exponent and a Hausdorff dimension whose measured value
is compatible with 4. Also, the transition to the crumpled phase is no
longer discontinuous but at least of third order, or even a crossover.
This was further confirmed in \cite{confirm}.

We also found that essentially the same effects can be generated by
the inclusion of an additional measure factor $\prod o^\beta$ in
the partition function. This seems to confirm the assumption that
the correlations between the field variables are weak. Further evidence
for this was given in \cite{dualgauge}, where it was shown that the
measure factor appears naturally when transforming a model of gauge
fields living on the dual links of the lattice to the version of the
model we examined here. For the model with fields on the dual links,
without the additional factor $o_{ijk}$ in the action, the observed
effects of the fields on the geometry are weak.

Despite its similarities to the Liouville phase, the crinkled phase also
exhibits a few less appealing features that are not present in the
two-dimensional case, such as an integrated curvature that is close to
the upper kinematic bound. Also, we cannot yet exclude the possibility
that the transition point between crumpled and crinkled phase might move
to infinity with growing $n_4$. For a decisive answer, we should take the
simulations to much larger systems. It might well turn out that vector
fields alone are not sufficient to make the model well-defined; if so,
an intriguing thought is that it might take an exactly balanced mixture
of scalar, vector, and fermion fields to create a physical universe.
In any case, we can say that we have shown, contrary to earlier
investigations, that the inclusion of matter fields can have a drastic
influence on four-dimensional simplicial gravity, making further studies
in this direction worthwhile.

In the matter of applying the strong coupling expansion to a four-dimensional
triangulated model, we can claim a clear success. We have found a reliable
method of calculating successive terms of the series expansion that, apart
from one final uncertainty in the form of `irreducible' configurations that
cannot be directly constructed from smaller surfaces, does not depend on
statistical methods. Calculation of the terms has been taken to $n_4 = 38$,
and could conceivably be taken even farther. Problems of increasing $CPU$
time and storage space should even now present much less of a difficulty
than they were three years ago, when this project was undertaken. The task
of finding an ever-more-sensitive distinguishing characteristic could be
avoided by a slight modification of the algorithm; rather than demanding
that our choice of characteristic be capable of separating every single
distinct triangulation, we could use it as only a rough instrument for
classifying configurations, and then directly compare all configurations
within each category so as to weed out the redundancies.

One practical problem that remains in the application of this method
is finding a way of calculating observables while systematically
reducing finite size effects. As we saw, the ratio method used here
has only a limited range of observables that it can work on; one
should certainly try to find other methods that could allow us to
access other quantities, such as the Hausdorff dimension, as well.

\chapter{One-dimensional structures in models with an area action}

Quantized superstring theory is nowadays believed by many to constitute the
best bet for a description of quantum gravity. Differently from other
attempts, it seems to be not just renormalizable but even finite to each
order of its perturbative expansion, and it contains a spin-2 particle
that can be interpreted as a graviton in a natural way. Nevertheless,
it suffers from the fact that it contains a path integral over surfaces
that in general cannot be solved. The only exceptions are the cases
$d \le 1$, where we actually have a prescription for how to perform the
integration \cite{onedimstring, moreonedimstring}; and the critical case
$d = 10$, where the surface integral decouples completely from the rest
of the theory and can therefore be dropped \cite{tendimstring}. The
generic solution at this point is to accept the ten-dimensional theory
as correct and argue that some of these dimensions are `compactified'
in our physical universe, making it appear four-dimensional at large
enough distances. So far, however, there exists no rigorous prescription
for how to perform this compactification.

A different problem comes from the fact that even though all individual
terms in the perturbative expansion appear to be finite, the series as a
whole is still divergent. In other words, there exists as yet no
non-perturbative definition of string theory. A relatively recent proposal
to solve this problem comes from the {\em IKKT model} \cite{ikkt, moreikkt,
evenmoreikkt}, which provides a sort of quantization prescription where
the bosonic and fermionic fields of a supersymmetric string theory are
replaced by $n \times n$ Hermitian matrices. This prescription leads to
an apparently well-defined partition function that exhibits some
striking similarities to the original string theory, and was therefore
conjectured to provide an actual constructive definition of same.
Consequently, the model was greeted with a lot of enthusiasm, and has since
then been investigated in great detail \cite{ikktstudies, moreikktstudies,
evenmoreikktstudies, yetmoreikktstudies, countlessikktstudies,
neverendingikktstudies}.

A much earlier attempt at providing a non-perturbative definition of string
theory comes from a discretization of the string world-sheet in terms of
dynamical triangulations. In its original formulation, however, this approach
turned out to have difficulties. A model of bosonic strings is known to lead
to an ill-defined partition function, due to geometrical defects called
spikes \cite{moredynatri}. Attempts to discretize a model with space-time
supersymmetry \cite{superfail}, which was hoped to cure these defects, have
mostly failed; this model also has a local world-sheet symmetry, called
$\kappa$ symmetry, that is explicitly broken on a triangulated lattice, and
it is not known how to ensure that it becomes restored in the continuum limit
\cite{moresuperfail}. At this point, research on this approach was more or
less abandoned -- until the development of the $IKKT$ model led, as a sort
of by-product, to a new formulation of the supersymmetric model in which the
$\kappa$ symmetry had been gauged away. Thus, a discretization of quantum
string theory once more appeared possible as well \cite{dynaschild}.

In this chapter, we will study some characteristics of both proposals,
although the main focus will be on the triangulated model, and the question
whether a supersymmetric model with an area action can avoid the singularities
encountered in the purely bosonic case. Once we know what is going on in the
surface model, we will try to apply our findings to the $IKKT$ model -- or
rather its lower-dimensional counterpart -- as well.

\section{The surface model}

Our starting point is the Green-Schwarz $I\!IB$ superstring in the
Nambu-Goto form, which is re-written to (a) gauge fix the local
$\kappa$ symmetry and (b) switch the integration to one over internal
metrics, which then allows us to do a discretization in terms of
dynamical triangulations. The complete calculation can be found in
\cite{ikkt}; here I will just sketch the major steps.

The action is
\begin{equation}
S_{I\!IB} = \int d^2 \sigma \left( \sqrt{-\frac{\Sigma^2}{2}}
+ i \sum_{j = 1}^2 \epsilon^{\alpha\beta} \partial_\alpha X^\mu
\bar{\theta}^j \Gamma_\mu \partial_\beta \theta^j
+ \epsilon^{\alpha\beta} \bar{\theta}^1 \Gamma^\mu \partial_\alpha \theta^1
\bar{\theta}^2 \Gamma_\mu \partial_\beta \theta^2 \right)
\end{equation}
where $X^\mu$, $\mu = 1, \ldots, 10$, are the bosonic fields;
$\theta^1_a, \theta^2_a$, $a = 1, \ldots, 16$, are the fermions;
$(\sigma_1, \sigma_2)$ is a parametrization of the world-sheet;
and $\Sigma$ is defined as
\begin{equation}
\Sigma^{\mu\nu} \equiv \epsilon^{\alpha\beta} \left(
\partial_\alpha X^\mu\!- i
\bar{\theta}^1 \Gamma^\mu \partial_\alpha \theta^1\!+ i
\bar{\theta}^2 \Gamma^\mu \partial_\alpha \theta^2 \right) \left(
\partial_\beta X^\nu\!- i
\bar{\theta}^1 \Gamma^\nu \partial_\beta \theta^1\!+ i
\bar{\theta}^2 \Gamma^\nu \partial_\beta \theta^2 \right)
\end{equation}
Note that this differs from the usual formulation of the model
\cite{greenschwarz} in that one of the fermionic fields has been replaced
by an analytic continuation, $\theta^2_a \to i \theta^2_a$. The action has
two symmetries, the ${\cal N} = 2$ space-time supersymmetry
\begin{equation}
\delta_S \theta^1 = \epsilon^1 \qquad
\delta_S \theta^2 = \epsilon^2 \qquad
\delta_S X^\mu = i \bar{\epsilon}^1 \Gamma^\mu \theta^1
                 - i \bar{\epsilon}^2 \Gamma^\mu \theta^2
\end{equation}
and the $\kappa$ symmetry
\begin{equation}
\delta_\kappa \theta^1 = \alpha^1 \qquad
\delta_\kappa \theta^2 = \alpha^2 \qquad
\delta_\kappa X^\mu = i \bar{\theta}^1 \Gamma^\mu \alpha^1
                 - i \bar{\theta}^2 \Gamma^\mu \alpha^2
\end{equation}
where
\begin{equation}
\alpha^1 \equiv \left( 1 + \frac{\Sigma^{\mu\nu} [ \Gamma_\mu, \Gamma_\nu ]}
{4 \sqrt{-\frac{\Sigma^2}{2}}} \right) \kappa_1 \qquad
\alpha^2 \equiv \left( 1 - \frac{\Sigma^{\mu\nu} [ \Gamma_\mu, \Gamma_\nu ]}
{4 \sqrt{-\frac{\Sigma^2}{2}}} \right) \kappa_2
\end{equation}
The $\kappa$ symmetry can be gauge fixed by setting
$\theta^1 = \theta^2 = \Psi$, which simplifies the action to
\begin{equation}
S'_{I\!IB} (X, \bar{\Psi}, \Psi) = \int d^2 \sigma
\left( \sqrt{-\frac{\sigma^2}{2}} + 2 i \epsilon^{\alpha\beta}
\partial_\alpha X^\mu \bar{\Psi} \Gamma_\mu \partial_\beta \Psi \right)
\label{nokappa}
\end{equation}
where
\begin{equation}
\sigma^{\mu\nu} \equiv \epsilon^{\alpha\beta} \partial_\alpha X^\mu
\partial_\beta X^\nu
\end{equation}
This action still has the ${\cal N} = 2$ supersymmetry, but the
transformation laws have to be adjusted to preserve the gauge condition.
They can now be written as
\begin{equation}
\begin{array}{cc}
\vspace{0.5cm} \delta_S^{(1)} \Psi
= - \frac{\sigma^{\mu\nu} [ \Gamma_\mu, \Gamma_\nu ] \epsilon}
{2 \sqrt{-\frac{\sigma^2}{2}}} \qquad & \qquad \delta_S^{(2)} \Psi = \xi \\
\delta_S^{(1)} X^\mu = 4 i \bar{\epsilon} \Gamma^\mu \Psi
\qquad & \qquad \delta_S^{(2)} X^\mu = 0
\end{array}
\end{equation}
As shown by Schild \cite{schildaction}, we can write down an action that
is at least classically equivalent to (\ref{nokappa}) if we replace the
integration over surfaces expressed in the external coordinates by an
integration over internal metrics:
\begin{equation}
S''_{I\!IB} (X, \bar{\Psi}, \Psi) = \int d^2 \sigma \sqrt{|g|}
\left( \frac{1}{4} \left\{ X^\mu, X^\nu \right\}^2
- \frac{i}{2} \bar{\Psi} \Gamma_\mu \left\{ X^\mu, \Psi \right\} \right)
\label{internaction}
\end{equation}
where $|g|$ is the determinant of the world-sheet metric,
and the Poisson bracket is defined as
\begin{equation}
\left\{ X, Y \right\} \equiv \frac{1}{\sqrt{|g|}}
\epsilon^{\alpha\beta} \partial_\alpha X \partial_\beta Y
\end{equation}
The supersymmetry now becomes
\begin{equation}
\begin{array}{cc}
\vspace{0.5cm} \delta_S^{(1)} \Psi = - \frac{1}{2}
\left\{ X^\mu, X^\nu \right\} [ \Gamma_\mu, \Gamma_\nu ] \epsilon \qquad
& \qquad \delta_S^{(2)} \Psi = \xi \\
\delta_S^{(1)} X^\mu = i \bar{\epsilon} \Gamma^\mu \Psi
\qquad & \qquad \delta_S^{(2)} X^\mu = 0
\end{array}
\end{equation}
At least formally, then, we can write down a quantum theory of the
$I\!IB$ superstring as a path integral over this action,
\begin{equation}
Z = \int {\cal D} g_{\mu\nu} {\cal D} X {\cal D} \bar{\Psi}
{\cal D} \Psi e^{-S''_{I\!IB} (X, \bar{\Psi}, \Psi)}
\label{surfpath}
\end{equation}
It should be noted that while, as mentioned above, the string theory is
defined only in $d = 10$ dimensions, we can in general formulate the
integral as a supersymmetric surface model in $d = 3$, 4, 6, and 10
dimensions. As we will see, many important properties, especially the long
distance behaviour, turn out to be essentially the same in any dimension.

\subsection{Discretization of the surface model}

We now want to discretize the integral (\ref{surfpath}) in terms of
dynamical triangulations, following essentially the suggestions in
\cite{dynaschild}.

For simplicity, I will fix the number of dimensions from the start to be
$d = 4$, since that is where the numerical simulations will be performed.
Generalization of our arguments to higher dimensions is straightforward,
and will be addressed once we understand what happens in four dimensions.

The choice of $d = 4$ also determines the fermionic variables to be either
Majorana or Weyl spinors, at least if we want to have supersymmetry.
Either choice will lead to the same results \cite{yetmoreikktstudies};
we will here take them to be Weyl spinors. We also choose to put both
the bosonic and the fermionic fields on the vertices of the lattice.

With these choices, we can write down the canonical partition function
for the triangulated model as
\begin{equation}
Z (n_2) = \sum_T \frac{1}{C(T)} \int \sideset{}{'}\prod_{i = 1}^{n_0} d^4 X_i
\sideset{}{'}\prod_{i = 1}^{n_0} d^4 \bar{\Psi}_i d^4 \Psi_i \
e^{-S_B (X_i) - S_F (X_i, \bar{\Psi}_i, \Psi_i)}
\end{equation}
Here, the products of integration variables have primes attached to them
because, as we will see, there are zero modes that have to be removed to get
a well-defined partition function. Note that since we are studying
triangulations with an internal dimension of 2, the numbers of triangles
and vertices are dependent on each other; choosing a specific value of $n_2$
automatically fixes $n_0$ (see section 3.1.3 for details).

\subsubsection{The action of the discrete model}

To write down the discretized action $S = S_B + S_F$, we first need a
discrete version of the Poisson bracket $\{ X, Y \}$. In \cite{dynaschild},
this is defined as an average value for each triangle,
\begin{equation}
\{ X, Y \}_{ijk} \equiv \frac{1}{2 A} \epsilon^{ijk}
\left( X_{jk} Y_{ki} - Y_{jk} X_{ki} \right)
\end{equation}
where $X_{ij} \equiv X_i - X_j$ and $A \equiv \frac{\sqrt{3}}{4} a^2$
is the triangle area. Using this, we find
\begin{eqnarray}
S_B (X_i) & = & \frac{2}{A} \sum_{\langle ijk \rangle}
\Big\{ - \left( \left( X_{ij}^2 \right)^2
                 + \left( X_{jk}^2 \right)^2
                 + \left( X_{ki}^2 \right)^2 \right) \nonumber \\
& & \hspace{1cm}
        + 2 \left( X_{ij}^2 X_{jk}^2
                   + X_{jk}^2 X_{ki}^2
                   + X_{ki}^2 X_{ij}^2 \right) \Big\} \\
S_F (X_i, \bar{\Psi}_i, \Psi_i) & = & \frac{i}{12} \sum_{\langle ij \rangle}
\bar{\Psi}_i^a \Gamma_\mu^{ab} \Psi_j^b
\left( X_{\omega_{ij}}^\mu - X_{\omega_{ji}}^\mu \right)
\end{eqnarray}
Here, $\langle \omega_{ij} \rangle$ and $\langle \omega_{ji} \rangle$
denote the remaining two vertices of the two triangles that contain the
link $\langle ij \rangle$. Specifically, $\langle \omega_{ij} \rangle$
is defined to be the neighbouring vertex of $\langle i \rangle$ that
comes after $\langle j \rangle$ when going counterclockwise around
$\langle i \rangle$:

\begin{center}
\psfrag{i}{\small{$\langle i \rangle$}}
\psfrag{j}{\small{$\langle j \rangle$}}
\psfrag{k}{\small{$\langle \omega_{ij} \rangle$}}
\psfrag{n}{\small{$\langle \omega_{ji} \rangle$}}
\includegraphics[height=2.5cm]{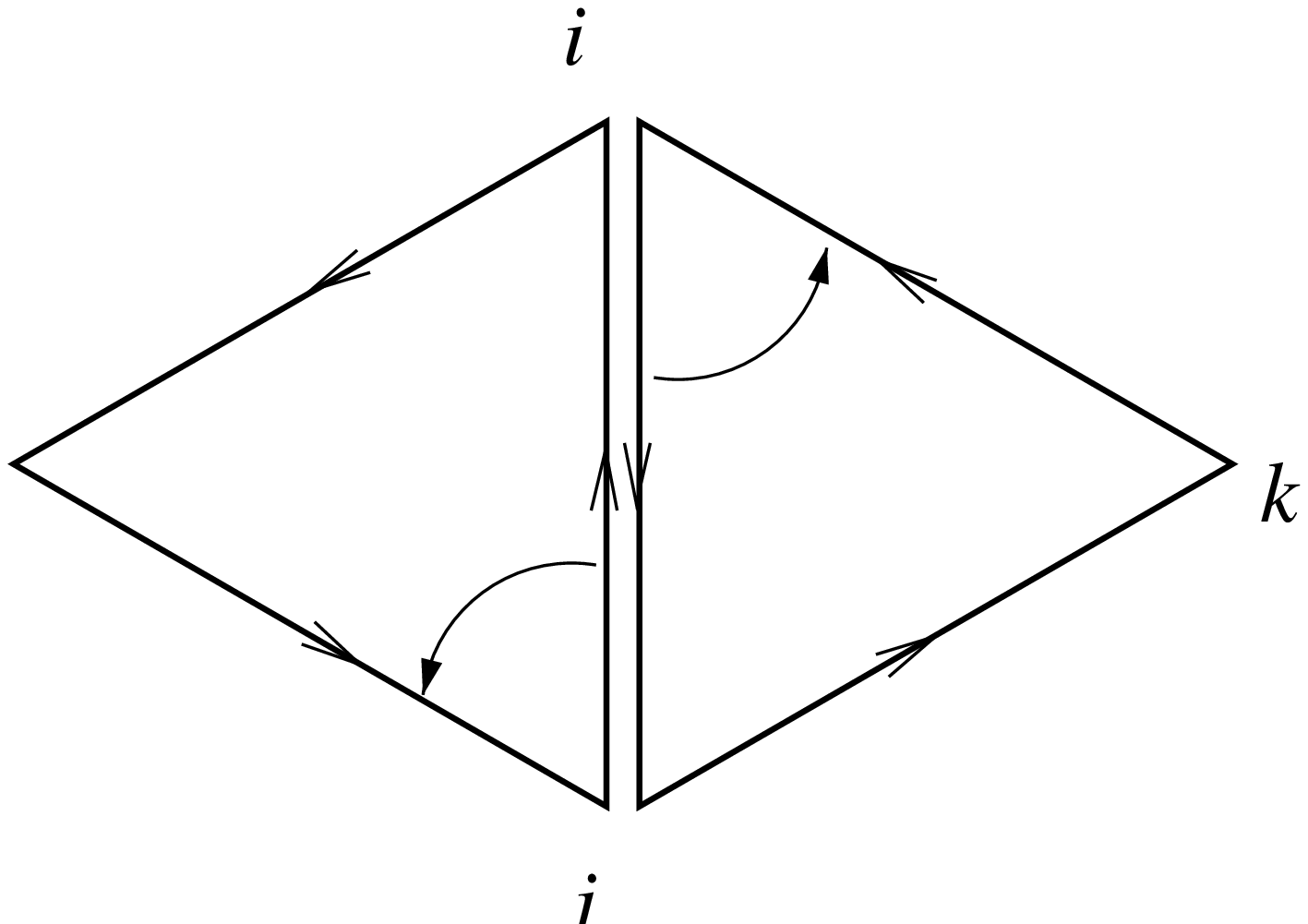}
\end{center}

\noindent
So far, we have not yet used the fact that $\bar{\Psi}_i$, $\Psi_i$
are Weyl fermions, $i.\,e.$ that they have a fixed handedness.
Mathematically, this means that we can write
\begin{equation}
\Psi_i = \frac{1}{2} (1 + \Gamma_5) \Psi_i \qquad
\bar{\Psi}_i = \bar{\Psi}_i \frac{1}{2} (1 - \Gamma_5)
\label{handweyl}
\end{equation}
If we choose the following definition of the Dirac matrices (chiral
representation):
\begin{equation}
\Gamma_0 = \left(
\begin{array}{cc} 0        & -\sigma_0 \\
                  \sigma_0 & 0         \end{array} \right) \qquad
\Gamma_i = \left(
\begin{array}{cc} 0        & \sigma_i \\
                  \sigma_i & 0        \end{array} \right) \qquad
\Gamma_5 = \left(
\begin{array}{cc} \mathbbm{1} & 0          \\
                  0           & - \mathbbm{1} \end{array} \right)
\end{equation}
where $\sigma_i$ are the Pauli matrices and $\sigma_0 \equiv i \mathbbm{1}$,
then (\ref{handweyl}) becomes
\begin{equation}
\Psi_i = \left( \begin{array}{cc}
\mathbbm{1} & 0 \\ 0 & 0 \end{array} \right) \Psi_i \qquad
\bar{\Psi}_i = \left( \begin{array}{cc}
0 & 0 \\ 0 & \mathbbm{1} \end{array} \right) \bar{\Psi}_i
\end{equation}
which means we can express the fermionic variables in terms of
two-component spinors $\bar{\psi}_i$, $\psi_i$:
\begin{equation}
\bar{\Psi}_i = (0, \bar{\psi}_i) \qquad
\Psi_i = \binom{\psi_i}{0}
\end{equation}
We can therefore re-write the fermionic part of the action as
\begin{equation}
S_F (X_i, \bar{\psi}_i, \psi_i)
= i \sum_{\langle ij \rangle} \bar{\psi}_i^a f_{ij}^\mu
\sigma_\mu^{ab} \psi_j^b
\label{defmat}
\end{equation}
where we have also defined
\begin{equation}
f_{ij}^\mu \equiv \frac{1}{12}
\left( X_{\omega_{ij}}^\mu - X_{\omega_{ji}}^\mu \right)
\end{equation}
Obviously, the $f_{ij}^\mu$ are real and antisymmetric in the lattice indices
$ij$. Also, it follows from the definition of $\langle \omega_{ij} \rangle$
that for each vertex $\langle i \rangle$, the sum of $f_{ij}^\mu$ over all
neighbouring vertices $\langle j \rangle$ vanishes:
\begin{equation}
\sum_{ \langle j \rangle : \langle ij \rangle } f_{ij}^\mu = 0
\label{nosumf}
\end{equation}
Here, the expression `$\langle j \rangle : \langle ij \rangle$' is just
a shorthand notation for `all vertices $\langle j \rangle$ such that
$\langle ij \rangle$ is an existing link on the triangulation', a convention
that I will use throughout the remainder of this chapter.

\subsubsection{Zero modes in the action}

Now that we know the discretized form of the action, we can see that there
are two zero modes that we have to deal with.

One follows from the invariance of $S$ under translations of the bosonic
fields, $X_i \to X_i + \delta$. This is because the coordinates never enter
directly into the action but only as differences between vertices $X_i - X_j$,
so that any global translation drops right out. We can fix this zero mode
by keeping the system's center of mass fixed at some given point, preferably
the origin. Technically, this introduces a delta function
$\delta^4 (\sum X_i)$ in the partition function.

The other zero mode comes from a similar effect in the fermionic sector,
namely invariance of the action under a change
$\psi_i \to \psi_i + \epsilon$,
$\bar{\psi}_i \to \bar{\psi}_i + \bar{\epsilon}$.
This follows from the properties of the $f_{ij}^\mu$, in particular equation
(\ref{nosumf}). Specifically, the change in the action under a translation
of the fermions is
\begin{eqnarray}
\delta S_F & = & i \sum_{\langle ij \rangle}
\bar{\epsilon} f_{ij}^\mu \sigma_\mu \psi_j
+ i \sum_{\langle ij \rangle}
\bar{\psi}_i f_{ij}^\mu \sigma_\mu \epsilon
+ i \sum_{\langle ij \rangle}
\bar{\epsilon} f_{ij}^\mu \sigma_\mu \epsilon \nonumber \\
& = &
\frac{i}{2} \sum_{\langle j \rangle} \bar{\epsilon} \sigma_\mu \psi_j
\sum_{ \langle i \rangle : \langle ij \rangle } f_{ij}^\mu
+ \frac{i}{2} \sum_{\langle i \rangle} \bar{\psi}_i \sigma_\mu \epsilon
\sum_{ \langle j \rangle : \langle ij \rangle } f_{ij}^\mu
+ \frac{i}{2} \bar{\epsilon} \sigma_\mu \epsilon \sum_{\langle i \rangle}
\sum_{ \langle j \rangle : \langle ij \rangle } f_{ij}^\mu
\nonumber \\
& = & 0
\end{eqnarray}
We can remove this zero mode by skipping the integration over one pair of
fermionic variables in the partition function, for example $\psi_{n_0}$
and $\bar{\psi}_{n_0}$. Technically, this amounts to inserting an
additional product $\bar{\psi}_{n_0} \psi_{n_0}$, which for Grassmann
variables acts like a delta function.

\subsubsection{The fermionic determinant}

As the final step, we can integrate out the fermions. This results in a
factor $|{\cal M}_{ij}^{ab}|$, where
\begin{equation}
{\cal M}_{ij}^{ab} \equiv i f_{ij}^\mu \sigma^{ab}_\mu
\qquad i, j = 1, \ldots, n_0 - 1 \quad a, b = 1, 2
\end{equation}
is the matrix that appears in (\ref{defmat}), with the two rows and columns
that correspond to the vertex $\langle n_0 \rangle$ crossed out because of
the fermionic `delta function'. Note that despite having four indices this
is just a two-dimensional matrix; each combination of indices $\binom{a}{i}$,
$\binom{b}{j}$ actually corresponds to only a single matrix index $I$, $J$.
The size of the matrix can be read off as $2 (n_0 - 1) \times 2 (n_0 - 1)$.

If we want to perform numerical simulations, it is essential to have a
non-negative integrand; otherwise we would encounter the so-called sign
problem, which makes any attempt at a numerical treatment all but
futile. Fortunately, it is possible to show that in $d = 4$ the determinant
$|{\cal M}_{ij}^{ab}|$ is in fact non-negative. To see this, take any
eigenvalue $\lambda$ of ${\cal M}_{ij}^{ab}$, together with its
corresponding eigenvector $l_i^a$:
\begin{equation}
\left( i f_{ij}^0 \sigma^{ab}_0 + i f_{ij}^1 \sigma^{ab}_1
+ i f_{ij}^2 \sigma^{ab}_2 + i f_{ij}^3 \sigma^{ab}_3 \right) l_j^b
= \lambda l_i^a
\end{equation}
Now take the complex conjugate of this equation to get
\begin{equation}
\left( i f_{ij}^0 \sigma^{ab}_0 - i f_{ij}^1 \sigma^{ab}_1
+ i f_{ij}^2 \sigma^{ab}_2 - i f_{ij}^3 \sigma^{ab}_3 \right) (l_j^b)^*
= \lambda^* (l_i^a)^*
\end{equation}
because $\sigma_1$ and $\sigma_3$ are real, whereas $\sigma_0$ and
$\sigma_2$ are purely imaginary. Next, multiply the equation by $\sigma_2$
from the left:
\begin{equation}
\left( i f_{ij}^0 \sigma^{ca}_2 \sigma^{ab}_0
- i f_{ij}^1 \sigma^{ca}_2 \sigma^{ab}_1
+ i f_{ij}^2 \sigma^{ca}_2 \sigma^{ab}_2
- i f_{ij}^3 \sigma^{ca}_2 \sigma^{ab}_3 \right) (l_j^b)^*
= \lambda^* \sigma^{ca}_2 (l_i^a)^*
\end{equation}
Finally, use the fact that $\sigma_2$ commutes with $\sigma_0$ but
anticommutes with $\sigma_1$ and $\sigma_3$ to get
\begin{equation}
\Big( i f_{ij}^0 \sigma^{ca}_0 + i f_{ij}^1 \sigma^{ca}_1
+ i f_{ij}^2 \sigma^{ca}_2 + i f_{ij}^3 \sigma^{ca}_3 \Big)
\sigma^{ab}_2 (l_j^b)^* = \lambda^* \sigma^{ca}_2 (l_i^a)^*
\end{equation}
In other words, if $\lambda$ is an eigenvalue of ${\cal M}_{ij}^{ab}$,
then so is $\lambda^*$, to the eigenvector $(\sigma^{ca}_2 (l_i^a)^*)$.
But since the determinant is just the product of the eigenvalues, and
$\lambda \lambda^* = ({\rm Re} \lambda)^2 + ({\rm Im} \lambda)^2$
is both real and non-negative, we find $|{\cal M}_{ij}^{ab}| \ge 0$
for any possible configuration.

Note that all this is true {\em only} in four dimensions, which is the main
reason for our choice of $d$.

\subsubsection{The partition function of the discrete model}

We can now write down the final form of the partition function that we
want to analyze:
\begin{equation}
Z (n_2, \gamma) = \sum_{T} \frac{1}{C(T)} \int \prod_{i = 1}^{n_0}
d^4 X_i^\mu \ \delta^4 \left( \sum X_i^\mu \right) \
|{\cal M}_{IJ}|^\gamma \ e^{-S_B (X_i^\mu)}
\label{finalpart}
\end{equation}
Here, we have introduced the additional parameter $\gamma$ as just a
convenient way to distinguish between several cases that we want to look at.
Namely, we will be interested in the choices $\gamma = 0$, 1, 2, which
correspond to the purely bosonic model, the supersymmetric model, and a
model with Dirac fermions instead of Weyl fermions, respectively.

\section{Spikes and needles}

As it turns out, the model as defined above has serious problems both
without and with `too many' fermions; in fact, we will show that the
partition function (\ref{finalpart}) can be well-defined only in the
supersymmetric case $\gamma = 1$. The reason for this can be
found in the appearance of so-called `flat directions' in the action
-- valleys of small and constant values of the action that extend outward
to very large and deformed-looking configurations, which can cause the
partition function to blow up.

In this particular case, the bosonic action $S_B$ is just the sum of all
triangle areas squared. Therefore, if we can find ways of deforming some
or all of the triangles into pathological shapes without affecting their
areas, we are heading towards trouble, because we can then expect the
partition function to pick up large contributions from these configurations
as they are not suppressed by the action.

\subsection{The bosonic case}

We will start by looking at possible singularities in the purely bosonic
model. As we will see, there are two different ways in which the flat
directions of the action can affect the partition function.

\subsubsection{Spikes}

One way of creating deformed triangles has actually been known for a
long time, going back to \cite{moredynatri}. There, it was shown that
for the purely bosonic model these configurations lead to a divergence
of the partition function. I will not reproduce the rigorous proof here,
but rather try to give a more intuitive argument that will tie in nicely
with the arguments presented in the next section.

\begin{center}
\includegraphics[height=0.5cm]{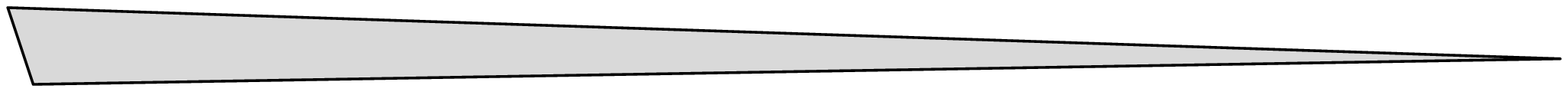}
\end{center}

\noindent
Choose a vertex $\langle i \rangle$ anywhere on the configuration, and
find all its neighbouring triangles. Take all the links that lie opposite
to $\langle i \rangle$ on these triangles ($i.\,e.$, those that form the
boundary of $\langle i \rangle$'s neighbourhood), and shrink them to
lengths of order $\sim 1/l$, where $l$ is large. Simultaneously, we can
now stretch all the links meeting in $\langle i \rangle$ to lengths of
order $\sim l$ and still leave all triangles in the neighbourhood of
$\langle i \rangle$ with an area of order $\sim 1$.\footnote{Obviously,
shrinking the link lengths on the boundary also affects triangle areas
outside the neighbourhood of $\langle i \rangle$, but only in a favourable
direction, $i.\,e.$ if anything making them smaller.} Since we can
vary $l$ freely, this means we can remove $\langle i \rangle$ arbitrarily
far from the rest of the configuration without affecting the action.
The result is a {\em spike}:

\begin{center}
\includegraphics[height=1cm]{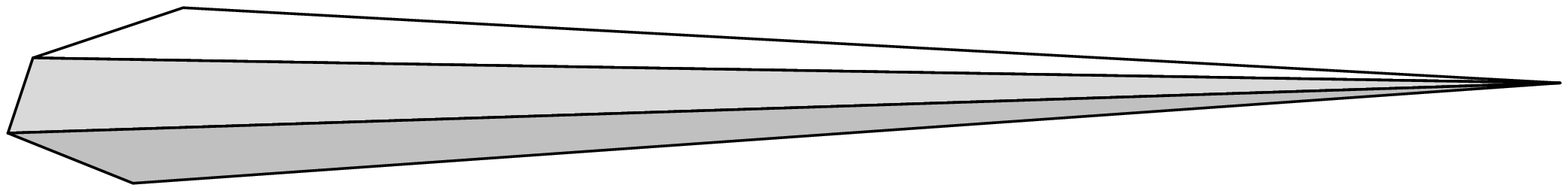}
\end{center}

\noindent
Of course, just because it looks strange does not mean that a spike is
necessarily a problem. We can examine this point in more detail by
trying to estimate a spike's contribution to the partition function.
Since the action
remains almost fixed during the creation of a spike, all we have to do
is look at the entropy factor. We will see that a simple power-counting
argument is enough for this purpose.

Assume that we have a spike $\langle i \rangle$ of maximum length $l$,
$i.\,e.$ the coordinates of $\langle i \rangle$ are all of order $\sim l$.
Then the entropy factor for the spike itself is $\sim l^4$, since we
are in $d = 4$ dimensions. On the other hand, all the vertices on the
boundary must be restricted to be within a distance of $\sim 1/l$ from
their neighbouring boundary vertices. In other words, we can choose one
of them freely, but then all the others have positions that are fixed
to within $\sim 1/l$. This provides an additional factor
$\sim l^{-4 (o_i - 1)}$. If we now integrate over all but one vertex
coordinate, we are left with
\begin{equation}
Z_{spike} \sim \int dl \ l^{-4 (o_i - 2) - 1}
\end{equation}
We can interpret the integrand as a probability distribution for the
spike length:
\begin{equation}
p(l) \sim l^{-4 (o_i - 2) - 1}
\end{equation}
Since the order of a vertex on a two-dimensional triangulation is
always at least 3, this gives us a worst-case distribution of
$p(l) \sim l^{-5}$. So the spike's contribution to the partition
function is still integrable, and the partition function itself still
well-defined. However, because they are suppressed only by a power law,
we can expect spikes of arbitrary length to appear in the simulations,
and as a result the distribution's higher moments will have no finite
expectation values.

Furthermore, the situation can become worse if we consider not just
one spike, but several in close proximity to each other. For example,
take a configuration as it has been drawn below, with four spikes
of order 3 grouped around a central vertex:

\begin{center}
\includegraphics[height=4cm]{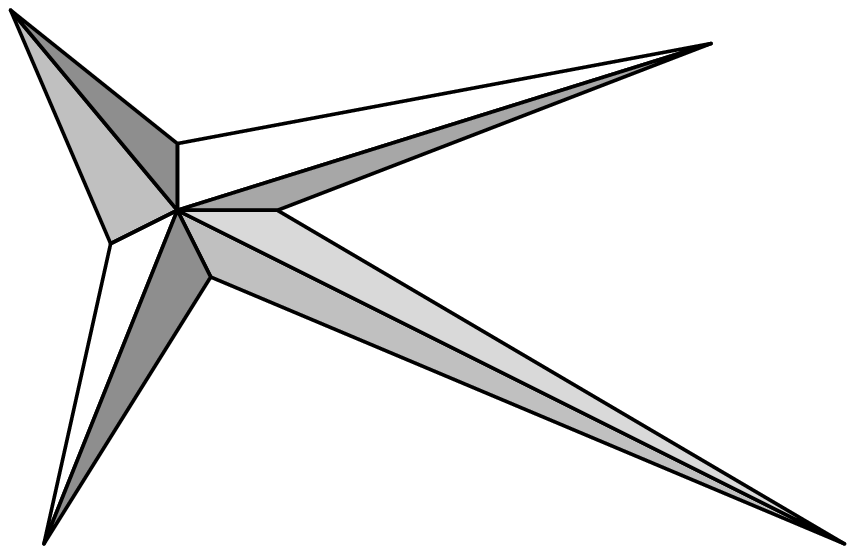}
\end{center}

\noindent
In this case, the entropy factor for the spikes is $l^{16}$, and we
have to restrict four vertices to be within a distance of $\sim 1/l$
from the central one, giving rise to another factor $l^{-16}$. If we
again integrate over all but one coordinate, we end up with a distribution
\begin{equation}
p(l) \sim l^{-1}
\end{equation}
that leads to a logarithmic divergence of the partition function. This
is exactly the result of the rigorous proof given in \cite{moredynatri}.

\subsubsection{Needles}

In the last section we created a deformed triangle of constant area by
shrinking its base while simultaneously increasing its height. We can
create another kind of deformation by doing exactly the reverse:
lengthening the base of a triangle while decreasing its height.
Differently from a spiky triangle with two long links and a short one,
this results in a triangle with three long links, as pictured below.

\begin{center}
\includegraphics[height=0.5cm]{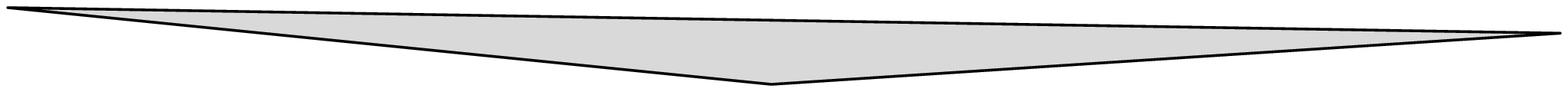}
\end{center}

\noindent
On the surface, it would seem that such a triangle should be even more
likely than a spike, because there is no need for restricting the
distance between vertices and thus no corresponding entropy `penalty'.
However, the fact that there is no short link on this kind of triangle
implies that, differently from a spike, a deformation like this cannot
appear as only a local phenomenon. Namely, an elongated triangle without
any short links can only have neighbouring triangles that are all
elongated in the same way; otherwise, there would be at least one
triangle with a long base and a not-so-small height, resulting in a
large area that is exponentially suppressed by the action. Since any
two triangles on a simplicial manifold can be connected by a path of
neighbouring triangles, it immediately follows that if just one triangle
is elongated, then all of them have to be. The result is a configuration
that is elongated as a whole -- a basically one-dimensional construct
that we call a `needle'.

\begin{center}
\includegraphics[height=1.75cm]{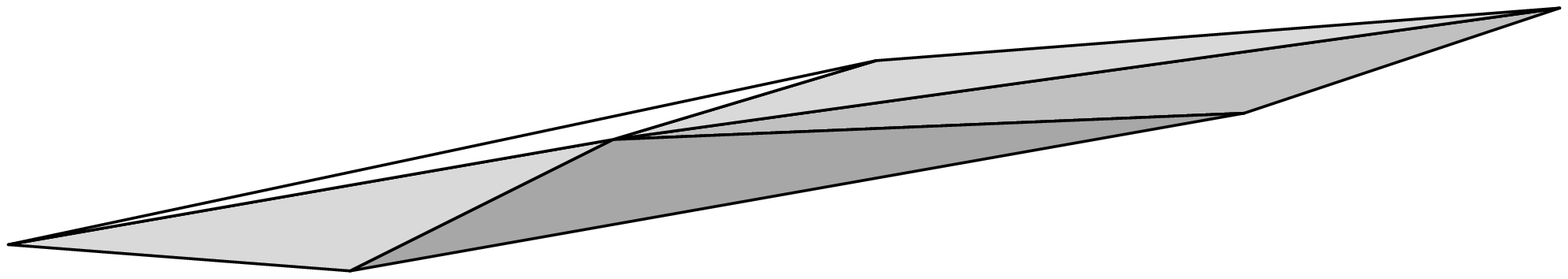}
\end{center}

\noindent
We can estimate the likelihood of a needle being created by the same
kind of power-counting argument as we used before. Since we now consider
the entire configuration, we have to take into account the model's
bosonic zero mode; we can deal with it here by keeping one vertex position
fixed. Consequently, this vertex then contributes only a factor of 1.
One other vertex can be freely chosen from within a sphere of
radius $l$ around the fixed one, thus contributing an entropy factor
of $\sim l^4$. Once this vertex has been chosen, the direction of the needle
has been determined; the $n_0 - 2$ other vertices will have one `free'
coordinate of order $\sim l$ corresponding to this direction, and three
coordinates that must be restricted to lie within a distance of
$\sim 1/l$ from the needle's principal axis so that the triangle areas
can all remain approximately constant. Thus, we get another factor
$\sim l^{-2(n_0 - 2)}$. Putting them all together and integrating as
before, we end up with a distribution
\begin{equation}
p(l) = l^{-2n_0 + 7}
\label{bosoneedle}
\end{equation}
Obviously, this means that needles will always be strongly suppressed
except on some of the smallest configurations -- starting with $n_0 = 7$,
even a single spike will be more likely than a needle configuration --,
which is, of course, why needles have not been observed so far, spikes
causing far greater singularities. Even so, it should be noted that just
like spikes, needles are suppressed only by a power law.

\subsection{The supersymmetric case}

Next, we can try to predict what will change in the model if we add
fermionic variables to it.

The physical reason behind doing this is the hope that fermionic repulsion
should keep the vertices from moving too close to each other, thus preventing
the creation of spikes that rely on the possibility of arbitrarily shortening
links between vertices. On the other hand, given the arguments above, it
might be expected that the presence of fermions could enhance needles, since
a needle configuration consists of vertices that are all far removed from
each other.

From a more technical point of view, the inclusion of fermions results
in an extra factor $| {\cal M}_{IJ} |$ that is a product of link lengths.
Since spikes and needles are both governed by powers of link lengths, it
seems likely that the distributions of both could be substantially altered
by the fermionic determinant. We can try to predict the changes by including
the determinant in our power-counting arguments. Even at this point,
however, it seems already obvious that the results can only be power laws
again; we will not see an exponential suppression of either spikes or
needles. Nevertheless, we can at least hope for a well-defined partition
function.

\subsubsection{Spikes in the model with fermions}

As can be seen from the definition of ${\cal M}_{IJ}$, it is structured
in such a way that with each vertex $\langle i \rangle$ can be associated
two rows and two columns that contain only distances between neighbours
of $\langle i \rangle$. If $\langle i \rangle$ is a spike of length $l$,
we therefore have two rows and columns with entries that are all of order
$\sim 1/l$. Since the diagonal elements of ${\cal M}_{IJ}$ are all zero,
each non-vanishing term in the sum that constitutes the determinant
$| {\cal M}_{IJ} |$ must contain exactly four entries from these two rows
and columns. Thus, each spike contributes a factor $l^{-4}$ to the
fermionic determinant.

At a first glance, we would conclude from this that a single spike should
now have a distribution $p(l) \sim l^{-9}$ rather than $\sim l^{-5}$ as
before, and that the formerly worse case of having $k$ spikes grouped
around a central vertex should have $p(l) \sim l^{-4k - 1}$ rather than
$\sim l^{-1}$, thus rendering spikes, if not harmless, then at least no
longer capable of causing actual divergences of the partition function.

However, this argument is a bit too naive, since the fermionic matrix
also contains entries of order $\sim l$ corresponding to links on the
boundary of $\langle i \rangle$. Therefore it seems likely that at least in
its leading terms the determinant should be more singular than the power
$\sim l^{-4k}$ we just claimed. Nevertheless, the end result should still
be one of spikes being more strongly suppressed than in the purely bosonic
model. We will examine this point further in the numerical part.

\subsubsection{Needles in the model with fermions}

Since needles are a global phenomenon, we will for an estimate of their
link length distribution have to consider the entire determinant and its
contribution to the partition function as a whole. More precisely, since
we want to see whether the determinant turns needles into a more severe
singularity, we need to know the behaviour of the determinant's leading
term.

We already know that on a needle configuration every vertex position has
one `large' component of order $\sim l$ and three `small' components of
order $\sim l^{-1}$. From the definition of ${\cal M}_{IJ}$, it is clear
that this means we can find terms of order $\sim l$ in each row and each
column of the determinant. At a first glance, it would therefore seem
obvious that the leading term should simply be a product of some of these
large components, and since ${\cal M}_{IJ}$ is a
$2(n_0 - 1) \times 2(n_0 - 1)$ matrix, the result should be of order
$\sim l^{2(n_0 - 1)}$. Inserting this into (\ref{bosoneedle}) would give us
\begin{equation}
p(l) \sim l^{-2n_0 + 7 + 2(n_0 - 1)} = l^5
\end{equation}
leading to a fatal singularity and an ill-defined partition function.

Fortunately, however, this turns out to be wrong, in that some of the
largest terms of $| {\cal M}_{IJ} |$ can be shown to cancel out. More
precisely, we will see that both the leading and the next-to-leading
terms are identically zero, leaving us with a contribution
$\sim l^{2(n_0 - 5)}$ from the determinant and thus a total distribution
\begin{equation}
p(l) \sim l^{-2n_0 + 7 + 2(n_0 - 5)} = l^{-3}
\label{fermineedle}
\end{equation}
which is perfectly integrable.

To prove the disappearance of the first two terms of $| {\cal M}_{IJ} |$,
we will use arguments from perturbation theory. Define a more general
matrix
\begin{equation}
{\cal M}_{IJ} (\epsilon) \equiv - {\cal M}_{IJ}^0 + i \epsilon {\cal M}_{IJ}^1
\equiv - \left( \begin{array}{cc} f_{ij}^0 & 0 \\ 0 & f_{ij}^0 \end{array}
\right) + i \epsilon \left( \begin{array}{cc} f_{ij}^3 & f_{ij}^1
+ i f_{ij}^2 \\ f_{ij}^1 - i f_{ij}^2 & -f_{ij}^3 \end{array} \right)
\label{perturbed}
\end{equation}
For $\epsilon = 1$, this describes a `typical' configuration from the bulk
of the distribution, whereas for $\epsilon = 0$, it defines an entirely
one-dimensional construct. (Specifically, a construct pointing in the
0-direction; but the direction is irrelevant since the model is rotationally
invariant.) An {\em almost} one-dimensional configuration, then, can be
described as a small perturbation around ${\cal M}_{IJ} (0)$. We will look
successively at the zeroth, first, and second orders of the perturbation
expansion. For now, we assume $n_0$ to be even.

To $0th$ order, note that $f_{ij}^0$ has at least one vanishing eigenvalue,
since for an entirely one-dimensional configuration
$X_i^\mu = (X_i^0, 0, 0, 0)$ the action in (\ref{finalpart}) gains an
additional invariance, namely the transformation
$\psi_i \to \psi_i + \alpha X_i^0$, and thus also an additional zero
mode.\footnote{This is a discrete remnant of a much more general invariance
$\psi (\xi) \to \psi (\xi) + \alpha h (X^0 (\xi))$ of the continuum action
(\ref{internaction}), where $h$ is an arbitrary function.} The corresponding
zero eigenvector of $f_{ij}^0$ is
\begin{equation}
v_i^0 = X_i^0 - X_{n_0}^0
\end{equation}
Consequently, ${\cal M}_{IJ}^0$ must have at least two zero eigenvectors
$a_I^0$ and $b_I^0$, which we can write as
\begin{equation}
a_I^0 = \binom{v_i^0}{0} \qquad b_I^0 = \binom{0}{v_i^0}
\label{dandufromw}
\end{equation}
Thus, we have $| {\cal M}_{IJ}^0 | = 0$, and therefore no terms of order
$\epsilon^0$ in the determinant of ${\cal M}_{IJ} (\epsilon)$.

Now consider a small perturbation $i \epsilon {\cal M}_{IJ}^1$, which will
change the (twice degenerate) eigenvalues $\lambda_i$ of ${\cal M}_{IJ}^0$
to $\lambda_i + i \epsilon \Delta_i$ and $\lambda_i - i \epsilon \Delta_i^*$,
respectively. In particular, we have a first order correction to the zero
eigenvalues of ${\cal M}_{IJ}^0$ that I will call $\Delta_0$. It obeys the
equation
\begin{equation}
\left| \begin{array}{cc}
a_I^0 {\cal M}_{IJ}^1 a_J^0 - \Delta_0 & a_I^0 {\cal M}_{IJ}^1 b_J^0 \\
b_I^0 {\cal M}_{IJ}^1 a_J^0 & b_I^0 {\cal M}_{IJ}^1 b_J^0 - \Delta_0
\end{array} \right| = 0
\end{equation}
or if we use (\ref{perturbed}) and (\ref{dandufromw}) to re-write it in
terms of $v_i^0$,
\begin{equation}
\left| \begin{array}{cc}
v_i^0 f_{ij}^3 v_j^0 - \Delta_0
& v_i^0 f_{ij}^1 v_j^0 + i v_i^0 f_{ij}^2 v_j^0 \\
v_i^0 f_{ij}^1 v_j^0 - i v_i^0 f_{ij}^2 v_j^0
& - v_i^0 f_{ij}^3 v_j^0 - \Delta_0
\end{array} \right| = 0
\end{equation}
But we already know that the $f_{ij}^\mu$ are all antisymmetric, which implies
$v_i^0 f_{ij}^\mu v_j^0 = 0$ and thus $\Delta_0 = 0$ from the equation above.
In other words, both zero eigenvalues remain intact to first order, which
means there will be no terms proportional to $\epsilon^2$ either. Therefore,
the leading terms of $| {\cal M}_{IJ} |$ must be at least of order
$\epsilon^4$. As the numerical simulations show, terms of this order do
indeed show up generically.

Accordingly, we can now predict that the determinant will contribute a
leading term that consists of $2n_0 - 6$ `long' components and 4 `short'
ones, $i.\,e.$ it will be of order $l^{2n_0 - 6} l^{-4} = l^{2n_0 - 10}$.
Inserting this into our power-counting argument leads to the distribution
(\ref{fermineedle}).

We still have to deal with the case of odd $n_0$, which turns out to lead
to a somewhat different result. In this case, $f_{ij}^0$ is an even by even
antisymmetric matrix, which means that it has pairs of eigenvalues
$\lambda$ and $-\lambda$. In particular, this means that it must have an
even number of zero eigenvectors. Since we know, by the same arguments as
above, that it has at least one of these eigenvectors, it therefore must
have at least two, which we call $v_i^0$ and $w_i^0$. Accordingly, we
find that ${\cal M}_{IJ}^0$ has not two but four zero eigenvectors, which
we write as
\begin{equation}
a_I^0 = \binom{v_i^0}{0} \qquad b_I^0 = \binom{w_i^0}{0} \qquad
c_I^0 = \binom{0}{v_i^0} \qquad d_I^0 = \binom{0}{w_i^0}
\label{dandutwicefromw}
\end{equation}
This leads to a somewhat more complicated expression for the first order
correction $\Delta_0$, namely
\begin{equation}
\left| \begin{array}{cccc}
a_I^0 {\cal M}_{IJ}^1 a_J^0\!-\!\Delta_0 & a_I^0 {\cal M}_{IJ}^1 b_J^0
& a_I^0 {\cal M}_{IJ}^1 c_J^0 & a_I^0 {\cal M}_{IJ}^1 d_J^0 \\
b_I^0 {\cal M}_{IJ}^1 a_J^0 & b_I^0 {\cal M}_{IJ}^1 b_J^0\!-\!\Delta_0
& b_I^0 {\cal M}_{IJ}^1 c_J^0 & b_I^0 {\cal M}_{IJ}^1 d_J^0 \\
c_I^0 {\cal M}_{IJ}^1 a_J^0 & c_I^0 {\cal M}_{IJ}^1 b_J^0
& c_I^0 {\cal M}_{IJ}^1 c_J^0\!-\!\Delta_0 & c_I^0 {\cal M}_{IJ}^1 d_J^0 \\
d_I^0 {\cal M}_{IJ}^1 a_J^0 & d_I^0 {\cal M}_{IJ}^1 b_J^0
& d_I^0 {\cal M}_{IJ}^1 c_J^0 & d_I^0 {\cal M}_{IJ}^1 d_J^0\!-\!\Delta_0
\end{array} \right| = 0
\end{equation}
which after insertion of (\ref{perturbed}) and (\ref{dandutwicefromw})
can be reduced to
\begin{equation}
\left[ \Delta_0^2
+ \sum_{k = 1}^{3} \left( v_i^0 f_{ij}^k w_j^0 \right)^2 \right]^2 = 0
\end{equation}
By definition, the $f_{ij}^\mu$ are linear functions of
$X_{\omega_{ij}}^\mu$ and $X_{\omega_{ji}}^\mu$, so that we can write
$f_{ij}^\mu \equiv f (X_{\omega_{ij}}^\mu, X_{\omega_{ji}}^\mu)$; and
the same is true for the eigenvectors, where we have
$w_i^\mu \equiv w (X_i^\mu)$. Using this, we find
\begin{eqnarray}
v_i^0 f_{ij}^k w_j^0
& \equiv & v (X_i^0) \ f (X_{\omega_{ij}}^k, X_{\omega_{ji}}^k) \ w (X_j^0)
\nonumber \\
& = & v (X_i^0) \ f (X_{\omega_{ij}}^k, X_{\omega_{ji}}^k)
      \left( w (X_j^0) + w (X_j^k) \right) \nonumber \\
& = & v (X_i^0) \left( f(X_{\omega_{ij}}^0, X_{\omega_{ji}}^0)
      + f (X_{\omega_{ij}}^k, X_{\omega_{ji}}^k) \right)
      \left( w (X_j^0) + w (X_j^k) \right) \nonumber \\
& = & v (X_i^0) \ f (X_{\omega_{ij}}^0 + X_{\omega_{ij}}^k,
      X_{\omega_{ji}}^0 + X_{\omega_{ji}}^k) \ w (X_j^0 + X_j^k) \nonumber \\
& = & 0
\end{eqnarray}
where we first used $f_{ij}^k w_j^k = 0$ to add a term
$v (X_i^0) f (X_{\omega_{ij}}^k, X_{\omega_{ji}}^k) w (X_j^k)$;
then used $v_i^0 f_{ij}^0 = 0$ to add
$v (X_i^0) f (X_{\omega_{ij}}^0, X_{\omega_{ji}}^0) (w (X_j^0) + w (X_j^k))$;
and finally applied the linearity of $f$ and $w$.

Thus, with some extra effort, we again find $\Delta_0 = 0$ for odd $n_0$
just as we did for even $n_0$. The situation is nevertheless different,
since ${\cal M}_{IJ}$ now has {\em four} zero eigenvalues to both zeroth
and first order rather than two, meaning that the second-order terms are
now proportional to $\epsilon^8$ rather than $\epsilon^4$.

We thus find that for odd $n_0$ the leading term of $| {\cal M}_{IJ} |$
should be of order $l^{2n_0 - 18}$, making the overall distribution of
the link length
\begin{equation}
p(l) \sim l^{-2n_0 + 7 + 2n_0 - 18} = l^{-11}
\end{equation}
This, however, makes a long needle far less likely than even a single
spike (which we argued above should have at least $p(l) \sim l^{-9}$
in the fermionic case). Thus, it would seem that we should not expect
to see needles for odd $n_0$.

\subsection{Higher powers of the determinant}

Once we have established the leading behaviour of the determinant, the
effect of having higher powers $\gamma > 1$ of ${\cal M}_{IJ}$ in the
partition function is not difficult to predict. For even $n_0$, the
determinant should contribute a factor $\sim l^{(2n_0 - 10) \gamma}$,
making the distribution
\begin{equation}
p(l) \sim l^{2n_0 (\gamma - 1) + 7 - 10 \gamma}
\end{equation}
Even for a $\gamma$ only marginally larger than 1, the result is obviously
catastrophic, since the power of the distribution starts to grow linearly
with $n_0$ and will sooner or later reach $p(l) \sim l^{-1}$, at which point
the partition function will blow up.

For odd $n_0$, the result will be essentially the same, since the exponents
simply differ by a constant; for large enough $n_0$, the partition function
will still become ill-defined.

In other words, we expect an ill-defined partition function if we have no
fermions in the model, but also if we include too many fermions; it seems
that the only existing model should be the one in which bosonic and
fermionic fields are susy-balanced.

\subsection{Higher dimensions}

As promised in the beginning of this chapter, we can easily generalize
our arguments for the large $l$ region of the model to the
higher-dimensional versions $d = 6$ and $d = 10$.

On a needle with a length scale of order $\sim l$ in $d$ dimensions, we again
have one vertex with entirely fixed coordinates, because of the zero mode;
one vertex whose $d$ coordinates can all vary freely within a sphere of
radius $l$; and $n_0 - 2$ vertices with one `long' component in the
direction of the needle and $d - 1$ `short' components perpendicular to
this direction. The fermionic matrix now has
$(d - 2)(n_0 - 1) \times (d - 2)(n_0 - 1)$ entries; by the same
arguments as before, the leading term of the determinant will pick up
`long' components of order $\sim l$ from all but four rows and `short'
components of order $\sim 1/l$ from the remaining ones, so that the
contribution from $| {\cal M}_{IJ} |^\gamma$ becomes
$l^{\gamma (d - 2)(n_0 - 5)}$. Putting it all together and integrating
over all but one component gives us the distribution of the link length
\begin{equation}
p(l) = l^{d - 1} l^{n_0 - 2} l^{-(d - 1)(n_0 - 2)} l^{\gamma (d - 2)(n_0 - 5)}
= l^{-2d + 5 + (\gamma - 1)(d - 2)(n_0 - 5)}
\end{equation}
For $\gamma > 1$, we again find a partition function that will inevitably
blow up if we just take $n_0$ to be large enough. In the supersymmetric case,
however, the $n_0$-dependent term once more drops out completely and we
find
\begin{equation}
p(l) = l^{-2d + 5}
\label{disthigher}
\end{equation}
In particular, we predict $p(l) = l^{-7}$ for $d = 6$ and
$p(l) = l^{-15}$ for $d = 10$.

In general, then, we would expect the same qualitative picture in any
dimension: the model should have a well-defined partition function in
the supersymmetric case, but with a power-law tail of the link length
distribution that is dominated by one-dimensional, needle-like
configurations, independently of the system size.

\subsection{Exact solution for the tetrahedron}

To further support the power-counting arguments of the preceding sections,
we can, as a last point before moving on to the numerical part, solve the
model exactly for the smallest possible configuration, $i.\,e.$ the
tetrahedron, and show directly that in this case the distribution of the
link length is indeed $p(l) \sim l^{-3}$.

First we have to choose a suitable coordinate system. Because of the
zero mode, we can fix one of the vertices at the origin,
$X_1 = (0, 0, 0, 0)$. Also, we can define the axes, which we will call
$l$, $x$, $y$, and $z$, in such a way that we have one vertex lying on
the $l$ axis, one in the $(l, x)$ plane, and one in the $(l, x, y)$
hyperplane: $X_2 = (l_2, 0, 0, 0)$, $X_3 = (l_3, x_3, 0, 0)$,
$X_4 = (l_4, x_4, y_4, 0)$.

The bosonic action and fermionic determinant can in these coordinates be
calculated as \cite{dynaschild, largenlimit}
\begin{eqnarray}
S_B & = & l_2^2 x_3^2 + \left( x_3 l_4 - l_3 x_4 \right)^2 + x_4^2 l_2^2
+ \left( \left( l_2 - l_3 \right) x_4 - \left( l_2 - l_4 \right) x_3 \right)^2
\nonumber \\
& & \hspace{1cm} + 2 y_4^2 \left( l_2^2 + l_3^2 + x_3^2 - l_2 l_3 \right) \\
| {\cal M}_{IJ} | & = & l_2^2 \, x_3^2 \, y_4^2
\end{eqnarray}
Inserting this into (\ref{finalpart}), we find for the partition function
\begin{equation}
Z (4, 1) = \int_{-\infty}^\infty d l_3 d l_4 d x_4
 \int_0^\infty d l_2 d x_3 d y_4 \ l_2^5 x_3^4 y_4^3 \ e^{-S_B}
\end{equation}
The integrations over $l_3$, $l_4$, $x_4$, and $y_4$ can all be performed,
leading to
\begin{equation}
Z (4, 1) = \int_0^\infty dl_2 dx_3 \
\frac{l_2^4 \, x_3^3}{\sqrt{\frac{3}{4} l_2^2 + x_3^2}^3} \
e^{-\frac{4}{3} l_2^2 x_3^2}
\end{equation}
While the last two integrations can also be carried out by an appropriate
substitution of variables, thus solving the integral completely, we are
more interested in the large $l_2$ behaviour of the integrand. Because of
the exponential part, for a given value of $l_2$ the second integration
variable will make a significant contribution only in the interval
$0 \le x_3 \le 1/l_2$. If $l_2$ is large, this interval
becomes small, and
we can regard the integrand as approximately constant across this range.
Thus, we can simply replace $x_3 \to l_2^{-1}$ and $\int dx_3 \to l_2^{-1}$,
and find that the integral becomes
\begin{equation}
Z (4, 1) \sim Z_{bulk} + \int_L^\infty dl_2 \ l_2^{-3}
\end{equation}
where $Z_{bulk}$ is the contribution of the $l_2 < L$ region to the partition
function, and $L$ is large.
We can interpret the integrand as a probability distribution for $l_2$,
valid in the range $l_2 > L$. But $l_2$ is nothing but the length of the
link that connects the vertices $\langle 1 \rangle$ and $\langle 2 \rangle$.
Since there is nothing to distinguish this link from any other on the
triangulation, we can interpret $p(l) \sim l^{-3}$ as the distribution
for {\em all} link lengths, thus arriving at the same result as before.

\section{Simulations of the surface model}

Our main objective for the numerical simulations will now be to show
whether or not our arguments from the preceding sections are correct.
Basically, this means we want to prove (or disprove) the following three
claims for the model with fermions: spikes are no longer dominant in this
case; needles become dominant in the large $l$ region, and actually cause
the partition function to diverge when $\gamma > 1$; and needles
obey a link length distribution $p(l) \sim l^{-3}$ in the supersymmetric
model, or $p(l) \sim l^{2n_0 (\gamma - 1) + 7 - 10 \gamma}$ in general.

\subsection{Observables}

To these ends, we first need to define at least two observables that
can function as indicators for the presence of spikes and of needles,
respectively. We also have to choose a convenient way of measuring
the link length distribution.

\subsubsection{Gyration radius and the minimal circle around a vertex}

As described in section 4.2.1, the presence of a spike of length $l$ implies
that the circle of links around this spike will have a length of order
$\sim l^{-1}$. Also, on a configuration dominated by spikes, we can expect
the gyration radius (which measures the average extent of the system) to
be proportional to the length of the largest spike. Therefore, we would
for a spiky triangulation expect an inverse proportion between the gyration
radius, which because of the delta function $\delta^4 (\sum X_i)$ has the
simple form
\begin{equation}
R \equiv \sqrt{\frac{1}{N} \sum X_i^2}
\end{equation}
and the length of the smallest circle of links around a vertex
\begin{equation}
l_c \equiv \min_{\langle i \rangle}
\sum_{\langle jk \rangle: \langle ijk \rangle}
\sqrt{(X_j - X_k)^2}
\end{equation}
Also, $l_c$ should become arbitrarily small. Conversely, when spikes are
absent we would expect no such inverse proportion, and small values of
$l_c$ should be suppressed. (In fact, for a needle
configuration we would expect a {\em positive} correlation between $l_c$
and $R$, since on a needle all link lengths should be of the same order
and the gyration radius should be proportional to the average link length.)

\subsubsection{Principal component analysis}

To decide whether a given configuration is a needle, we have to determine
whether it behaves like a one-dimensional structure as a whole. To this end,
we first define the correlation matrix
\begin{equation}
C^{\mu\nu} \equiv \frac{1}{N} \sum_{\langle i \rangle} X_i^\mu X_i^\nu
\end{equation}
The eigenvalues $r^\mu$ of this matrix can be interpreted as the system's
square extent in four independent directions. (Note that the sum of these
eigenvalues equals the gyration radius squared,
$\sum r^\mu = {\rm Tr} \, C^{\mu\nu} = \frac{1}{N} \sum X_i^2 = R^2$.)
The eigenvector belonging to the largest eigenvalue is called the principal
axis and designates the direction along which the configuration has its
greatest extent.

We can now take each vertex $\langle i \rangle$ on the configuration and
calculate its projection on the principal axis, $l_i$, as well as its
distance from it, $d_i$. From these, we can define the system's overall
length and thickness as, respectively,
\begin{equation}
L \equiv \max_{\langle i \rangle \langle j \rangle} \left( l_i - l_j \right)
\end{equation}
\begin{equation}
D \equiv \max_{\langle i \rangle} d_i
\end{equation}
The ratio of length to thickness can be used to determine the dimensionality
of the system; for sufficiently large values of $L/D$ we would say that the
configuration is essentially one-dimensional. In fact, we can put this in
more quantitative terms by remembering that, according to our construction,
a needle of length $L$ should have a thickness of $\sim L^{-1}$. Since the
gyration radius of a one-dimensional configuration is just the same as its
length, we can say that for a needle with gyration radius $R$, we expect
a length-to-thickness ratio of order $\sim R^2$. Conversely, for a
configuration with spikes, or indeed any four-dimensional system, we should
see a ratio of order $\ge 1$.

\subsubsection{The distribution of link lengths}

While it would certainly be possible to directly measure the distribution
of link lengths, we can make our job easier by noting that, as mentioned
before, on a needle there should be an approximate proportionality
between a typical link length and the gyration radius, $l \sim R$. Since
the gyration radius is something we measure anyway, we can simply create
a distribution of $R$ and compare it to the power law $p(R) \sim R^{-3}$.

\subsection{Implementation}

For the most part, implementation of the model is straightforward and not
too difficult. The greatest problem comes from the rapid increase of $CPU$
time required for the calculation of the fermionic determinant, which
forces us to keep the system size relatively small.

The part of the algorithm that describes and updates the triangulation
itself is based on an earlier program that was written for four-dimensional
manifolds \cite{simplicomp}. It was re-written in C++ for this project, and
becomes much simpler in its two-dimensional version; but the principles
remain the same, and I will therefore not describe it here in detail.

The fermionic determinant $| {\cal M}_{IJ} |$ is calculated with a standard
algorithm based on decomposition of ${\cal M}_{IJ}$ into two triagonal
matrices that was taken from \cite{recipes}. This algorithm requires of order
$o(n_0^3)$ operations for each calculation of the determinant; this is the
fastest general method available for this kind of problem.\footnote{There
exists a more specialized method for calculating the determinant of a matrix
that differs only in a few elements from a matrix with a known determinant,
which could in principle be used here. This method requires only of order
$o(n_0^{5/2})$ operations; however, it would have to be performed several
times in succession for each updating step, leading to a larger pre-factor
for the actual number of operations. Thus, while this method would eventually
win over the more general one, it is in fact slower on the small systems that
we have to restrict ourselves to.} Furthermore, we have to perform of order
$o(n_0)$ geometrical transformations and field updates to change every part
of the triangulation. Finally, we have to repeat this cycle often enough for
these local changes to propagate across the entire configuration; some short
test runs show that this leads to yet another factor $o(n_0)$. All in all,
then, we find that the $CPU$ time required to generate a given amount of
measurements grows with the system size at least like $\sim n_0^5$. In
practice, this limited our studies to surfaces with up to 28 triangles
($n_0 = 16$), plus some preliminary measurements for surfaces with 60
triangles ($n_0 = 32$). The matrix ${\cal M}_{IJ}$ itself is stored and
updated as required during the simulations, so that it, at least, does not
have to be completely re-calculated in each step.

We have to perform two separate kinds of updates: changes in the geometrical
structure, and updates of the vertex coordinates. The geometry is altered
using the $(p, q)$ moves described in section 2.2.2. As noted there, as long
as we are studying a canonical partition function in two dimensions, the
only move we actually need is the flip $(2, 2)$.

Apart from the symmetry factors $C(T)$, which are automatically taken care
of by the way the $(p, q)$ moves work, there is no weight associated with
the geometrical structure that would have to be included in the Metropolis
question. However, both the bosonic action and the fermionic matrix are
changed by the move. In particular, if we call the original triangles
involved in the move $\langle ijk \rangle$ and $\langle jil \rangle$, we find
for the change in the bosonic action
\begin{eqnarray}
\Delta S_B & = & \frac{4}{A} \left\{ \left( X_{kl}^2 - X_{ij}^2 \right)
\left( X_{ij}^2 + X_{ik}^2 + X_{il}^2 + X_{jk}^2 + X_{jl}^2 + X_{kl}^2 \right)
\right. \nonumber \\
& & \hspace{1cm} + \left. \left( X_{il}^2 - X_{jk}^2 \right)
\left( X_{ik}^2 - X_{jl}^2 \right) \right\}
\end{eqnarray}
The fermionic matrix has to be updated in twelve different $2 \times 2$
`blocks'. First of all, since the link $\langle ij \rangle$ is removed by
the transformation, all elements associated with it have to be set to zero,
${\cal M}_{ij}^{ab} = {\cal M}_{ji}^{ab} = 0$. Secondly, the newly created
link $\langle kl \rangle$ generates the new elements
${\cal M}_{kl}^{ab} = \frac{i}{12} (X_i^\mu - X_j^\mu) \sigma^{ab}_\mu$ and
${\cal M}_{lk}^{ab} = \frac{i}{12} (X_j^\mu - X_i^\mu) \sigma^{ab}_\mu$.
Finally, the elements associated with the remaining links of the two
triangles involved in the move have to be changed by
\begin{eqnarray}
\Delta {\cal M}_{ik}^{ab} = \frac{i}{12} \Big( X_j^\mu - X_l^\mu \Big)
\sigma^{ab}_\mu & &
\Delta {\cal M}_{ki}^{ab} = \frac{i}{12} \Big( X_l^\mu - X_j^\mu \Big)
\sigma^{ab}_\mu \nonumber \\
\Delta {\cal M}_{il}^{ab} = \frac{i}{12} \Big( X_k^\mu - X_j^\mu \Big)
\sigma^{ab}_\mu & &
\Delta {\cal M}_{li}^{ab} = \frac{i}{12} \Big( X_j^\mu - X_k^\mu \Big)
\sigma^{ab}_\mu \nonumber \\
\Delta {\cal M}_{jk}^{ab} = \frac{i}{12} \Big( X_l^\mu - X_i^\mu \Big)
\sigma^{ab}_\mu & &
\Delta {\cal M}_{kj}^{ab} = \frac{i}{12} \Big( X_i^\mu - X_l^\mu \Big)
\sigma^{ab}_\mu \nonumber \\
\Delta {\cal M}_{jl}^{ab} = \frac{i}{12} \Big( X_i^\mu - X_k^\mu \Big)
\sigma^{ab}_\mu & &
\Delta {\cal M}_{lj}^{ab} = \frac{i}{12} \Big( X_k^\mu - X_i^\mu \Big)
\sigma^{ab}_\mu
\end{eqnarray}
For a coordinate update, we choose a vertex $\langle i \rangle$ at random
and propose a change generated from the interval $[-c, c]$ with a flat
distribution, where the parameter $c$ is adjusted so as to produce reasonable
acceptance rates (about $60\% - 70\%$). If we leave out the summation
indices $\mu$ for simplicity and use the notation
$\Delta f (X_i) \equiv f (X_i^{new}) - f (X_i^{old})$, we can calculate the
change in the bosonic action as
\begin{eqnarray}
\Delta S_B & = & \frac{4}{A} \Bigg\{ o_i \left( \left( \Delta X_i^2 \right)^2
- \Delta X_i^4 \right) \nonumber \\
& & \hspace{1cm} - 2 \sum_{\langle j \rangle : \langle ij \rangle} \Big(
X_j^2 \Delta X_i^2 - 2 X_j \Delta (X_i^2 X_i) \\
& & \hspace{2.8cm} + 2 X_j \Delta X_i^2 \Delta X_i
+ 2 \Delta (X_i X_j)^2 - X_j^2 X_j \Delta X_i \Big) \nonumber \\
& & \hspace{1cm} + 2 \sum_{\langle jk \rangle: \langle ijk \rangle}
\Big( X_{jk}^2 \Delta X_i^2 + 2 (\Delta X_i)^2 X_j X_k
- (X_j + X_k) X_{jk}^2 \Delta X_i \Big) \Bigg\} \nonumber
\end{eqnarray}
The fermionic matrix has to be updated in all elements whose lattice
indices form a triangle with $\langle i \rangle$, and becomes
\begin{eqnarray}
\Delta {\cal M}_{jk}^{ab} = \frac{i}{12} \Delta X_i^\mu \sigma^{ab}_\mu & &
\Delta {\cal M}_{kj}^{ab} = -\frac{i}{12} \Delta X_i^\mu \sigma^{ab}_\mu
\end{eqnarray}
where $ijk$ is a clockwise arrangement of the vertices.

As for our chosen observables, the gyration radius $R$, the smallest circle
around a vertex $l_c$, and the correlation matrix $C_{\mu\nu}$ can all
be straightforwardly calculated, since they are all simple functions of
the bosonic coordinates $X_i^\mu$. The eigenvalues and eigenvectors of
$C_{\mu\nu}$ are determined using a standard algorithm from \cite{recipes}.

A very first test that we can run on the program is a measurement of
the average bosonic action, which we can predict from a simple scaling
argument. Say we introduce an additional scale factor $\lambda$ in front
of the bosonic action. Symbolically, counting only powers of $X$, we can
then write the partition function as
\begin{equation}
Z (n_2, \gamma, \lambda) = \int (dX)^{4(n_0-1)} \ X^{2(n_0-1) \gamma} \
e^{-\lambda X^4}
\end{equation}
If we now make the substitution $X \to Y \equiv \lambda^{1/4} X$, we find
\begin{equation}
Z (n_2, \gamma, \lambda) = \int \frac{(dY)^{4(n_0-1)}}{\lambda^{n_0-1}} \
\frac{Y^{2(n_0-1) \gamma}}{\lambda^{\frac{n_0-1}{2} \gamma}} \
e^{-Y^4} = \lambda^{-\frac{\gamma + 2}{2} (n_0-1)} \ Z (n_2, \gamma, 1)
\end{equation}
From this we can calculate the average bosonic action as
\begin{equation}
\langle S_B \rangle
= - \left. \frac{\partial \ln Z}{\partial \lambda} \right|_{\lambda = 1}
= \frac{\gamma + 2}{2} (n_0 - 1)
\end{equation}
The action can be easily measured in a short numerical run and compared
to this formula; we find that they do indeed agree.

\begin{figure}
\begin{center}
\psfrag{R}{$R$}
\psfrag{p}{$p(R)$}
\includegraphics[height=7.5cm]{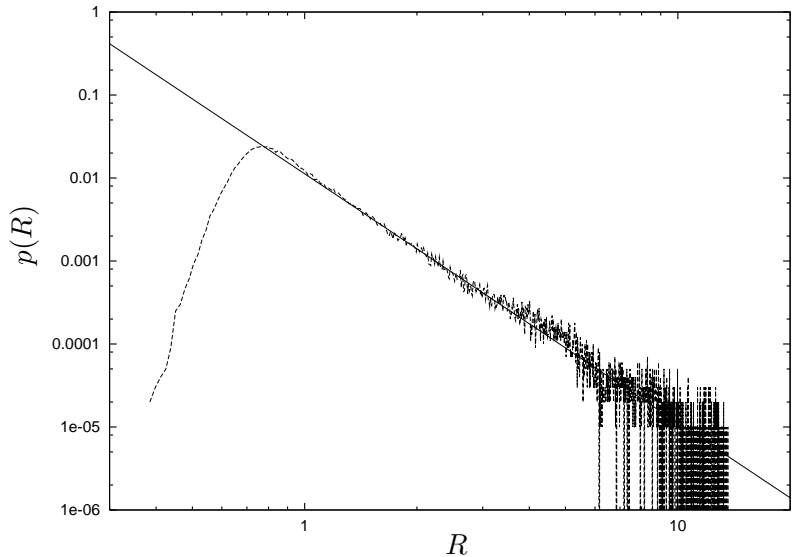}
\end{center}
\caption{\label{tetradist}The distribution of the gyration radius on
the tetrahedron, together with a best fit to a power law
$p(R) = a R^{-3}$.}
\end{figure}

Another possible test comes from the analytic calculations for the
tetrahedron, from which we actually know for certain that we should find
a gyration radius distribution $p(R) \sim R^{-3}$ for large enough $R$.
This also is reproduced by the program, as can be seen in figure
\ref{tetradist}.

\subsection{Results}

Simulations were performed for values of $n_2$ up to 28 and
$\gamma = 0, 1, 2$. On average, 100000 measurements were performed for
each combination $(n_2, \gamma)$.

\subsubsection{Appearance of spikes}

\begin{figure}
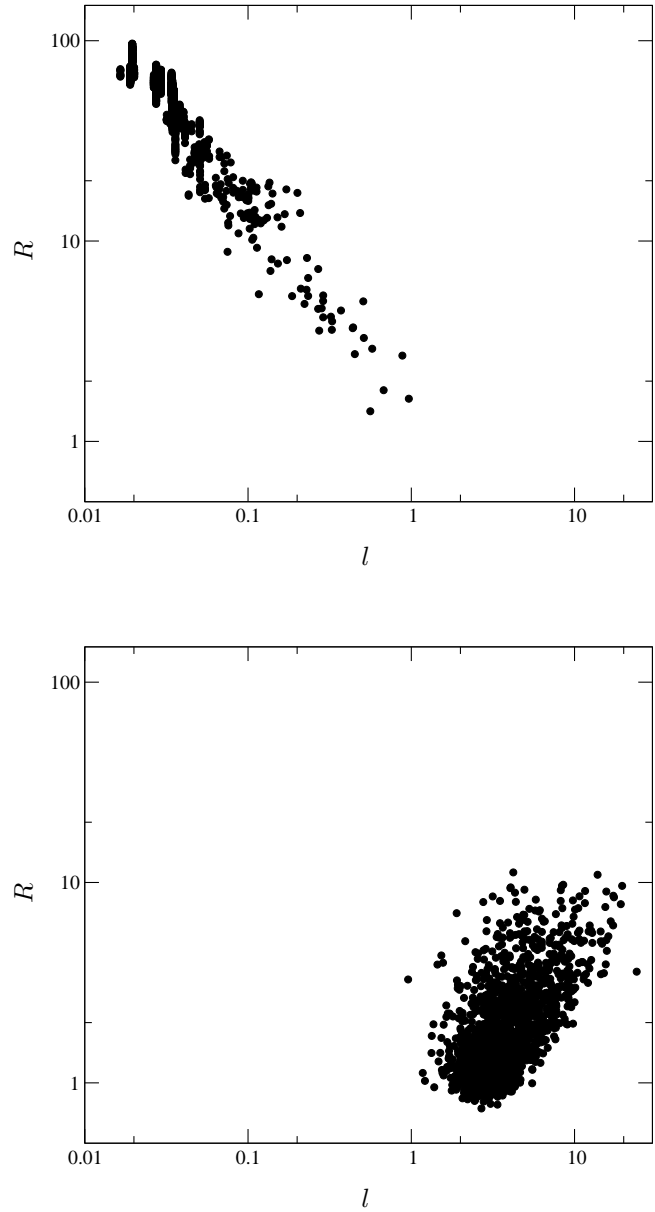

\begin{center}
\psfrag{xl}{\small $l$}
\psfrag{yl}{\small $R$}
\includegraphics[height=7.5cm]{loop_spikes_0.eps}\vspace{1cm}
\includegraphics[height=7.5cm]{loop_spikes_1.eps}
\end{center}
\caption{\label{loopcirc}The gyration radius $R$ as a function of the
shortest loop length $l_c$, both for the bosonic case (upper figure) and
the supersymmetric case (lower figure). The number of triangles is
$n_2 = 28$.}
\end{figure}

Figure \ref{loopcirc} shows the gyration radius $R$ versus the length of the
shortest circle around a vertex $l_c$ in logarithmic scale, once for
the purely bosonic model $(\gamma = 0)$ and once for the supersymmetric
model $(\gamma = 1)$, with the triangulation size $n_2 = 28$ in both cases.

The upper figure shows an obvious inverse proportionality between
$R$ and $l_c$, so we can say that the bosonic model is in fact dominated
by spikes. We already knew that, of course, but it does assure us that
our method works.

In the supersymmetric case, we see that, as expected, there is no inverse
proportionality and, in fact, really small circles are not present at all.
Instead, we see the expected positive correlation.

We can therefore conclude that the fermionic determinant does indeed
successfully suppress the appearance of spikes.

\subsubsection{Appearance of needles}

To test for the appearance of needles, we examined configurations that
were selected randomly from the region of large gyration radius ($R > 8$),
since this is the only regime where we can expect needles.

\begin{figure}
\begin{center}
\includegraphics[height=7.5cm]{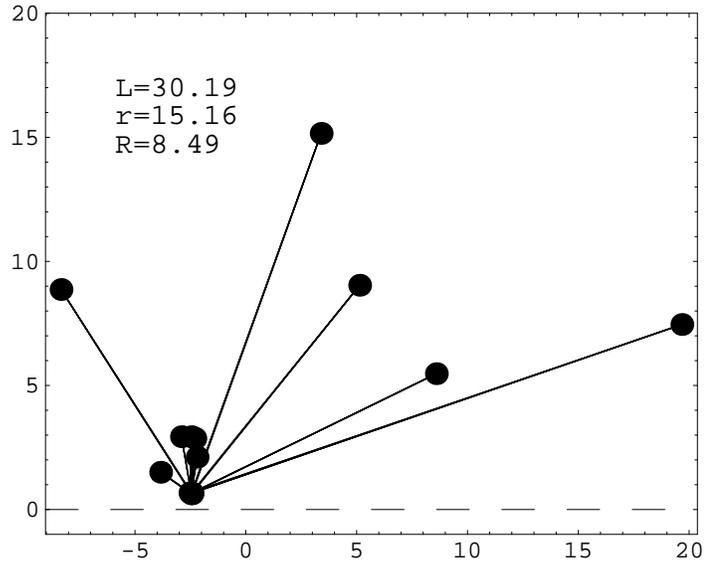}\vspace{1cm}
\includegraphics[height=7.5cm]{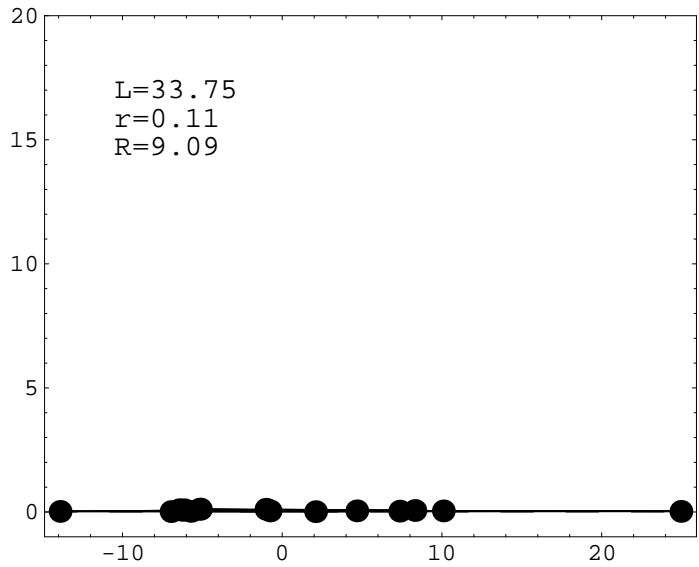}
\end{center}
\caption{\label{spikesnneedles}Snapshots of large $R$ configurations
projected onto their principal axes, both for the bosonic case (upper
figure) and the supersymmetric case (lower figure). The $x$ axis shows
the position of each point on the principal axis, whereas the $y$ axis
shows the point's distance from it. The number of triangles is $n_2 = 28$.}
\end{figure}

Figure \ref{spikesnneedles} shows snapshots of single configurations
after projection onto their respective principal axes. The purely bosonic
configuration is clearly not one-dimensional; in fact, we can see directly
that this is, not surprisingly, a configuration of spikes. The configuration
from the supersymmetric model, on the other hand, shows all vertices lying
practically on the principal axis.

Of course, these are only single examples presented here to
better illustrate the differences between bosonic and fermionic
configurations; by themselves they do not prove anything. But the estimates
of the length-to-thickness ratios for the entire runs show us the same
thing: we find $L/D = 2.26(8)$ for the bosonic case, and $L/D = 212(20)$
for the fermionic one. These should be compared to our predictions, which
were $L/D \sim o(1)$ for the bosonic case and $L/D \sim o(R^2)$ for the
fermionic one, with $R$ being restricted to $8 \le R \le 10$ for this run.
Given that these are only rough estimates of the order of magnitude, the
agreement is quite reasonable.

The conclusion, then, is that needles are indeed absent (or more correctly,
strongly suppressed) in the bosonic case, while being dominant in the
large $R$ region of the fermionic model.

\subsubsection{The distribution of $R$}

Finally, we wanted to measure the distribution of the gyration radius.
Here we encounter a few technical difficulties due to the power-law
tail of the distribution. On the one hand, the tail contains only a few
percent of all possible configurations, meaning that if we let the
algorithm range freely it will spend most of its time in the bulk of
the distribution, drastically increasing the time needed for measurements
since we are only really interested in the behaviour of the tail. On the
other hand, once the system does enter the tail, it tends to make long
excursions to extremely large $R$ every now and then because the
power-law behaviour means that there is no natural cut-off. The result
is a very large autocorrelation time that, again, means a dramatic
increase in the time requirement for each run. The simplest solution to
both problems is to set upper and lower limits on $R$ to both prevent
uncontrollable excursions and keep the system out of the bulk. The lower
limit does have the disadvantage of reducing the acceptance rate of the
algorithm because the system constantly tries to go back to the bulk,
but we found this to be a much less severe problem than the non-existence
of such a limit would be (a factor of about 3 in the acceptance rate as
opposed to the system spending about $97 \%$ of its time in the bulk if
we let it go wherever it wants).

We face a slightly different problem in those cases where the system
simply tries to run away to larger and larger $R$ because of a positive
power law, as happens when $\gamma > 1$. In these cases, we found it
sufficient to simply include an upper limit, since the system does not
really try to leave the tail in the direction of lower $R$ anyway. Of
course, the system trying to run away is just an indication of the fact
that the model is not defined in these cases; we just include them here
to show that our predictions for the power laws seem to work universally
for any $\gamma \ge 1 $.

\begin{figure}
\begin{center}
\psfrag{xl}{\small $R$}
\psfrag{yl}{}
\psfrag{xl2}{\small $R$}
\psfrag{yl2}{}
\includegraphics[height=7.5cm]{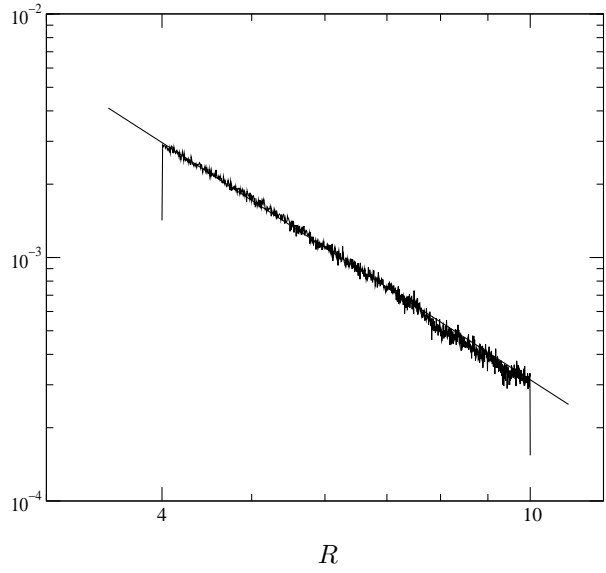}\vspace{1cm}
\includegraphics[height=7.5cm]{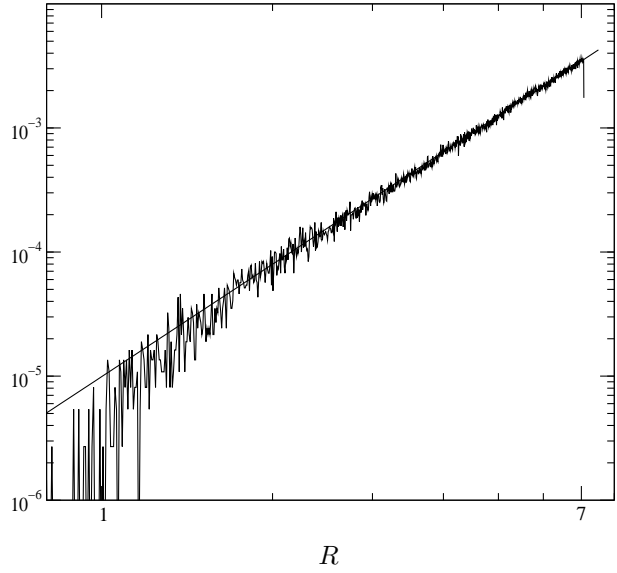}
\end{center}
\caption{\label{distofr}The distribution of the gyration radius $p (R)$
for the supersymmetric case (upper figure) and the model with Dirac
fermions (lower figure). The numbers of triangles are $n_2 = 28$ and
$n_2 = 12$, respectively.}
\end{figure}

In figure \ref{distofr}, I again present two specific results as pictures
to show what the distributions look like. The upper figure shows $p (R)$
for $n_2 = 28$ triangles in the supersymmetric case, with lower and upper
limits at $R = 4$ and $R = 10$, respectively. The only thing that
should be observed is that the distribution does indeed obey a power law.
(Note the logarithmic scale.) The same is true for the lower figure, which
shows the distribution for the model with Dirac fermions ($\gamma = 2$) and
$n_2 = 12$ triangles. The only limit here is an upper boundary at $R = 7$.

Assuming a general power law $p(R) \sim R^{\alpha}$, we can estimate
$\alpha$ for various values of $\gamma$ and $n_2$. The results, together
with our theoretical predictions, are presented in the following table: \\

\begin{center}
\begin{tabular}{|c|c|c|c|}
\hline
$\gamma$ & $n_2$  & $\alpha_{num}$ & $\alpha_{the}$ \\
\hline
1        & 20     & -2.39(4)       & -3              \\
1        & 28     & -2.41(3)       & -3              \\
\hline
1        & 10     & -5.20(1)       & -11             \\
1        & 18     & -3.25(1)       & -11             \\
\hline
2        & 8      & -0.92(2)       & -1              \\
2        & 12     & 3.02(5)        & 3               \\
2        & 16     & 7.12(12)       & 7               \\
\hline
\end{tabular}
\end{center}

\noindent \\
First of all, we can unequivocally say that in those cases where we expect
needles to be dominant to the point of causing actual divergences --
$i.\,e.$ for $\gamma > 1$ -- the agreement between prediction and
simulation is perfect, showing that our mechanism does indeed provide a
correct description of what happens in the large $R$ region.

In the supersymmetric case $\gamma = 1$ with $n_2 = 4k$ (corresponding to
an even number of vertices), the agreement is obviously
less perfect, but we are still in the general vicinity of what we expected.
There are several possible explanations for the deviation from the
predictions. Most simply, it could just be finite size effects, given
that these are really not very large systems. Another possibility is that
the results could be influenced by contributions from other configurations,
for example because our chosen range of $R$ might still be too near the
bulk of the distribution. Finally, there is of course the possibility that
our mechanism does work only as a first approximation even for large values
of $R$, but this seems to be contradicted by the positive results for
$\gamma = 2$.

As for $\gamma = 1$ and $n_2 = 4k + 2$, the results are obviously totally
off, but given the very strong suppression of needles expected from our
theoretical arguments in these cases, this apparently just means that
there are indeed yet other configurations that are more singular than
$R^{-11}$.

Overall, we would again conclude that the numerical results support our
hypothesis, although it should certainly be tried to push simulations of
the supersymmetric model to higher values of $n_2$ so as to get better
estimates of the power of the distribution.

\section{The matrix model}

We now turn to an investigation of the $IKKT$ model, which was suggested
in \cite{ikkt} as a constructive definition of (\ref{surfpath}). This
model corresponds to a complete reduction of supersymmetric Yang-Mills theory
to 0 dimensions and is therefore also called a Yang-Mills integral. It is
defined by
\begin{equation}
Z (n) = \int {\cal D} A {\cal D} \bar{\Psi} {\cal D} \Psi
e^{-S (A, \bar{\Psi}, \Psi)}
\label{matpath}
\end{equation}
where the action is again a sum of a bosonic and a fermionic part,
\begin{equation}
S (A, \bar{\Psi}, \Psi) = S_B (A) + S_F (A, \bar{\Psi}, \Psi)
\end{equation}
\begin{equation}
S_B (A) = - \frac{1}{4} {\rm Tr} \, \left[ A^\mu, A^\nu \right]^2
\end{equation}
\begin{equation}
S_F (A, \bar{\Psi}, \Psi) = - \frac{1}{2} {\rm Tr} \, \bar{\Psi}^a
\Gamma_\mu^{ab} \left[ A^\mu, \Psi^b \right]
\end{equation}
Here, $A^\mu$, $\bar{\Psi}^a$, and $\Psi^b$ are $n \times n$ Hermitian
matrices, with the entries of $\bar{\Psi}$ and $\Psi$ being Grassmann
variables. The model can again be defined as a supersymmetric model in
$d = 3$, 4, 6, or 10 dimensions. We will choose $d = 4$ for the same
reasons as above.

\subsection{Needles in the matrix model}

Remarkably, the link length distributions (\ref{disthigher}) that we
found by applying the needles scenario to the surface model are exactly
the same that have been conjectured for the eigenvalue distributions
of the matrix model (\ref{matpath}), based on the only available exact
solution that exists for $SU(2)$ \cite{moreikkt, neverendingikktstudies}.
The complete formulas are
\begin{eqnarray}
\rho_4 (\lambda) & = & \frac{6}{\sqrt{2 \pi}} \lambda^2
U_\lambda^5 \nonumber \\
\rho_6 (\lambda) & = & \frac{105}{2 \sqrt{2 \pi}} \lambda^2
\left\{ U_\lambda^9 - \frac{33}{16} U_\lambda^{13} \right\} \\
\rho_{10} (\lambda) & = & \frac{27027}{64 \sqrt{2 \pi}} \lambda^2
\left\{ 13 U_\lambda^{17}
- \frac{2261}{16} U_\lambda^{21}
+ \frac{334305}{512} U_\lambda^{25}
- \frac{5014575}{4096} U_\lambda^{29} \right\} \nonumber
\label{eigendist}
\end{eqnarray}
where
\begin{equation}
U_\lambda^a \equiv U \left( \frac{a}{4}, \frac{1}{2}, 4 \lambda^4 \right)
\end{equation}
and
\begin{equation}
U (a, b, z) = \frac{1}{\Gamma (a)} \int_0^\infty dt t^{a - 1}
(1 + t)^{b - a - 1} e^{-zt}
\end{equation}
is the Kummer U-function. While these formulas look rather unwieldy,
their large $\lambda$ behaviour can be expressed as a simple power law,
which is exactly that of the surface model:
\begin{equation}
\rho_d (\lambda) \sim \lambda^{- 2d + 5}
\end{equation}
While no solution for higher $n$ exists, it was conjectured that this
power law should apply independently of $n$, again just as we found for
the surface model.

The obvious question, then, is whether the mechanism that is
responsible for this behaviour might also be the same as in the
surface model, $i.\,e.$ whether large $\lambda$ configurations
might be one-dimensional needles. Even though the analogies are
obvious, it is not {\em a priori} clear whether we should expect
this, since we now lack the geometrical structure of the triangulations,
which in our arguments was essential for the formation of needles as
a global phenomenon. To take a different example, it is known that
spikes do {\em not} occur in the matrix model -- $i.\,e.$ the model
is well-defined in the purely bosonic case \cite{countlessikktstudies}
-- exactly because there is no concept of a local geometry or neighbourhood
relations between vertices. Nevertheless, a closer look at the behaviour
of the matrix model in the large $\lambda$ region seems definitely
worthwhile.

\subsection{Simulations of the matrix model}

Treatment of the model (\ref{matpath}) to prepare it for numerical
simulations is mostly analogous to what we did with the surface model,
and I will just briefly sketch the necessary steps.

Since we are still in $d = 4$ dimensions, we can once more use the Weyl
condition to replace the four-component spinors $\bar{\Psi}$, $\Psi$ by
two-component ones $\bar{\psi}$, $\psi$. This makes the fermionic part
of the action
\begin{equation}
S_F (A, \psi, \bar{\psi}) = - \frac{1}{2}
{\rm Tr} \, \bar{\psi}^a \sigma_\mu^{ab} \left[ A^\mu, \psi^b \right]
\end{equation}
with $\sigma_0 \equiv i \mathbbm{1}$ as before. Expanding the commutator
leads to
\begin{equation}
S_F (A, \psi, \bar{\psi}) = - \frac{1}{2}
\bar{\psi}_{ij}^a \left( A_{jk}^\mu \sigma^{ab}_\mu \delta_{li}
- A_{li}^\mu \sigma^{ab}_\mu \delta_{jk} \right) \psi_{kl}^b
\end{equation}
As with the surface model, we find that the action as it stands has
two zero modes coming from a translational symmetry both in the bosonic
and in the fermionic sector. We deal with them as before: the bosonic
zero mode is fixed by inserting a factor
$\delta^4 (\sum_i A_{ii}) \equiv \delta^4 ({\rm Tr} A)$, whereas the
fermionic one is removed by skipping the integration over one pair of
fermions, which we arbitrarily choose to be $\bar{\psi}_{nn}, \psi_{nn}$.

Integrating over the fermions gives us the determinant of the matrix
\begin{equation}
{\cal M}^{ab}_{ijkl} = \frac{1}{2}
\left( A_{li}^\mu \sigma^{ab}_\mu \delta_{jk}
- A_{jk}^\mu \sigma^{ab}_\mu \delta_{li} \right)
\end{equation}
where $i, j, k, l = 1, \ldots, n$ (but excluding the combinations
$ij = nn$, $kl = nn$), and $a, b = 1, 2$. As before, I will usually
combine the indices $\binom{a}{ij}$, $\binom{b}{kl}$ and write the
matrix as just ${\cal M}_{IJ}$. The matrix size is now
$2 (n^2 - 1) \times 2 (n^2 - 1)$.

As long as we stay in four dimensions, the determinant $|{\cal M}_{IJ}|$
is again real and non-negative. The partition function that we want to
simulate is thus
\begin{equation}
Z = \int \prod_{i < j} d^4 {\rm Re} A_{ij} \, d^4 {\rm Im} A_{ij}
\prod_i d^4 A_{ii} \ \delta^4 ({\rm Tr} A) \ |{\cal M}_{IJ}| \ e^{-S_B (A)}
\end{equation}

\subsubsection{Observables}

Since, as noted, spikes do not appear in the matrix model, our main
objective here is simply the search for needle-like configurations. We can
do this in a way similar to what we did before; however, some changes have
to be made due to the lack of a geometrical structure in this case.

Namely, this lack of a geometrical interpretation means we will not be
able to discern needles by the use of an intuitive quantity like the
length-to-thickness ratio $L/D$ that we employed for the surface model.
At least, to do so would mean that we would first have to tackle the
problem of introducing some sort of commuting coordinates to the model,
where for the moment we have only non-commuting matrices. While there
exist suggestions for how to do this \cite{evenmoreikktstudies}, we would
prefer a more direct approach. This can be found in the form of the
correlation matrix, which can be defined in analogy to the surface model
except that it now also contains quantum fluctuations:
\begin{equation}
C^{\mu\nu} \equiv \frac{1}{N} {\rm Tr} A^\mu A^\nu
\end{equation}
On a one-dimensional configuration, we would have all but one of the
eigenvalues $r_\mu$ of $C^{\mu\nu}$ going towards 0. Our main observable
will therefore be the {\em asymmetry parameter}
\begin{equation}
\eta = \frac{r_2^2 + r_3^2 + r_4^2}{R^2}
\end{equation}
where the eigenvalues have been ordered by size, $r_i \ge r_{i+1}$.
If we can show $\eta (R) \to 0$ for $R \to \infty$, and show that this
holds even with increasing $n$, we can say that, once again, we see a
domination of one-dimensional configurations in the large $R$ regime
of the model.

\subsubsection{Implementation}

In principle, implementation of the matrix model is even easier than
it was for the surface theory, since all the geometrical aspects of the
triangulations can simply be dropped. The only variables we have to
store are the bosonic matrices $A^\mu_{ij}$ and the fermionic matrix
${\cal M}_{IJ}$, and there will be no geometrical updates to
perform.

The greatest difficulty that we face is again the $CPU$ time required to
calculate the fermionic determinant. In fact, this problem becomes much
worse in the matrix model, since the size of ${\cal M}_{IJ}$ now grows
with $n^2$, meaning that the calculation of the determinant takes of order
$o(n^6)$ operations. Also, it will now take of order $o(n^2)$ operations
to update all the matrix elements, and we find in test runs that the system
now requires of order $o(n^2)$ updates for a change to spread across the
entire matrix, so that we end up with a $CPU$ demand that grows at least
with $n^{10}$. Obviously, we will not be able to go very far with this kind
of burden; in practice, we managed to perform measurements up to $SU(8)$
within reasonable timespans.

\begin{figure}
\begin{center}
\psfrag{rho}{$\rho(\lambda)$}
\psfrag{lambda}{$\lambda$}
\includegraphics[height=7.5cm]{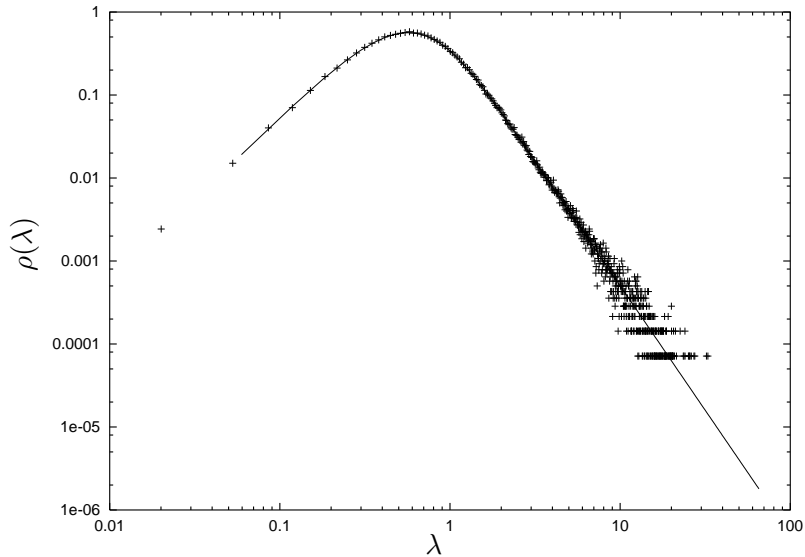}
\caption{\label{eigenmeas1}The distribution of the eigenvalues of
$A^\mu$ from the theoretical prediction (solid line) and the numerical
data (crosses).}
\end{center}
\end{figure}

The program can be tested in ways similar to those we employed in the surface
model. The average bosonic action can again be predicted from scaling
arguments to be $\langle S_B \rangle = \frac{\gamma + 2}{2} (n^2 - 1)$,
which is reproduced correctly by the program. Another possibility is
measuring the distribution of eigenvalues of $A^\mu$ for the case $n = 2$,
and comparing it to the theoretical formula (\ref{eigendist}). As figure
\ref{eigenmeas1} shows, the agreement is perfect.

\subsubsection{Results}

We will start this section with a look at the case of $SU(2)$, where an
analytical treatment of the model is possible. The calculation has been
performed in detail elsewhere \cite{moreikkt}, and I will not repeat it
here. Suffice to say that we end up with the following distribution for the
eigenvalues of $C^{\mu\nu}$:
\begin{equation}
\rho (r_\mu) \sim \delta (r_4) r_1^\alpha r_2^\alpha r_3^\alpha
(r_1^2 - r_2^2) (r_2^2 - r_3^2) (r_3^2 - r_1^2)
e^{-2 (r_1^2 r_2^2 + r_2^2 r_3^2 + r_3^2 r_1^2)}
\end{equation}
As written, this formula holds in $d = 4$ dimensions, but it remains the
same in $d = 6$ and $d = 10$ except for some additional delta functions
$\delta (r_5) \ldots \delta (r_d)$. The exponent is $\alpha = 2d - 5$ in
the fermionic case, or $\alpha = d - 3$ in the bosonic one.

From this formula we can now extract the behaviour of the model for
large values of $R^2 = r_1^2 + r_2^2 + r_3^2$. These large values will
be exponentially suppressed unless we can arrange the eigenvalues in
just such a way that the term $r_1^2 r_2^2 + r_2^2 r_3^2 + r_3^2 r_1^2$
remains of order 1. Since at least one eigenvalue must become large in
order for $R^2$ to grow, this is only possible if $r_1^2 \sim R^2$,
$r_{2, 3}^2 \le R^{-2}$. As before, we can approximately describe this
behaviour by replacing $r_1 \to R$, $r_{2, 3} \to R^{-1}$, and
$\int dr_{2, 3} \to R^{-1}$, which makes the distribution for $R$
\begin{equation}
p (R) \sim R^{-\alpha}
\end{equation}
as expected. Again, then, we find a power-law tail of the distribution
as a consequence of a flat direction in the exponential part of the
action. Also, from the arguments above we can expect the asymmetry
parameter to behave as
\begin{equation}
\eta = \frac{r_2^2 + r_3^2}{R^2} \sim \frac{1}{(R^2)^2}
\to 0
\end{equation}
so we should once more see a dominance of one-dimensional configurations
in the large $R$ region.

\begin{figure}
\begin{center}
\psfrag{RR}{$R^2$}
\psfrag{r}{$\eta$}
\includegraphics[height=7.5cm]{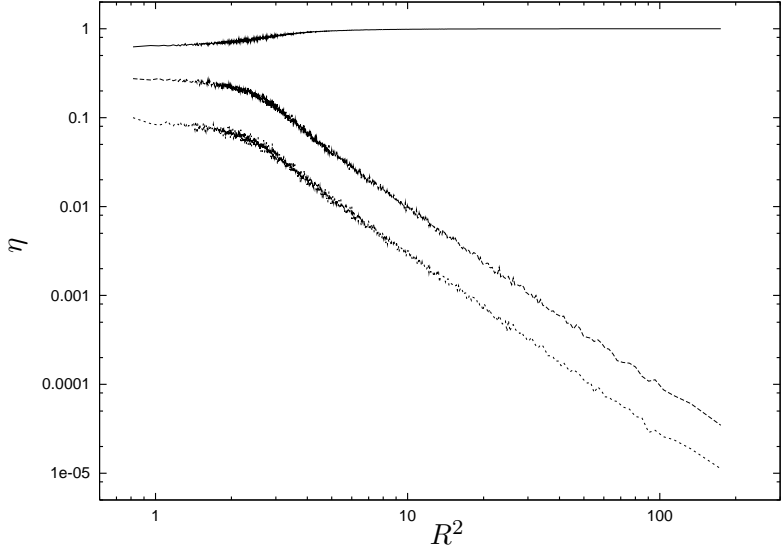}
\caption{\label{eigenmeas2}The eigenvalues $r_{1, 2, 3}^2 / R^2$ as functions
of $R^2$ for $n = 2$. Note that the data has been smoothed so that each point
actually represents an average $(\overline{R^2}, \overline{r^2/R^2})$, taken
over 100 successive values of $R^2$.}
\end{center}
\end{figure}

This is exactly what we find in numerical simulations of the $SU(2)$
case. The results of the measurements are presented in figure
\ref{eigenmeas2}; we see that only the largest eigenvalue does not go
to zero with increasing $R^2$. For the others we find a power law
$r_i^2 / R^2 \sim a_i (R^2)^{b_i}$; a best fit of the exponents
yields $b_2 = -2.04(2)$ and $b_3 = -2.02(3)$, to be compared to the
prediction $b_{2, 3} = -2$.

\begin{figure}
\begin{center}
\psfrag{R}{$R^2$}
\psfrag{eta}{$\eta$}
\includegraphics[height=7.5cm]{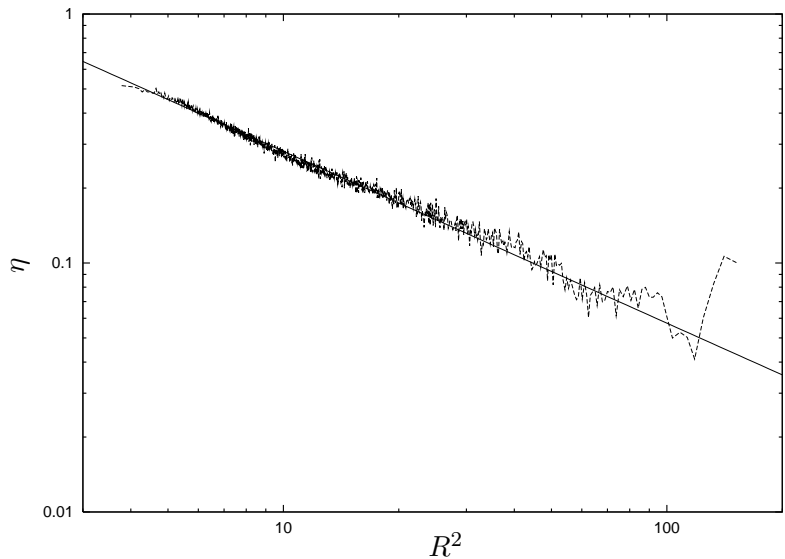}
\caption{\label{onedimens}The asymmetry parameter
$\eta = \frac{r^2_2 + r^2_3 + r^2_3}{R^2}$ as a function of $R^2$,
for $n = 4$. The data has been smoothed as explained for figure
\ref{eigenmeas2}. The solid line represents the best fit to a power law.}
\end{center}
\end{figure}

For $n > 2$, no analytic solutions exist, and we have to contend
ourselves with numerical measurements of $\eta$ and its distribution.
One typical plot, for $n = 4$, is shown in figure \ref{onedimens};
it is obvious that we again have a power law $\eta = a (R^2)^b$.
A best fit of the parameters to the data yields
$a = 1.378(7)$ and $b = -0.690(4)$. As a test to see whether this power
law remains stable when we go with $R^2$ to infinity, we repeated the
run with a lower bound imposed on the gyration radius, $R^2_{min} = 120$,
and found that the resulting curve was well described by the same
power law.

Overall, we performed simulations for matrix sizes up to $n = 8$.
Repeating the calculations from above leads to the following exponents
for the power law:

\begin{center}
\begin{tabular}{|c|c|}
\hline
$n$ & $b$        \\
\hline
3   & -0.825(2)  \\
4   & -0.690(4)  \\
5   & -0.606(4)  \\
6   & -0.562(5)  \\
8   & -0.503(29) \\
\hline
\end{tabular}
\end{center}

\noindent
While we do see a negative power in all cases, its absolute value clearly
becomes smaller with $n$, so it is vital to know whether it always stays
negative or goes towards 0 in the large $n$ limit. We can try to make a
prediction by assuming various types of finite size corrections
$b(n) = b_\infty + c/n^p$ to the limiting value; taking the power of
$n$ to be $p = 1$, $1.5$, and 2, we find $b_\infty = -0.30(3)$,
$-0.41(2)$, and $-0.47(2)$, respectively. Alternatively, if we include
$p$ as a free variable in the fit, we find $b_\infty = -0.37(11)$ and
$p = 1.26(40)$.

With some due caution because of our limited data, we would from these
values conclude that $b$ does indeed remain negative in the large $n$
limit. Thus we can say that we again see one-dimensional configurations
in the large $R$ regime, and that this phenomenon seems to be preserved
even when going to larger $n$.

\section{Discussion}

When we look at the results found for the surface and matrix models
from both theoretical conjectures and numerical simulations, we see a
consistent picture emerging for both models. The crucial point lies
in the fact that we have an area action -- a sum over triangle areas
in the discretized surface model; a commutator of matrices as the
corresponding quantity in the matrix model. This action successfully
suppresses configurations with a large two- or higher-dimensional extent,
but is insensitive to a deformation of surfaces (or matrix eigenvalues)
to an elongated state in only one dimension, provided that the other
directions are simultaneously contracted in a way that leaves the area
unchanged. The result is a distribution of link lengths (or eigenvalues)
with a power-law tail $p(l) \sim l^{-2d+5}$ for any
finite system size, a prediction that, while admittedly a conjecture
without rigorous proof, has been arrived at independently in both models
and approximately reproduced in numerical simulations. In the case of
bosonic surface models, the impact of these needle-like configurations
is blotted out by the appearance of an even more singular phenomenon,
namely multiple spikes; otherwise, the mechanism seems to be essentially
the same in both models.

An interesting point is that we can change the dimensionality of the
elongated structures by a simple adaptation of the action. For example,
if we replace the `area action' of the matrix model
$S_B \sim \rm{Tr} [A^\mu, A^\nu]^2$ by another term
\begin{equation}
S_B \sim \rm{Tr} [A^\mu, A^\nu] [A^\nu, A^\rho] [A^\rho, A^\mu]
\end{equation}
we find a new model with the same symmetries as the old one, but also
with the possibility of two-dimensional elongated structures. Namely,
the distribution of eigenvalues of the correlation matrix now becomes,
at least in the case of $SU(2)$ which we can solve,
\begin{equation}
\rho (r_i) \sim \delta (r_4) \ldots \delta (r_d) r_1^\alpha r_2^\alpha
r_3^\alpha \left( r_1^2 - r_2^2 \right) \left( r_2^2 - r_3^2 \right)
\left( r_3^2 - r_1^2 \right) e^{-24 r_1^2 r_2^2 r_3^2}
\end{equation}
If we again look for possibilities of going to large values of
$R^2$ while keeping the exponential part of the distribution constant,
we see that now a choice of $r_{1, 2} \sim R$, $r_3 \sim R^{-2}$ will
do the trick -- a choice that obviously describes a two-dimensional
surface rather than a one-dimensional needle. Taking this argument even
further, we could probably create $k$-dimensional surfaces in the large
$R$ region for any $k < d$. In particular, we might build a
ten-dimensional model that at large distances contracts to four
dimensions. While this is an intriguing possibility, for the moment we
do not yet know how to motivate such a change of the action.

An as yet unsolved question concerning our discoveries is the behaviour
of the model if we were to actually take the system size ($n_2$ or $n$,
respectively) to infinity. We can state with fair confidence that the
exponent in the power law is independent of $n$, but that still leaves
the possibility of a prefactor that might vanish with $n \to \infty$.
In fact, if we were to assume all link lengths as independent, then the
central limit theorem of statistics would tell us that the distribution
of $R$, which can be written as a sum of all link lengths, should become
Gaussian in the large $n$ limit. On the other hand, it is not
at all clear whether we should take independence of the link lengths
as a given; actually, one basic assumption in our power counting arguments
was that all link lengths should be of the same order, which directly
contradicts independence. Also, the model that presumably forms the
large $n$ limit of both models has itself an area action, so it actually
seems quite plausible that it should allow the same kind of one-dimensional
structures. Nevertheless, for a definite answer this point clearly has to
be examined further, which at the time of writing of this thesis was still
a work in progress.

\chapter{Summary}

In this thesis I have examined two different models of random geometries
and their discretizations in terms of dynamical triangulations. Both models
have been studied in simpler forms before, but turned out to have
difficulties; here, we applied modifications designed to stabilize their
behaviour. In both cases, we observed considerable changes as a result of
these modifications.

In four-dimensional simplicial quantum gravity, we followed a suggestion,
based on a study of the effective action for the conformal factor, that
the model might need a minimum amount of matter for a continuum limit to
exist. We studied the model with a variable amount of vector
fields, and indeed found that starting with $n_V = 3$ the phase structure
becomes significantly altered. Namely, the strong coupling phase, which
in the pure gravity model is dominated by essentially one-dimensional
branched polymers, is replaced by a four-dimensional universe with a
negative string susceptibility exponent. We were also able to show that
the same effect can be achieved by the inclusion of an additional measure
term $\prod o_i^\beta$ if $\beta$ is taken to be large enough and negative.
This is the first time that such a fundamental deviation from the pure
gravity behaviour has been observed.

The new so-called crinkled phase that is created in this way still faces
some uncertainties that we were unable to resolve on the comparatively
small surfaces examined here; in particular, the pseudo-critical value of
the coupling constant $\kappa_2^c (n_4)$ grows with the system size in a
way that makes it difficult to decide whether it will converge to a finite
value in the large $n_4$ limit. This question can only be addressed in
further studies of much larger systems. Alternatively, an examination of a
model that includes both vector and fermion fields also seems promising.

This project also allowed us to test the applicability of the strong
coupling expansion as a non-statistical method to four-dimensional
simplicial quantum gravity. We were able to show that for large enough
values of the coupling constant, the predictions for the susceptibility
exponent from the strong coupling expansion are well compatible
with the numerical results. This allowed us to perform much more thorough
studies of the effects of varying parameters such as $n_V$ or $\beta$,
since the series terms have to be calculated only once, after which the
re-calculation of observables for different parameter values takes almost
no extra time. A very practical problem we still face, however, is to
find more general methods for extracting observables from the expansion
while reducing the finite size effects.

The other project we explored was the discretization of quantum
string theory based on a re-formulation of the $I\!IB$ superstring, in
the hope that the addition of fermions might cure the diseases observed
in the model of bosonic strings. This did in fact turn out to be true;
our numerical studies show that the geometrical defects known as spikes
are much more strongly suppressed in the supersymmetric case, and in
particular are no longer able to create divergences in the partition
function of the model. Instead, we see a different kind of geometrical
deformation emerging that in the case of `too many' fermions again leads
to divergent contributions to the partition function. In the supersymmetric
case, however, we can show that the model does indeed become well-defined.

Differently from spikes, the `new' kind of geometrical deformation that
is enhanced by the fermions turns out to be a global phenomenon of
thin, needle-like surfaces. Being essentially one-dimensional, these
structures are `invisible' to the model's area-type action, which means
that they are not exponentially suppressed as one would usually suspect.
Instead, from power-counting arguments we were able to predict that they
should be governed by a power law with an exponent that is independent
from the system size; this we were also able to confirm numerically. While
our numerical simulations extend only to the four-dimensional version of the
model, our theoretical arguments predict the same kind of behaviour in any
dimension, with the only difference being a change in the exponent of the
power law.

A curious parallel between this model and the $d$-dimensional
generalization of the $IKKT$ matrix model, where exactly the same
power laws were observed for the distribution of the matrix eigenvalues,
led us to repeat our simulations in the context of this model. Indeed,
we were again able to show the emergence of one-dimensional structures
in the region of large eigenvalues. This is not very surprising, since the
model has a sort of matrix analogue to the area-type action that was
responsible for these structures in the surface model. A question that
still has to be investigated further is whether or not these things can
be expected to survive in the large $n$ limit, although it seems plausible
that they should.

\subsubsection{Acknowledgments}

First and foremost, I would like to thank Prof. B. Petersson for, well,
everything really; without his constant support and advice, helpful
ideas, and the occasional friendly prodding for results when I needed it,
this thesis never would have been written at all. Many thanks must also go
to Z. Burda, for being infinitely patient in the face of a thousand silly
and repeated questions. Finally, I would like to thank all the people that
I worked with over the last four years, in particular P. Bialas, S. Bilke,
G. Thorleifsson, and M. Wattenberg. It has been fun.

\backmatter

\end{document}